\documentclass[final,3p,twocolumn]{elsarticle}

\usepackage{url,amsmath,amsfonts,amssymb,mathrsfs}
\usepackage{lscape,adjustbox,comment,wrapfig,hyperref,float,soul,gincltex,bbold,mathtools,framed,placeins} 
\usepackage[rightcaption,ragged]{sidecap}
\usepackage{xcolor,bm}
\usepackage[short]{optidef}

\usepackage{float} 
\newfloat{addendumbox}{}{myfloats}

\usepackage[english]{babel}

\setlocalecaption{english}{contents}{\Large Index}
\AtBeginDocument{
  \addtocontents{toc}{\footnotesize}
  \addtocontents{lof}{\footnotesize}
}

\usepackage{tikz,pgfplots}
\usetikzlibrary{mindmap,decorations.pathreplacing,fit,arrows,shapes,positioning,automata,backgrounds,petri,trees}

\usepackage[innerleftmargin=0.2em, innerrightmargin=0.2em,skipabove=0.2em, innertopmargin=0.2em, innerbottommargin=0.2em,skipbelow=-0.05em,framemethod=tikz]{mdframed}
\mdfdefinestyle{sidefig}{backgroundcolor=white}
\usepackage{xpatch}

\usepackage{cuted} 


\allowdisplaybreaks

\def\fnom{g_{ \text{nom}}}

\def\R{\mathbb{R}}
\def\Ker{\mathscr{K}}
\def\GP{\mathcal{G}\mathcal{P}}
\def\indupo{\bar{Z}}
\def\induvar{g_{\indupo}}
\def\KL{\mathcal{K}\mathcal{L}}
\def\Tr{\text{Tr}}
\definecolor{teal}{rgb}{0.0, 0.5, 0.5}

\DeclarePairedDelimiterX{\infdivx}[2]{(}{)}{%
  #1\;\delimsize\|\;#2%
}
\newcommand{\infdiv}{\KL\infdivx}

\DeclareMathOperator*{\argmin}{argmin}
\DeclareMathOperator*{\argmax}{argmax}

\newtheorem{remark}{Remark}
\makeatletter
\def\@opargbegintheorem#1#2#3{\trivlist
   \item[]{\bfseries #1\ #2\ (#3)} \itshape}
\makeatother

\journal{Annual Reviews in Control}

\begin{document}

\begin{frontmatter}



\title{Gaussian processes for dynamics learning in model predictive control}


\author[inst1]{Anna Scampicchio\corref{cor1}}
\ead{ascampicc@ethz.ch}
\cortext[cor1]{Corresponding author}
\author[inst1]{Elena Arcari}
\author[inst1]{Amon Lahr\fnref{label2}}
\fntext[label2]{A.~Lahr is supported by  the European Union’s Horizon 2020 research and innovation programme,  Marie Skłodowska-Curie grant agreement No.~953348, ELO-X.}
\ead{amlahr@ethz.ch}
\author[inst1]{Melanie N.~Zeilinger}
\ead{mzeilinger@ethz.ch}

\affiliation[inst1]{organization={Institute for Dynamic Systems and Control, ETH Zurich},
            addressline={Leonhardstrasse 21}, 
            city={Zurich},
            postcode={8092}, 
            country={Switzerland}}

\begin{abstract}
Due to its state-of-the-art estimation performance complemented by rigorous and non-conservative uncertainty bounds, Gaussian process regression is a popular tool for enhancing dynamical system models and coping with their inaccuracies. This has enabled a plethora of successful implementations of Gaussian process-based model predictive control in a variety of applications over the last years. 
However, despite its evident practical effectiveness, there are still many open questions when attempting to analyze the associated optimal control problem theoretically 
and to exploit the full potential of Gaussian process regression in view of safe learning-based control. \\
The contribution of this review is twofold. The first is to survey the available literature on the topic, highlighting the major theoretical challenges such as (i)~addressing scalability issues of Gaussian process regression; (ii)~taking into account the necessary approximations to obtain a tractable MPC formulation; (iii)~including online model updates to refine the dynamics description, exploiting data collected during operation. The second is to provide an extensive discussion of future research directions, collecting results on uncertainty quantification that are related to (but yet unexploited in) optimal control, among others. 
Ultimately, this paper provides a toolkit to study and advance Gaussian process-based model predictive control.

\end{abstract}


\end{frontmatter}




\section{INTRODUCTION}\label{sec:introduction}
\paragraph*{Framework} The last decades have witnessed the rise of Model Predictive Control (MPC) as the prime technique to regulate multi-variable dynamical systems subject to constraints, with applications ranging from robotics to chemical reactors, building control, and power electronics (see, e.g.,~\cite{mayne_model_2014,schwenzer_review_2021}). 
Its rationale consists in solving online a constrained optimal control problem in a receding-horizon fashion -- that is, one computes the solution of a finite-horizon optimal control problem, applies the first input of the obtained sequence, and repeats the procedure. Stability and recursive feasibility are commonly ensured by a suitable design of the terminal set, the terminal control law, and the cost function~\cite{rawlings_model_2017}.

As the fundamental ingredient of this control strategy is the prediction of the system trajectory, a reliable dynamics model is essential. The latter is typically obtained from first-principles laws of physics -- however, these might not be able to capture complex phenomena with sufficient accuracy. For instance, when providing the mathematical description of a robotic arm, the deterministic nominal model given by the kinematics does neither capture external disturbances, such as interactions with the environment, nor take into account possible mismatches due to an inaccurate knowledge of model parameters, such as moments of inertia and centers of mass; furthermore, the model might change over time due to mechanical stress. To account for these phenomena and act in their anticipation -- thus, to apply MPC to complex systems --, control design has been complemented with \textit{learning-based} approaches~\cite{hewing_learning-based_2020,mesbah_fusion_2022}.

The first paradigm is \textit{robust} learning-based MPC, which focuses on disturbances with bounded support and seeks for best performance in the worst-case scenario~\cite{bemporad_robust_1999,kouvaritakis_model_2016}, relying for instance on set-membership estimation~\cite{milanese_optimal_1991, milanese_set_2004, canale_nonlinear_2014}.  
Such a set-up typically does not consider probabilistic models for the dynamics, and might suffer from the fact that the assumption on noise boundedness can be restrictive; furthermore, its worst-case policy might result in overly conservative controllers. 
On the other hand, \textit{stochastic} learning-based MPC works with performance in expectation, thus reducing conservativeness, and can deal with unboundedness in the noise support by considering chance-constraints~\cite{mesbah_stochastic_2016,farina_stochastic_2016}. Within this framework, it is possible to consider probabilistic models of maximal generality: typically, these are given by a nominal model plus a residual term that captures disturbances and uncertainties entering the system and is to be learned from data. 
To ensure maximal flexibility, such a term can be estimated by black-box approaches such as Gaussian process (GP) regression~\cite{rasmussen_gaussian_2006}, neural networks~\cite{piche_nonlinear_2000,ren_tutorial_2022,bao_learning-based_2023,bao_learning-_2023} or kinky inference~\cite{limon_learning-based_2017,manzano_robust_2018}.

\paragraph*{Scope} In this paper, we focus on nonlinear, stochastic MPC, in which unknown dynamics are learned using Gaussian process regression.  
We choose such a non-parametric learning method because, compared with other estimation techniques, 
it has been shown to return state-of-the-art performance in function estimation, also in the low-data regime, while requiring little design intervention. Furthermore, being rooted in Bayesian statistics, it comes automatically with a rigorous uncertainty quantification paradigm, which can be explicitly incorporated in the optimal control problem as chance constraints~\cite{paulson_nonlinear_2018}. 
\begin{mdframed}[backgroundcolor=yellow!10]
\indent
The goal of this paper is to provide a self-contained review of Gaussian process-based model predictive control, highlighting its effectiveness, its theoretical challenges, and surveying the related topics that can be useful to advance it.
\end{mdframed}

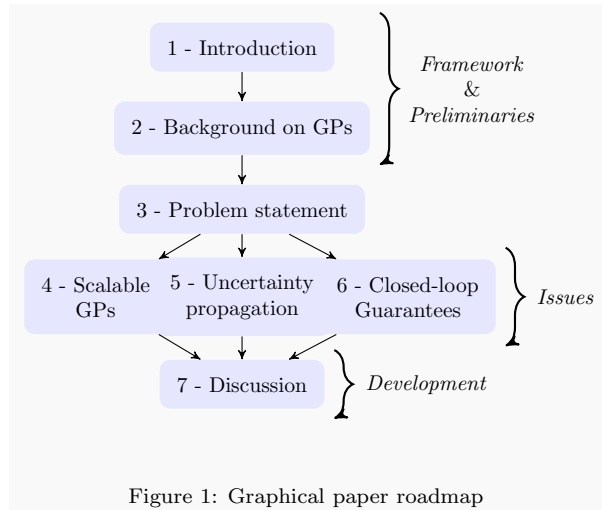
\begin{figure}[h!]
    \begin{mdframed}[backgroundcolor=gray!5, hidealllines=true]
    \centering
    \begin{tikzpicture}[->,>=stealth',shorten >=0.pt,auto, node distance=3.8em and 6.8em,>=stealth',bend angle=45,auto,scale=0.7]

\tikzstyle{block}=[rectangle,rounded corners,fill=blue!10,minimum width=3em,align=center,on grid,inner sep=0.8em]

\begin{scope}[scale=1.15,transform shape,yshift=5em]
\node [block] (1) {\ref{sec:introduction} - Introduction};
\node [block] (2) [below = of 1] {\ref{sec:background} - Background on GPs};
\node [block] (3) [below = of 2] {\ref{sec:problem} - Problem statement};

\node [block] (5) [below = of 3,yshift=-0.3em] {\ref{sec:uncertaintyprop} - Uncertainty\\ propagation};

\node [block] (4) [left = of 5] {\ref{sec:scalableGPs} - Scalable\\ GPs};

\node [block] (6) [right = of 5, xshift = 0.9em] {\ref{sec:safetyguar} - Closed-loop\\ Guarantees};

\node [block] (7) [below = of 5, yshift=-0.3em] {\ref{sec:discussion} - Discussion};

\path (1) edge node {} (2);
\path (2) edge node {} (3);
\path (3) edge node {} (4);
\path (3) edge node {} (5);
\path (3) edge node {} (6);
\path (4) edge node {} (7);
\path (5) edge node {} (7);
\path (6) edge node {} (7);

\node[fit=(1)(2)](group){};
\draw[line width=.8pt,black,decorate,decoration={amplitude=7pt,brace}]
  (group.north east) -- (group.south east);
  \node[right=of group,anchor=center,xshift=-2.5em,align=center]{\textit{Framework} \\ \& \\ \textit{Preliminaries}};
\node[fit=(4)(5)(6)](group){};

\draw[line width=.8pt,black,decorate,decoration={amplitude=7pt,brace}]
  (group.north east) -- (group.south east);
  \node[right=of group,anchor=center,xshift = -4em]{\textit{Issues}};

\node[fit=(7)](group){};
\draw[line width=0.8pt,black,decorate,decoration={amplitude=7pt,brace}]
  (group.north east) -- (group.south east);
  \node[right=of group,anchor=center,xshift = -2.5em]{\textit{Development}};

\end{scope}
\end{tikzpicture}
      \caption{Graphical paper roadmap}
    \label{fig:paper_roadmap}
\end{mdframed}
\end{figure}

\paragraph*{Outline} The paper structure is summarized in Figure~\ref{fig:paper_roadmap}. After a synopsis of Gaussian process regression, with particular attention to its use to learn dynamical systems (Section~\ref{sec:background}), we formulate the stochastic optimal control problem to be approximated by MPC in Section~\ref{sec:problem}. By displaying the structure of the latter control scheme, we highlight the three main challenges that are faced when dealing with nonlinear, stochastic MPC based on Gaussian process regression, and we present them in depth in the following sections. 
The first is the scalability of the learning approach, also accounting for streaming data (Section~\ref{sec:scalableGPs}); the second is how to propagate the uncertainty throughout the trajectory over the prediction horizon of the MPC problem (Section~\ref{sec:uncertaintyprop}); and the last is how to ensure closed-loop constraint satisfaction (thus, safety) and recursive feasibility in presence of chance constraints (Section~\ref{sec:safetyguar}). Note that Sections~\ref{sec:scalableGPs} and~\ref{sec:uncertaintyprop} may contain material that was derived outside the MPC literature, and thus has not yet been applied there, but is promising to advance the state-of-the-art control performance. Then, in Section~\ref{sec:discussion} we discuss the possible control pipeline considering the three issues mentioned above, surveying the methods available since the first study~\cite{murray-smith_adaptive_2003}, and pointing out their strengths and limitations. Next, we discuss the practical implementation of the MPC problem with dynamics learned with (scalable) Gaussian process regression. Furthermore, we also provide insights into alternative models for the dynamics and promising, alternative uncertainty quantification paradigms.

\paragraph*{Topics not covered} Finally, we would like to point out that in this review paper, we focus on Gaussian process regression just for learning the dynamics, and do not dwell on their use for learning the constraints (e.g., for safe active learning) or the cost function (for problems of inverse optimal control or Bayesian optimization), nor for approximating MPC laws -- we refer to~\cite{mesbah_fusion_2022} for insights on these issues. Finally, we also do not review other control strategies that deploy Gaussian process regression, such as model-based reinforcement learning (\cite{kober_reinforcement_2013,polydoros_survey_2017,garcia_comprehensive_2015,moerland_model-based_2023,plaat_high-accuracy_2023,deisenroth_pilco_2011,kamthe_data-efficient_2018,berkenkamp_safe_2017} and reviewed in~\cite{brunke_safe_2022}), sliding mode control~\cite{lima_sliding_2020}, model reference control~\cite{kim_path_2017}, adaptive control~\cite{chowdhary_bayesian_2015}, internal model control~\cite{gregorcic_gaussian_2005}, gain scheduling~\cite{azman_fixed-structure_2009}, and feedforward nonlinear control~\cite{nguyen-tuong_computed_2008}. 

\section{BACKGROUND ON GAUSSIAN PROCESS REGRESSION}\label{sec:background}
\begin{figure*}[h!]
\begin{mdframed}[backgroundcolor=gray!5, hidealllines=true]
    \centering
\begin{tikzpicture}[mindmap, concept color=blue!20]
\begin{scope}[scale=0.7,transform shape]
 \node [concept] {\ref{sec:background} Background on GPs}
    child[concept color=red!10,grow=180] {node[concept] {\ref{subsec:overview} Overview of static GPs}
    child[grow=45] {node[concept] {\ref{subsub:frameworkGP} Framework}}
    child[grow=90] {node[concept] {\ref{subsub:ideaGP} The idea }}
    child[grow=150] {node[concept] {\ref{sec:subsuncertainty} Bayesian inference and uncertainty quantification}}
    child[grow=205] {node[concept] {\ref{subsub:hyperparameter} Hyper-parameters selection}}     
    }
    child[concept color=red!10,grow=0]  {node[concept] {\ref{sec:gp4dyn} GPs for dynamic systems} 
    child[grow=120] {node[concept] {\ref{subsub:iossGP} Input-output vs.~state-space models}}
    child[grow = 50] {node[concept] {\ref{subsubsec:latent_state_opt} State-space models: latent state optimization}}
    child[grow=-20] {node[concept] {\ref{subsubsec:alt_funct_learning_state_inf} State-space models: alternating function learning and state inference}}
    };
    \end{scope}
\end{tikzpicture}
\caption{Graphical overview of Section~\ref{sec:background}.}
\label{fig:mindmap_sec2}
\end{mdframed}
\end{figure*}
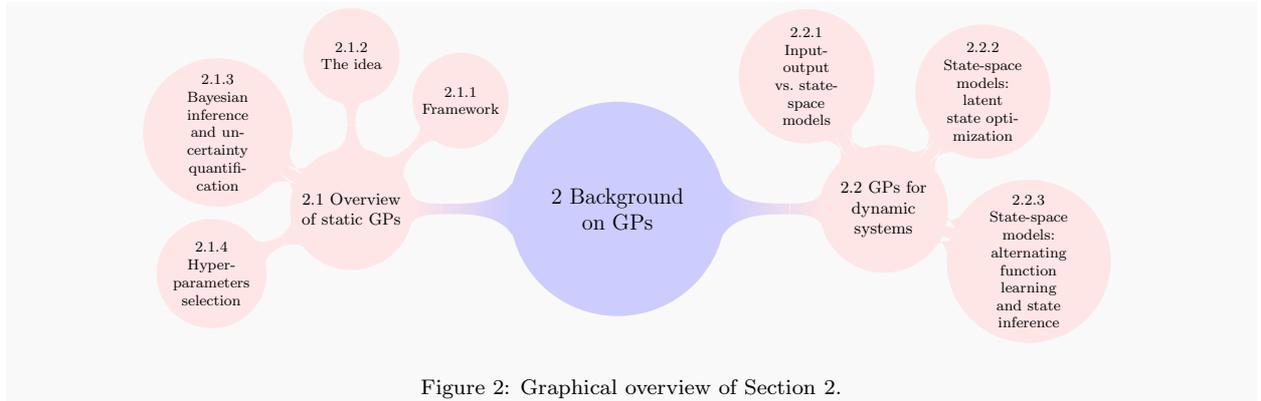

This section is aimed at providing a self-contained review of the tools for  
Gaussian process regression. After framing such an approach within the class of non-parametric, black-box methods for (static) function estimation, we present its rationale and review the computations needed to perform the desired statistical inference (i.e., predicting the value of the unknown function at arbitrary, possibly unseen input locations), also discussing the choice of hyper-parameters. Then, we present possible ways of using Gaussian processes for learning dynamical systems. In this part, we will delve into general formulations that have not yet been deployed in MPC, but show potential in view of Gaussian process-based control. 
The overall structure of the section is graphically summarized in Figure~\ref{fig:mindmap_sec2}.

\subsection{Overview of static Gaussian process regression}
\label{subsec:overview}

\subsubsection{Framework}\label{subsub:frameworkGP}
Consider a curve-fitting problem, where we assume to have access to the continuous function to be estimated, $g \colon \mathcal{Z} \subset \R^{n_z} \to \R$, through the measurement model
\begin{equation}
    y_i = g(z_i) + w_i, \label{eq:measmodgp}
\end{equation}
with $w_i \sim \mathcal{N}(0,\,\sigma^2_w)$ independent and identically distributed for each index $i$. The goal consists in estimating $g(\cdot)$ from a finite data-set $\{(z_i,y_i)\}_{i=1}^N$ generated according to~\eqref{eq:measmodgp}.\\

 To avoid ill-posedness of such a problem, a first approach consists in specifying a parametric
structure for the unknown function, typically reformulating it through linear regression to be solved via least squares~\cite{hastie_elements_2009,ljung_system_1986}. Such a solution yields an interpretable model, and is computationally appealing, because the number of features, i.e., the number of parameters to estimate, is usually much lower than the data-set cardinality; however, choosing them is usually an intricate problem, especially if not enough information on $g(\cdot)$ is available. 

An alternative to alleviate the burden of model selection consists in non-parametric approaches, determining $g(\cdot)$ as an element of an infinite-dimensional hypothesis space. Among these, a very popular class is the one of kernel methods, also known as kriging in geostatistics~\cite{matheron_principles_1963}, encompassing spline models~\cite{wahba_spline_1990}, 
regularization networks and support vector machines~\cite{poggio_networks_1990,evgeniou_regularization_2000, suykens_least_2002,scholkopf_learning_2018}, as well as Gaussian process regression~\cite{rasmussen_gaussian_2006}. 
In this paper, we focus on the latter, which was introduced in the statistics community by~\cite{ohagan_curve_1978}, and gained attention in the machine learning one after~\cite{neal_bayesian_1996} proved that they can be regarded as neural networks of infinite width. While most of the kernel methods treat the unknown function as a deterministic object, relying on the theory of Reproducing Kernel Hilbert Spaces~(RKHS) to define the hypothesis space and ensure well-posedness~\cite{aronszajn_theory_1950,berlinet_reproducing_2004,saitoh_theory_2016}, Gaussian process regression is grounded into Bayesian statistics~\cite{gelman_andrew_bayesian_2015} and allows for an automatic quantification of the uncertainty, as will be later discussed in this section. Further details on the connection with estimation in 
RKHSs is deferred to the textbox titled \textit{Kernel-based methods: the view from RKHSs} on page~\pageref{sec:boxkernel}.

\subsubsection{The idea}\label{subsub:ideaGP} The rationale of Gaussian process regression consists in modelling the unknown function $g(\cdot)$ as a Gaussian random field \cite{hristopulos_gaussian_2020} defined on the $z$-domain $\mathcal{Z}$ -- that is, $g(\cdot)$ is treated as an unknown realization of a stochastic process, whose prior mean and covariance operator are chosen by the user to encode available information, e.g., on spatial correlation between points in $\mathcal{Z}$. A typical choice is to introduce the prior model\footnote{Typically, the prior mean could be different from zero, but we focus on this choice in alignment with the problem formulation in Section \ref{sec:problem}.}
\begin{equation}\label{eq:prior}
    g \sim \GP(0,\, \Ker(\cdot,\cdot)),
\end{equation}
where the covariance $\Ker \colon \mathcal{Z} \times \mathcal{Z} \to \R$ is a symmetric, positive-definite operator. 
Due to its universal approximation properties~\cite{micchelli_universal_2006}, it is customarily set as the Gaussian, also known as squared-exponential, kernel
\begin{equation}\label{eq:gaussker}
    \Ker(z,z^{\prime}) = \lambda \exp\Bigg\lbrace -\frac{\|z - z^{\prime}\|^2}{2\eta} \Bigg\rbrace
\end{equation}
with scaling factor $\lambda > 0$, and $\eta >0$ denoting the kernel width. In this framework, $\lambda$ plays the role of model order: because it belongs to the positive real numbers, it is more flexibly tunable than the discrete value denoting the number of features in parametric models. There are many other possible choices for the kernel, depending on the prior information to be encoded: some popular options are the Laplacian,  Màtern, polynomial, and spline; see, e.g.,~\cite{muller_introduction_2001,duvenaud_automatic_2014} for a discussion on their properties.

\subsubsection{Bayesian inference and uncertainty quantification}\label{sec:subsuncertainty} Let us now derive the function estimate at an arbitrary set of input locations 
$Z^* = \{z_1^*,\,\cdots\,z_N^*\}$. Being a Gaussian random field implies that any vector of function evaluations is jointly Gaussian: that is, considering the input data $Z = \{z_1,\,\cdots\,z_N\}$, and denoting with $g_A$ the vector containing the function $g(\cdot)$ evaluated at every input point in the set $A \subset \mathcal{Z}$, we have that
\begin{equation}\label{eq:bayesprior}
    \begin{bmatrix}
        g_Z\\ g_{Z^*}
    \end{bmatrix} \sim \mathcal{N}\Bigg(\begin{bmatrix}
        0\\0
    \end{bmatrix},\, \begin{bmatrix}
        \Ker_{Z,Z} & \Ker_{Z,Z^*} \\ \Ker_{Z^*,Z} & \Ker_{Z^*,Z^*}
    \end{bmatrix} \Bigg),
\end{equation}
where $\Ker_{A,B}$ is such that $[\Ker_{A,B}]_{a,b} = \Ker(A_a,B_b)$, with indices $a$ and $b$ ranging in the cardinality of elements in $A$ and $B$. We can now consider the posterior distribution of $[g_Z^{\top},\, g_{Z^*}^{\top}]^{\top}$ given the observations $Y = [y_1,\,\cdots ,\, y_N]^{\top}$: its density, according to Bayes' rule, can be written as
\begin{equation}\label{eq:bayesfullpost}
    p(g_Z, g_{Z^*}|Y) = \frac{p(Y|g_Z)p(g_Z,g_{Z^*})}{p(Y)}.
\end{equation}
Note that $p(Y|g_Z,g_{Z^*}) = p(Y|g_Z)$ because the data $Y$ are conditionally independent of all other function values $g_{Z^*}$ given $g_Z$, which are the ones generating $Y$.
The likelihood $p(Y|g_Z)$, using the measurement model~\eqref{eq:measmodgp} and assuming that the noise term $w_i$ is independent from the prior on $g(\cdot)$, is distributed as
\begin{equation}\label{eq:bayeslik}
    Y|g_Z \sim \mathcal{N}(g_Z,\, \sigma_w^2 I_N).
\end{equation}
Thus, with the Gaussian distribution being self-conjugate, the posterior~\eqref{eq:bayesfullpost} is Gaussian as well. The posterior of interest, $p(g_{Z^*}|Y)$, is retrieved by marginalization:
\begin{equation}\label{eq:bayespost}
    p(g_{Z^*}|Y) = \int_{\mathcal{Z}} \frac{p(Y|g_Z)p(g_Z,g_{Z^*})}{p(Y)}dg_Z.
\end{equation}
Hence, by computing the first and second order moments of $g_Z,g_{Z^*}|Y$ and considering the components associated with $g_{Z^*}$, we obtain  
\begin{equation}\label{eq:allpost}
    g_{Z^*}|Y \sim \mathcal{N}(\mu(Z^*),\Sigma(Z^*)), \quad \text{ where} 
\end{equation}
\begin{equation}\label{eq:postparams}
\begin{cases}
    &\mu(Z^*) = \Ker_{Z^*,Z}(\Ker_{Z,Z} + \sigma_w^2I_N)^{-1}Y \\ 
    &\Sigma(Z^*) = \Ker_{Z^*,Z^*} - \\ &\qquad \quad \quad \Ker_{Z^*,Z}(\Ker_{Z,Z} + \sigma_w^2I_N)^{-1}\Ker_{Z^*,Z}. 
    \end{cases}%
\end{equation}%
The estimated function values corresponding to the inputs collected in $Z^*$ are then taken as the posterior mean $\mu(Z^*)$ in~\eqref{eq:postparams}, according both to the Maximum-A-Posteriori and Minimum Mean Squared Error criteria (which coincide in the Gaussian case: see, e.g.,~\cite{anderson_optimal_2005}).

The information carried in the posterior covariance $\Sigma(Z^*)$ in~\eqref{eq:postparams} allows for uncertainty quantification. In fact, one can extract the $\tilde{\alpha}\%$-level quantile around the estimate by integrating the corresponding posterior distribution, thus obtaining credible regions of probability $\tilde{\alpha}$ around the posterior mean.

\begin{remark}[Alternative paradigms for uncertainty quantification]\label{rmk:uncertainty_paradigms} Aside from the credible regions above mentioned, other uncertainty bounds have been derived for non-Bayesian kernel-based methods in statistical learning~\cite{cucker_learning_2007}. 
In such a framework, the estimate is assumed to belong to a Reproducing Kernel Hilbert Space (see textbox titled \emph{Kernel-based methods: the view from RKHSs}), and bounds are derived leveraging concentration inequalities~\cite{boucheron_concentration_2013}. Some results for bounded additive noise are, e.g.,~\cite{smale_learning_2007,maddalena_deterministic_2021}, while the ones for unbounded sampling can be found in~\cite{wang_optimal_2011,guo_concentration_2013}. 
These results are mostly aimed at studying the estimation performance in the asymptotic regime, finding conditions that ensure the fastest convergence to the true function with respect to the number of data points: yet, despite their great theoretical value, these bounds tend to be conservative in the finite-data regime, and sometimes they cannot even be evaluated in practice because they depend on the unknown function to be estimated. 

Other bounds have been derived using information-theoretic tools in the context of multi-armed bandits: see, e.g.,~\cite{srinivas_information-theoretic_2012,chowdhury_kernelized_2017,fiedler_practical_2021}. Their typical form reads as follows: denoting 
$\mathbb{P}_{g|Y}$ the probability associated 
with the posterior Gaussian process,
with mean and covariance presented in~\eqref{eq:postparams}; 
and defining $\beta_{\mathcal{H}}$ a constant depending on the probability level $p$ and quantities related to the hypothesis space, its associated kernel and to the additive noise; 
the bounds are expressed as
\begin{equation}
\label{eq:boundKrause} 
\mathbb{P}_{g|Y}\Big(|g(z) - \mu(z)| \leq \beta_{\mathcal{H}}\Sigma(z) \quad \forall z \in \mathcal{Z}\Big) \geq  p.
\end{equation}
These bounds typically depend on quantities that are usually unknown, such as the maximum information gain or the maximum RKHS norm -- yet, they have the advantage of being uniform over the domain. Therefore, a careful (yet, typically heuristic) choice of $\beta_{\mathcal{H}}$ returns non-conservative, yet expressive and data-dependent bounds that are very useful in practice -- see their role Gaussian process-based, robust-in-probability model predictive control in Section~\ref{subsec:RIP}.

Finally, we mention two alternative approaches returning informative, non-asymptotic bounds that are very useful for calibration. The first computes confidence bounds in the frequentist setting (yet, possibly including prior information, following the Bayesian rationale) and is presented in~\cite{baggio_bayesian_2022}; the second given by conformal prediction~\cite{shafer_tutorial_2008,  vovk_algorithmic_2022, lei_distribution-free_2014, tibshirani_conformal_2019, angelopoulos_gentle_2022,  fontana_conformal_2023, cauchois_robust_2024}, where results are derived in the distribution-free setting (thus, do not leverage available information on posterior updates) but are computationally fast to obtain. 
\label{rmk:uncertainty}
\end{remark}

\subsubsection{Hyper-parameter selection}\label{subsub:hyperparameter} Computing the posterior~\eqref{eq:allpost} requires knowledge of the values of $\lambda$, $\eta$ and $\sigma_w^2$, which are also known as hyper-parameters and are collected in a vector $\xi$. Typically, they are estimated from a subset of data $(Z_h,Y_h)$ by optimizing the marginal likelihood, i.e., by solving
\begin{align}\label{eq:maxmarglik}
    &\argmax_{\xi}\, p(Y_h|\xi) \notag \\
    = &\argmax_{\xi}\,\int_{\R^N}p(Y_h|g_{Z_h},\xi)p(g_{Z_h}|\xi)dg_{Z_h}.
\end{align}
In the case studied in this section (i.e., under i.i.d.~measurements), it turns out that ${Y_h|\xi \sim \mathcal{N}(0, \Ker_{Z_h,Z_h} + \sigma^2_w)}$. Thus,~\eqref{eq:maxmarglik} can be re-written as a negative-log-likelihood minimization problem:
\begin{align}\label{eq:negloglik}
    \argmin_{\xi} \: &Y_h^{\top}(\Ker_{Z_h,Z_h} + \sigma^2_w)^{-1}Y_h \notag \\ &+ \log\det(\Ker_{Z_h,Z_h} + \sigma^2_w). 
\end{align}
Such a program can be numerically solved by using deterministic, gradient-based optimization routines; however, the cost in~\eqref{eq:negloglik} is non-convex, and such a strategy might return unreliable results due to local minima and sensitivity to initial conditions. A powerful  alternative consists in using Markov Chain Monte Carlo approaches~\cite{gilks_markov_1995}, resorting for instance to the Empirical Bayes method~\cite{maritz_empirical_2018} and performing numerical integration on the cost in~\eqref{eq:maxmarglik}.\\

\begin{addendumbox*}[ht!]
\begin{mdframed}[backgroundcolor=gray!10]
\subsection*{ADDENDUM -- Kernel-based methods: the view from RKHSs}\label{sec:boxkernel}%
Instead of recasting the problem in the framework of Bayesian statistics through Gaussian process regression, estimating an unknown continuous function $g \colon \mathbb{R}^{n_z} \to \mathbb{R}$ from noisy data $\{z_i,y_i\}_{i=1}^N$ can be formulated as a deterministic optimization problem, which typically takes the form
\begin{equation}\label{eq:tikhonov}
    \argmin_{g\in \mathcal{H}} \frac{1}{N}
    \sum_{i=1}^N \|y_i - g(z_i)\|^2 + \gamma \|g\|_{\mathcal{H}}^2, 
\end{equation}
where $\mathcal{H}$ is an infinite-dimensional hypothesis space.
Well-posedness of this estimation problem 
is restored thanks to the regularization term weighted by the positive scalar $\gamma$, trading off data fit with complexity~\cite{tikhonov_solutions_1977}. From the formulation in~\eqref{eq:tikhonov}, the space $\mathcal{H}$ is required to be a Hilbert space, with norm induced by the inner product $\langle \cdot, \cdot \rangle_{\mathcal{H}}$, and point-wise evaluation of $g(\cdot)$ has to be properly defined to uniquely identify the function. A hypothesis space with such a structure is called a Reproducing Kernel Hilbert space (RKHS), and it is in one-to-one correspondence with a symmetric, positive semi-definite operator $\Ker \colon \mathbb{R}^{n_z} \times \mathbb{R}^{n_z} \to \mathbb{R}$ by the Moore-Aronszajn Theorem~\cite{aronszajn_theory_1950}. The kernel is key in defining point-wise evaluation: specifically, at an arbitrary input $\bar{z}$, it holds that $g(\bar{z}) = \langle g, \Ker(\bar{z},\cdot) \rangle_{\mathcal{H}}$. In this way, point-wise evaluation is a linear and continuous functional.  A key consequence of this structure is the so-called Representer Theorem~\cite{wahba_spline_1990}, stating that the solution of~\eqref{eq:tikhonov} is a linear combination of kernel sections centered at the input locations $\{\Ker(z_i,\cdot) \}_{i=1}^N$. Consequently, calling $Z$ be the set of input locations and $Y$ the vector of observations, problem~\eqref{eq:tikhonov} can be equivalently written as
\begin{equation}\label{eq:tikhonovfindim}
    \argmin_{c\in\mathbb{R}^N} \|Y - \Ker_{Z,Z}c\|^2 + \gamma c^{\top}\Ker_{Z,Z}c.
\end{equation}
This is a finite-dimensional convex program, whose solution is computable in closed form as $\hat{c} = (\Ker_{Z,Z} + \gamma I_N)^{-1}Y$. Evaluating the estimated function at a new (set of) input locations $Z^*$ corresponds to computing $\Ker(Z^*,Z)\hat{c}$, which is the same result as the posterior mean in~\eqref{eq:postparams} by choosing $\gamma=\sigma_w^2$.  Such a correspondence is further explored in~\cite{kimeldorf_correspondence_1970,aravkin_connection_2015} and~\cite{kanagawa_gaussian_2018}, where it is discussed that, while the posterior mean of the Gaussian process with covariance $\Ker$ belongs to the RKHS with the same kernel, its general sample paths do not, with probability one (a result known as Driscoll's zero-one law~\cite{driscoll_reproducing_1973}).\\

The formulation as a deterministic decision problem based on RKHSs, typical in statistical learning~\cite{cucker_mathematical_2002,cucker_learning_2007}, does not have an intrinsic description of the uncertainty. In fact, solving~\eqref{eq:tikhonov} returns just the estimate (corresponding to the mean of a Gaussian posterior distribution), while the Bayesian framework automatically returns also the posterior covariance, which captures the uncertainty around it. Moreover, as reviewed in Section~\ref{sec:subsuncertainty}, the stochastic set-up also allows for a principled selection of hyper-parameters with the Empirical Bayes approach~\cite{maritz_empirical_2018}. However, results on uncertainty bounds outside Bayesian statistics are available, and are mentioned in Remark~\ref{rmk:uncertainty} as well as in Sections~\ref{sec:safetyguar} and~\ref{sub:alternativeuncertaintyprop}.
\end{mdframed}%
\end{addendumbox*}

\subsection{Gaussian processes for dynamical systems}\label{sec:gp4dyn}

 We now review models and methods where Gaussian process regression is deployed for representing dynamical systems. The topic can be framed in that of time series analysis, and has also been treated in~\cite{hartikainen_sequential_2013,mchutchon_nonlinear_2014, frigola-alcalde_bayesian_2015, kocijan_modelling_2016,svensson_machine_2018,sarkka_use_2021}. 
 In the following, we will present input-output and state-space models, and focus mostly on the latter because of their relevance in the context of MPC. For the same reason, we will consider the discrete-time case only -- however, the tools to study the continuous-time one remain largely the same.  

\subsubsection{Input-output vs.~state-space models}\label{subsub:iossGP}
A first option to describe a dynamical system is the Nonlinear, Auto-Regressive with eXogenous input (NARX) model~\cite{billings_nonlinear_2013}, extending the framework of~\cite{akaike_autoregressive_1971}.  Given observed output and input variables, denoted by $y_i$ and $u_i$ at time $i$, a NARX model is given by
\begin{equation}\label{eq:narx}
y_{i} = g_{NARX}(y_{i-1},...,y_{\tau_y}, u_{i-1},...,u_{\tau_u}) + w_i,
\end{equation}
where $w_i$ is an additive disturbance that drives the system dynamics, and $\tau_u$, $\tau_y$ are the lags in the inputs and outputs, respectively. In this situation, standard static Gaussian process regression can be applied to estimate the map $g_{NARX}(\cdot)$: see, e.g.,~\cite{gregorcic_gaussian_2002, girard_gaussian_2002,girard_learning_2003,kocijan_dynamic_2005}, and~\cite{groot_multiple-step_2011,gutjahr_sparse_2012,krivec_simulation_2021} for solutions that take into account scalability issues (see Section~\ref{sec:scalableGPs}). Note also that the same procedure can be applied to similar model structures such as the nonlinear versions of the Finite Impulse Response (NFIR), Output Error (NOE), or AutoRegressive, Moving Average with Exogenous Inputs (NARMAX) models~\cite{kocijan_modelling_2016,sarkka_use_2021}.\\  

Alternatively, one can formulate a state-space model. In its general form, it reads as
\begin{equation}
    \begin{cases}
        x_{i+1} = f(x_i,u_i) + v_i\\
        y_i = g_{SS}(x_i) + w_i.
    \end{cases}
    \label{eq:statespacegen}
\end{equation}
Here, the respective transition and emission maps $f(\cdot)$ and $g_{SS}(\cdot)$ can be known or unknown, and Gaussian process regression can be deployed to learn either or both of them. Typically, one has that $g_{SS}(\cdot)$ is known, and only $f(\cdot)$ has to be estimated from data; however, the case in which (also) the emission map is unknown can be equivalently re-written in terms of transition map estimation for an equivalent system with augmented state~\cite{frigola-alcalde_bayesian_2015}.

Model~\eqref{eq:statespacegen} could be simplified if the state could be perfectly measured, leading to a $g_{SS}(\cdot)$ equal to the identity and the absence of measurement noise $w_i$ (see Remark~\ref{rmk:gp_digression}). However, when the state-space model takes the general form~\eqref{eq:statespacegen}, it comes with two intertwined challenges: learning of the unknown transition and emission maps, and state inference. The proposed approaches to tackle them can be roughly divided into two categories: those that treat latent states as variables to be optimized (e.g., as hyper-parameters), and those that alternate learning and state inference. Such a division can also be read in terms of marginalization over the unknown maps and over the state sequence, respectively~\cite{frigola-alcalde_bayesian_2015}. We discuss these two choices in the following subsections~\ref{subsubsec:latent_state_opt} and~\ref{subsubsec:alt_funct_learning_state_inf}. \\

Finally, we would like to point out that auto-regressive and state-space representations have also been combined in the general framework of latent autoregressive Gaussian processes~\cite{mattos_latent_2016}, which were devised to address the feedback of noisy observations into the dynamics and enable uncertainty propagation (a thorough discussion on this topic is deferred to Section~\ref{sec:uncertaintyprop}). Building upon~\cite{damianou_deep_2013}, such a framework has been then further generalized to deep recurrent models~\cite{mattos_recurrent_2015}. 

\subsubsection{State-space models: Optimizing latent state variables}
\label{subsubsec:latent_state_opt}
The first class of methods stems from the latent variable models approach proposed in~\cite{lawrence_gaussian_2003,lawrence_probabilistic_2005}, which focused on a low-dimensional representation of data and can be seen as a particular case of Gaussian process autoencoders~\cite{jiang_gaussian_2014,takano_gaussian_2019}.
Originally, latent states were modeled as independent; later, such a set-up was generalized to encode correlation and enhance their representation. Works in this direction are~\cite{lawrence_local_2006}, which was deployed for a WiFi-SLAM in~\cite{ferris_wifi-slam_2007};~\cite{lawrence_hierarchical_2007}, where the latent states are treated as a temporal Gaussian process~\cite{hartikainen_kalman_2010}; and~\cite{wang_gaussian_2008}, where Markovianity in the state sequence is taken into account. Instead of treating states as hyper-parameters, possibly incurring over-fitting,~\cite{titsias_bayesian_2010} proposed a variational inference framework (see Section~\ref{sub:indupost}), in which pseudo-points are introduced to facilitate the marginalization of the transition function; such work was then refined in~\cite{damianou_variational_2011,damianou_variational_2016}, where the prior on the latent states was modeled as a temporal Gaussian process, and in~\cite{souza_learning_2021}, where general kernels can be dealt with thanks to the unscented transform. All of these contributions do not include control inputs, and predictions are mostly intended for simulation of the outputs (e.g., for human motion animation~\cite{wang_gaussian_2008}), not of states. To overcome this issue,~\cite{ko_learning_2011} extended the latent variable model approach to estimate the states and then apply Bayesian filtering~\cite{ko_gp-bayesfilters_2008}.

\subsubsection{State-space models: Alternating function learning and state inference}
\label{subsubsec:alt_funct_learning_state_inf}
The second class of approaches, which has a stronger focus on learning the unknown maps $f(\cdot,\cdot)$ or $g_{SS}(\cdot)$ in~\eqref{eq:statespacegen}, originates in works aimed at extending Bayesian techniques -- such as the Extended Kalman Filter (EKF)~\cite{shachter_linear_1990,simon_optimal_2006}, the Unscented Kalman Filter (UKF)~\cite{wan_unscented_2000,julier_unscented_2004}, the Assumed Density Filter (ADF)~\cite{maybeck_chapter_1979,brigo_approximate_1999}, and the Particle Filter (PF)~\cite{gordon_novel_1993} -- to non-parametric models. In particular, see~\cite{ko_gaussian_2007} for the UKF,~\cite{ko_gp-bayesfilters_2008} for EKF and PF, and~\cite{deisenroth_analytic_2009} for ADF, which was also investigated for smoothing in~\cite{deisenroth_robust_2012} and studied in the broader framework of expectation propagation in~\cite{deisenroth_expectation_2012}. Overall, approximations used in these works for inference can be grouped into three main branches: linearization using Taylor expansions, exact moment matching, and particle representations (see also Section~\ref{sec:gaussapproxup}). In all of these contributions, it is assumed that ground-truth observations of the latent states are available to perform system identification, which might be restrictive~\cite{turner_state-space_2010}. In case direct state measurements are not available, the dependence on the unknown maps needs to be integrated out to perform state inference. Proposed methods are typically based on Expectation Maximization~\cite{dempster_maximum_1977}, alternating between latent state inference (E-step) and model learning (M-step)~\cite{roweis_learning_2001}. To ensure tractability, they can rely on approximated analytic forms~\cite{turner_state-space_2010,mchutchon_nonlinear_2014} or sampling-based approaches~\cite{schon_system_2011,frigola_identification_2014}. A full Bayesian treatment of the problem can be found in~\cite{frigola_bayesian_2013}: the proposed particle Markov Chain Monte Carlo method~\cite{schon_sequential_2015} is capable of capturing temporal correlations, but suffers from computational inefficiency. In general, the state transition map in the methods outlined above is described using full Gaussian processes, which (as discussed in Section~\ref{sec:scalableGPs}) are encumbered by high computational complexity. A first way to address this issue consists in approximating the Gaussian process using truncated orthogonal basis functions expansions: works in this direction are, e.g.,~\cite{svensson_nonlinear_2015,svensson_computationally_2016,svensson_flexible_2017}. Another approach consists in using variational inference (discussed in Section~\ref{sub:indupost}) to approximate the posterior of the states. Some first attempts include assumed independence between the latent-state trajectory and the transition maps evaluation or pseudotargets in the variational posterior~\cite{frigola_variational_2014, eleftheriadis_identification_2017}: the resulting analytical tractability comes at the price of a less expressive model. In contrast, works that explicitly describe correlation between states and the transition map are~\cite{ialongo_closed-form_2017, doerr_probabilistic_2018,ialongo_overcoming_2019,lindinger_variational_2023}. Another usage of variational inference for state-space nonlinear system identification can be found in~\cite{courts_variational_2023}.

   \section{PROBLEM FORMULATION}\label{sec:problem}
Consider the discrete-time dynamical system described by the difference equation
\begin{equation}\label{eq:model}
x_{i+1} = \fnom(x_i,u_i) +  B_d g(x_i,u_i) + v_i,
\end{equation}
where $x_i \in \R^{n_x}$ and  $u_i \in \R^{n_u}$ are the system state and the control input at time index $i$, respectively. The process noise $v_i$ is assumed to be independent and identically distributed in time: typically, $v_i \sim \mathcal{N}(0, \sigma^2_v I_{n_x})$.

The transition map encompasses two terms: the first is a deterministic and known part $\fnom \,: \R^{n_x}\times \R^{n_u} \rightarrow \R^{n_x}$ encoding the nominal model; the second entails an unknown function $g \colon  \R^{n_x}\times \R^{n_u} \to \R^{n_d}$ living in a subspace spanned by the columns of the known $n_x\times n_d$ matrix $B_d$. The presence of the nominal term $\fnom$ finds applications, e.g., in robotics and mechanical systems, where dynamic or kinematic relations provide a first-principles description of the system, and $B_dg(\cdot)$ captures inaccuracies in model parameter specifications or interactions with the environment (see, e.g.,~\cite{ko_gaussian_2007, arcari_bayesian_2023}). Equation~\eqref{eq:model} includes also the case in which the dynamics is fully unknown, yielding $\fnom(\cdot,\cdot) \equiv 0$ and $B_d = I_{n_x}$. Moreover, available information about states (e.g., those related through time derivatives) can be embedded in $g(\cdot,\cdot)$ to simplify the estimation task, as proposed in~\cite{hall_modelling_2012}.

We consider the setting in which the unknown function $g(\cdot)$ is to be learned from data. These are gathered according to the measurement model
\begin{equation}\label{eq:measmod}
    y_i \doteq B_d^{\dagger}(x_{i+1} - \fnom(x_i, u_i)) = g(x_i,u_i)+ w_i 
\end{equation}
where $(\cdot)^\dagger$ denotes the Moore-Penrose pseudo-inverse, and 
$w_i = B_d^{\dagger}v_i$, yielding ${w_i \sim \mathcal{N}(0, B_d^{\dagger}\Sigma_v B_d)}$ by the assumptions on model~\eqref{eq:model}. Introducing the vector $z_i = [x_i^{\top}\; u_i^{\top}]^{\top}$ for any time index $i$, the overall data-set is then a set of pairs $\{(z_i,\: y_i) \}_{i=1}^N$. In this paper, the function $g(\cdot)$ will be estimated by applying Gaussian process regression. 
Since the image of $g(\cdot)$ is generally multi-dimensional, the most common modeling choice is to perform estimation on the single output components independently. Thus, in this paper Gaussian process regression will be applied to data-sets $\{(z_i, y_{ij})\}_{i=1}^N$ for each respective component $j=1,\dots,n_d$ according to the theory presented in Section \ref{sec:background}; alternative choices are discussed in Section~\ref{sub:multioutputGP}.

\begin{remark}[On the model choice]\label{rmk:gp_digression}
The scenario that is of interest in this paper, and is common in the learning-based MPC literature, is a particular case that connects both the NARX and the state-space representations outlined in Section \ref{sec:gp4dyn}. Specifically, we assume that states can be fully observed and are noise-free: in this way, the dynamics description in~\eqref{eq:model} can be regarded as a state-space model with emission map equal to the identity and without measurement noise, or as a NARX model with lags $\tau_u = \tau_y = 1$. Finally, we point out that in the main body of the paper we do not consider online learning: we discuss this aspect in Sections \ref{sec:online} and \ref{subsec:scalableonlinempc}.
\end{remark}

Overall, the problem at the core of this paper can be informally stated as follows.
\begin{mdframed}
 {($\mathcal{P}$):}
\emph{Given the dynamical system in~\eqref{eq:model}, the goal consists in designing a feedback control law $u_i = \pi_i(x_i)$ that minimizes a given cost over an arbitrary, finite time horizon $\bar{T}$, while enforcing that states and inputs in closed-loop satisfy constraints $h(x_i,u_i) \leq 0$, where $h: \mathbb{R}^{n_x}\times \mathbb{R}^{n_u} \rightarrow \mathbb{R}^{n_h}$, with a specified probability.}
\end{mdframed}

Constraints on state and input can, e.g., encode physical limitations as well as ensure that states do not exit a ``safe" region of the domain. Furthermore, since the system to be controlled is affected by noise and uncertainty in the dynamics, constraints are to be satisfied with a user-chosen probability level. Problem $(\mathcal{P})$ can then be formulated as a stochastic optimal control problem, which reads as follows:
\begin{subequations}
\begin{alignat}{2}
    &\!\text{minimize}_{\{ \pi_i \}} \quad \mathbb{E}\Big[\bar{\mathcal{L}}_{\bar{T}}(x_{\bar{T}}) + \sum_{i=0}^{\bar{T}-1}\bar{\mathcal{L}}_i(x_i,u_i)\Big] & &\\%
    &\text{subject to} & & \notag\\
    &x_{i+1} = \fnom(x_i,u_i) + B_d g(x_i,u_i) + v_i, & & \label{eq:OCdynamics}\\ 
    &u_i = \pi_i(x_i), \label{eq:OCpolicy} & & \\
    & \mathbb{P}(h_j(x_i,u_i) \leq 0, \, \forall i \geq 0) \geq p_j \;\: {\scriptstyle \forall j=1,...,n_h,} \label{eq:OCconstr} & & \\
    & x_0 = \bar{x}, \label{eq:OCinitialcond} & &
\end{alignat}%
\label{eq:SOCP}%
\end{subequations}%
where the constraints that appear in the problem are: in~\eqref{eq:OCdynamics} the system stochastic dynamics~\eqref{eq:model}; in~\eqref{eq:OCpolicy}, the adopted policy structure; in~\eqref{eq:OCconstr}, the system chance-constraints; and in~\eqref{eq:OCinitialcond} the initial condition. \\

The stochastic optimal control problem in equation~\eqref{eq:SOCP} is challenging to solve for several reasons, including: optimizing over general feedback laws; dealing with the presence of $g(\cdot)$, which is a stochastic process 
depending on both the \textit{random} state and input variables; 
and, finally, ensuring satisfaction of state and input chance-constraints jointly in time. In this paper, we consider the problem of computing an approximate solution to~\eqref{eq:SOCP} by formulating it as an MPC problem to be solved in a receding-horizon fashion. The resulting problem is structured, for each instance of time $k$ at which it is solved, as follows:
\begin{mini!}|l|[3]{\{\pi_{i|k}\}}{\! \! \! \! \mathbb{E}\Big[
    \mathcal{L}_T(x_{T|k}) + \sum_{i=0}^{T-1} \mathcal{L}_i(x_{i|k},u_{i|k})\Big] \label{eq:cost}}{\label{eq:optimization_problem}}{}
\addConstraint{{x_{i+1|k} = \fnom(x_{i|k},u_{i|k}) +  B_d g(x_{i|k},u_{i|k}) + v_{i|k}}   \label{eq:dynamicsmpc}}{}{}
\addConstraint{ u_{i|k} = \pi_{i|k}(x_{i|k}) \label{eq:policy}}{}{}
\addConstraint{ \text{system chance-constraints} \label{eq:constr}}{}{}
\addConstraint{  x_{0|k} = x_k\label{eq:initialcond}.}{}{}
\end{mini!}

We will detail the formulation of~\eqref{eq:optimization_problem} in Section~\ref{sub:solvingMPC}. The core idea consists in solving~\eqref{eq:optimization_problem} at each $k$, applying the obtained $\pi_{0|k}(x_{0|k})$ to the system, and then repeating the procedure with the new initial state $x_{k+1}$. Note that time horizons $\bar{T}$ and $T$ might differ: typically, $\bar{T}$ is large, and $T \ll \bar{T}$; moreover, also the MPC cost can differ from the one in the original optimal control problem to ensure, for instance,  differentiability.

Even after choosing the structure for the policy $\pi_{i|k}$, the problem as stated in~\eqref{eq:optimization_problem} is still not tractable due to the uncertainty in the dynamics model~\eqref{eq:model}. 
To address this fact, we discuss in Section~\ref{sec:uncertaintyprop} how to deploy Gaussian process regression to rewrite~\eqref{eq:dynamicsmpc} and simulate the predicted state sequence. The adopted model choice implies the following intertwined issues that will be addressed in the remainder of the paper:
\begin{itemize}
\item how to reconcile the high computational complexity of full Gaussian process regression with the stringent computational times required in MPC applications -- in other words, what are the available approximation methods for scalable Gaussian process regression; how to deal with streaming data (Section~\ref{sec:scalableGPs}); 
    \item how to evaluate a Gaussian process on uncertain inputs, and how to perform uncertainty propagation over the prediction horizon  (Section~\ref{sec:uncertaintyprop});
    \item how to choose the feedback policy in~\eqref{eq:policy}; how to rewrite state and input constraints in~\eqref{eq:constr} so that they capture the closed-loop guarantees of the optimal control problem~\eqref{eq:SOCP}; which actual safety guarantees  can be obtained (Section~\ref{sec:safetyguar}).
\end{itemize}

\section{SCALABLE METHODS FOR GAUSSIAN PROCESS REGRESSION}\label{sec:scalableGPs}
 \begin{figure*}[h]
 \begin{mdframed}[backgroundcolor=gray!5, hidealllines=true]
    \centering
\begin{tikzpicture}[mindmap, concept color=blue!20]
\begin{scope}[scale=0.7,transform shape]

\node [concept] {\ref{sec:scalableGPs} Scalable GPs}
    child[concept color=red!10,grow=210] {node[concept] {\ref{sec:subsetdata} Subset of data} 
    child [grow=125] {node[concept] {\ref{subsec:subsetdatasingle} Single model with pruned data-set}}
    child [grow=190] {node[concept] {\ref{sec:multiple_models} Multiple models: expert-based methods}}}
    child[concept color=red!10,grow=150]  {node[concept] {\ref{sec:methodsinducing} Methods based on inducing variables}
    child [grow=125] {node[concept] {\ref{sec:induprior} Approximating the prior}}
    child [grow=190] {node[concept] {\ref{sub:indupost} Approximating the posterior}}
    }
    child[concept color=red!10,grow=90] {node[concept] {\ref{sec:finitedim} Finite-dimensional representation of the kernel operator}
    child [grow=180] {node[concept] {\ref{sec:ssgp} Sparse-spectrum GPs}}
    child [grow=0] {node[concept] {\ref{sub:covserexp} Covariance series expansion}}
    }
    child[concept color=red!10,grow=10] {node[concept] {\ref{sec:gpcomptricks} Computational techniques}
   child [grow=45] {node[concept] {\ref{sec:nystrom} Nystr\"om approximation}}
    child [grow=0] {node[concept] {\ref{sub:conjugategradient} Conjugate gradient method}}    
    child [grow=-45] {node[concept] {\ref{sub:parallel}  Parallelization}}  
    }
    child[concept color=red!10,grow=-40] {node[concept] {\ref{sec:online} Online learning}};
   
    \end{scope}
    
\end{tikzpicture}
\caption{Graphical overview of Section~\ref{sec:scalableGPs}.}
\label{fig:mindmap_sec4}
\end{mdframed}
\end{figure*}
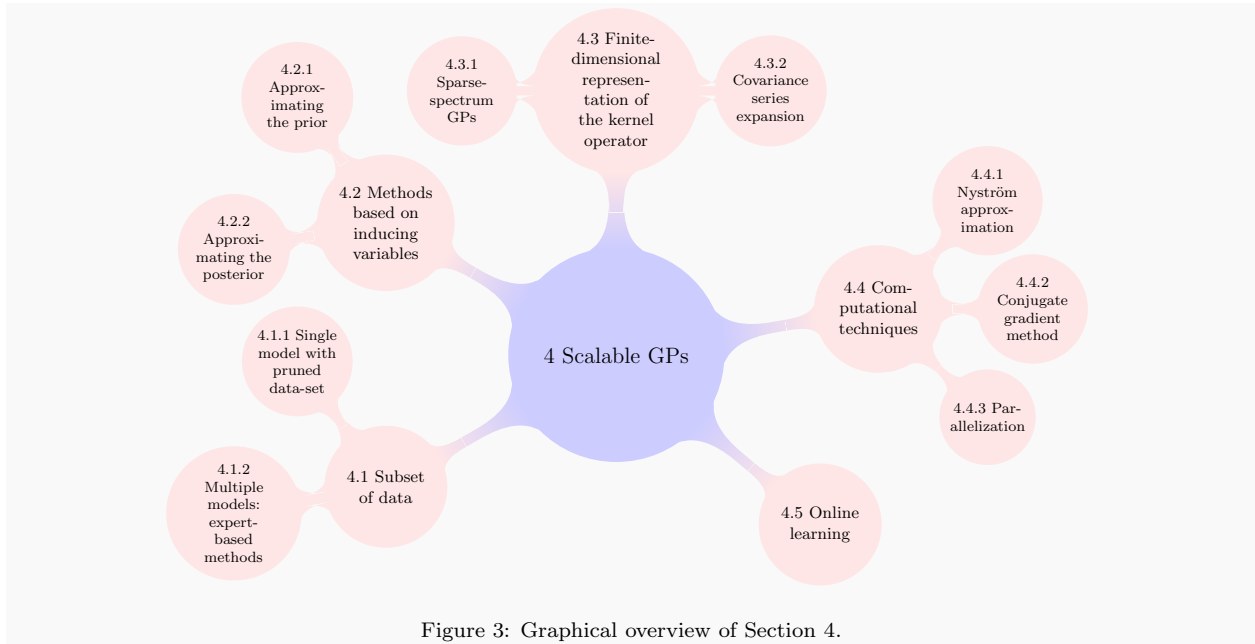
The benefits of Gaussian process regression are manifold: it requires very little user design choice compared to parametric models,  
relies mostly on data, and is naturally endowed with uncertainty quantification around the estimates. 
However, exactly evaluating and differentiating the expressions for the posterior mean and covariance in~\eqref{eq:allpost} for inference, 
as well as the log-marginal likelihood~\eqref{eq:negloglik} 
for training, 
incurs a restrictive computational cost with increasing data-set cardinality~$N$. 
Indeed, the computational complexity of a direct implementation is given by $\mathcal{O}(N^3)$, i.e., the cost of the required full-rank Gram matrix factorization, 
which becomes prohibitive 
even for a moderate number of data points.
To fully factorize the positive definite Gram matrix, usually a Cholesky decomposition is employed, which computes a lower triangular matrix $L$ such that $L L^T = \Ker_{Z,Z}$. 
Based on the precomputed Cholesky factors, 
inference according to~\eqref{eq:allpost} may be performed in $\mathcal{O}(N^2)$ operations, while
the required 
computations
for training, i.e., \mbox{$\log \det \left( \Ker_{Z,Z} \right)$} and \mbox{$\frac{\partial}{\partial \xi} \log \det \left( \Ker_{Z,Z} \right) = \mathrm{tr} \left( \Ker_{Z,Z}^{-1} \frac{\partial \Ker_{Z,Z}}{\partial \xi} \right)$}, require $\mathcal{O}(N)$ operations. 
Note that for predictions with a fixed data-set and hyper-parameters, it suffices to pre-factorize the Gram matrix once, as it does not depend on the test points.
For the posterior mean, the computational and memory footprint can further be reduced by pre-computing \mbox{$V^* = (\Ker_{Z,Z} + \sigma_w^2I_N)^{-1} Y$}, 
which allows for evaluating 
\mbox{$\mu(Z^*) = \Ker_{Z^*,Z} V^*$}  
with a single vector product. 
However, when updating kernel parameters during hyper-parameter optimization, or when data points are added during online inference, it is generally required to 
refactorize the Gram matrix.
Additionally, with an increasing number of input locations $N$, the Gram matrix tends to become ill-conditioned, potentially leading to numerical errors when attempting to factorize it fully.\\

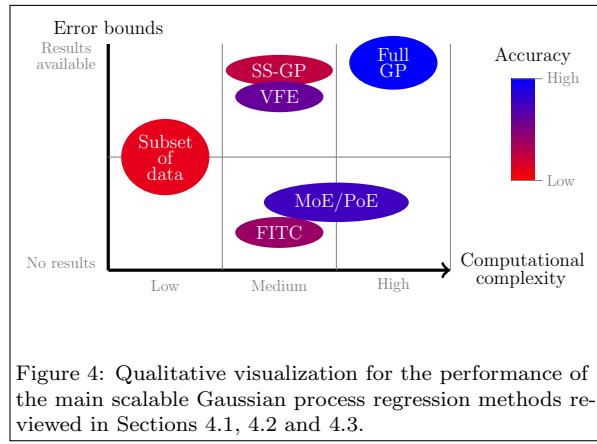
\begin{figure}[!t]
 \begin{mdframed}
    \centering
\begin{tikzpicture}[every node/.style={scale=1},domain=0:10]
\begin{scope}[scale=0.5,transform shape,framed]
  \draw[->, very thick] (0,0) -- (9,0) node[right,align=center, text width=9em,inner sep=0.7em] { \Large Computational complexity};
  \draw[-, very thick] (0,0) -- (0,6) node[above,inner sep=0.7em] { \Large Error bounds}; 
  
  \draw[-,very thin,color=gray] (3,0) -- (3,6);
  \draw[-,very thin,color=gray] (6,0) -- (6,6);
  \draw[-,very thin,color=gray] (9,0) -- (9,6);
  
  \draw[-,color=gray] (0,3) -- (9,3);

  \node[anchor=center,color=gray] at (1.5,-0.4) {\large Low};
  \node[anchor=center,color=gray] at (4.5,-0.4) {\large Medium};
  \node[anchor=center,color=gray] at (7.5,-0.4) {\large High};
  
  \node[anchor=east,color=gray, text width=5em, align=center] at (-0.1, 5.7) {\large Results available};
  \node[anchor=east,color=gray] at (-0.2,0.2) {\large No results};

 \begin{scope} [yshift=-1.2em,xshift=-1.1em]
  \shade[bottom color=red, top color=blue] (11,2.8) rectangle (11.6,5.5); 
  \node[anchor=center] at (11.55, 6.1) {\Large Accuracy};
  
  \draw[-,very thin,color=gray] (11,2.8) -- (11.8,2.8) node[right] {\large Low};  
  \draw[-,very thin,color=gray] (11,5.5) -- (11.8,5.5) node[right] {\large High};  
  
  \end{scope}
  
 \node[ellipse,fill=red!90!blue,text=white,anchor=center,text width=4em,align=center] at (1.5,3) {\Large Subset of data};
 
 \node[ellipse,fill=blue,text=white,anchor=center,text width=4em,align=center] at (7.5,5.5) {\Large Full GP};
 
 \node[ellipse,fill=red!75!blue,text=white,anchor=center,text width=5em,align=center] at (4.5,5.3) {\Large SS-GP};
 
 \node[ellipse,fill=red!40!blue,text=white,anchor=center,text width=4em,align=center] at (4.5,4.6) {\Large VFE};
 
 \node[ellipse,fill=red!60!blue,text=white,anchor=center,text width=4em,align=center] at (4.5,1) {\Large FITC};
 
   \node[ellipse,fill=red!25!blue,text=white,anchor=center,text width=7em,align=center] at (6,1.8) {\Large MoE/PoE};

\end{scope}
\end{tikzpicture}
\hrulefill
    \caption{Qualitative visualization for the performance of\\ the main scalable Gaussian process regression methods reviewed in Sections~\ref{sec:subsetdata}, \ref{sec:methodsinducing} and \ref{sec:finitedim}.}
\label{fig:qualitative_scalableGPs}
\end{mdframed}
\end{figure}

The goal of this section, whose structure is graphically presented in Figure~\ref{fig:mindmap_sec4}, is to provide a comprehensive overview of the available techniques for enhancing scalability in Gaussian process regression, including promising techniques that have not yet been exploited in the context of learning-based MPC.  We review the available literature on methods that alleviate these issues and improve scalability in Gaussian process regression, focusing on approaches based on 
subsets of data (Section \ref{sec:subsetdata}), 
inducing variables (Section \ref{sec:methodsinducing}),
finite-dimensional kernel operator approximations (Section \ref{sec:finitedim}), 
as well as numerical 
approximations~(Section~\ref{sec:gpcomptricks}).
We also provide references for results on uncertainty bounds around the estimates and for existing applications. For a more complete overview on the subject, we refer to the survey~\cite{liu_when_2020}; we also sketch the related results derived in the geostatistics community in the textbox titled \textit{Scalable Gaussian process regression in the geostatistics literature} on page~\pageref{sec:geostats}. We summarize the computational complexities of the best-known approaches in Table~\ref{tab:scalableGPs} on page \pageref{tab:scalableGPs}, and provide a qualitative visualization of their performance in Figure~\ref{fig:qualitative_scalableGPs} on page \pageref{fig:qualitative_scalableGPs}. 
Finally, we conclude the section with a review of the approaches that, instead of operating on the whole data-set at once, deal with the streaming data scenario (Section \ref{sec:online}).

\subsection{Subset of data, and its application to divide-and-conquer methods}\label{sec:subsetdata}
In this section, we review methods that rely on the idea of training a Gaussian process on a data-set that is smaller than the original one. They are divided into two classes: the first considers a single Gaussian process and uses a subset of the given data; the second consists in using multiple, local Gaussian processes (also known as experts), and can thus be regarded as a distributed set-up. We discuss these two classes in the following Sections~\ref{subsec:subsetdatasingle} and~\ref{sec:multiple_models}.

\subsubsection{Single model with pruned data-set}\label{subsec:subsetdatasingle} A first, na\"ive method to speed up computations consists in selecting $M \ll N$ data points, and performing exact inference with such a reduced data-set. The sub-sampling can be performed either completely randomly, or starting from a randomly chosen point and then computing $M$ cluster centers~\cite{gonzalez_clustering_1985,feder_optimal_1988}. The first approach has a computational complexity of $\mathcal{O}(M^3)$, while the second can be of $\mathcal{O}(MN)$ or $\mathcal{O}(N\log M)$ depending on the clustering algorithm implementation. 
An alternative to random selection consists in using information-theoretic criteria, such as the so-called informative vector machine introduced in~\cite{lawrence_fast_2002} for classification, updates, 
principal feature analysis~\cite{lu_feature_2007},
or matching pursuit~\cite{keerthi_matching_2005}. For a comprehensive overview of the available methods for sub-sampling, we refer also to~\cite[Chapter 4]{seeger_bayesian_2003}. \\

 Despite being quite drastic, possibly overestimating uncertainty in predictions, random sub-sampling has been recently complemented with rigorous bounds on the estimates and their generalization capability by leveraging the theory of graphons~\cite{hayashi_random_2020}, also showing that such a method yields a favorable trade-off between run-time and accuracy. Other error bounds have been provided in~\cite{oh_explorative_2010,xu_mobile_2011}, where it is argued that subset of data is the most suitable scalability paradigm for mobile sensor networks, and a comparison between information vector machine~\cite{lawrence_fast_2002}, principal feature analysis~\cite{lu_feature_2007} and mutual information~\cite{krause_near-optimal_2008} for data selection is presented. Otherwise, the subset of data method has been mostly taken as a baseline method for comparison and is mentioned, e.g., in application-oriented papers such as~\cite{bartok_gaussian_2015} and~\cite{wang_sparse_2022} for inter-atomic potentials fitting and wind gusts forecasting, respectively. 

\subsubsection{Multiple models: expert-based methods} \label{sec:multiple_models}
Instead of training a single Gaussian process, an alternative option is to train multiple models (also known as ``experts" in the literature) on subsets of the data-set, and then combine the single estimates. A large body of literature has been devoted to this kind of technique, mainly aimed at dealing with non-stationarity in the data and scalability. In the following, we focus on the latter for its relevance in Gaussian process-based model predictive control. \\

A first strategy is known as Mixture-of-Experts (MoE)~\cite{jacobs_adaptive_1991,tresp_mixtures_2000, yuksel_twenty_2012}. Given $N_{exp}$ Gaussian processes with likelihood  $p_k(y|g^k_z)$ for $k=1,...,N_{exp}$, denoting with $\{s_k(\cdot)\}_{k=1}^{N_{exp}}$ a set of gating functions yielding a probabilistic partition of the domain, the overall likelihood reads as $p_{MoE}(y|g^1_z,...,g^{N_{exp}}_z) = \sum_{k=1}^{N_{exp}} s_k(g^k_z)p_k(y|g^k_z)$. A na\"ive implementation of such a model, which requires learning both the Gaussian processes and the gating functions, typically requires Expectation-Maximization or Markov Chain Monte Carlo, and thus does not scale well with big data-sets. To overcome such an issue, the most flexible approaches are the infinite mixture-of- experts~\cite{rasmussen_infinite_2001,meeds_alternative_2005}, or the ones resorting to the approximation techniques reviewed in later Sections~\ref{sec:methodsinducing} and~\ref{sec:finitedim}~\cite{yuan_variational_2008,nguyen_fast_2014,nguyen_variational_2016}: these allow for optimal allocation of experts, but the process might fail because the optimization problem is intricate. Alternatively, one can pre-allocate the experts, at the price of losing the quantification of expert interaction~\cite{nguyen-tuong_model_2009}: see also~\cite{nguyen-tuong_local_2008,nguyen-tuong_incremental_2011} for online updates. Finally, we mention that this class of approaches is related to the method of ``bagging", which has been investigated 
in~\cite{chen_bagging_2009,das_fast_2018}, and has been applied to robotics, e.g., in~\cite{schneider_robot_2010,mckinnon_learning_2017}.\\

Instead of resorting to a linear combination of the Gaussian process models, another model averaging paradigm consists in the so-called Product-of-Experts (PoE), where the likelihood is expressed, for an arbitrary pair $(z,y)$, as ${p_{PoE}(y|g^1_z,...,g^{N_{exp}}_z) \propto \prod_{k=1}^{N_{exp}}p_k(y|g^k_z})$. Such a method is encumbered by the possible contribution of weak experts which, differently from the MoE, still impacts the prediction. There are two main routes to overcome this issue: the first consists in a weighted product rule for the predictive distribution~\cite{hinton_training_2002}, and the second is the so-called Bayesian Committee Machine~\cite{tresp_bayesian_2000}. Among all possible extensions and generalizations of these two approaches, we consider~\cite{deisenroth_distributed_2015}, which combines them in a unified framework. The resulting method allows for parallelization, which we discuss in Section~\ref{sub:parallel}; moreover, since all experts share the same prior, the optimization problem is easier to solve, but it might underestimate the uncertainty. Recently, such an approach has been built upon by~\cite{lederer_gaussian_2021}, allowing also to deal with streaming data.\\

Expert-based methods have been mostly deployed in geostatistics (see the textbox \textit{Scalable Gaussian process regression in the geostatistics literature} on page~\pageref{sec:geostats}) and in robotics for human pose inference~\cite{urtasun_sparse_2008,liu_kinect_2016} and model learning~\cite{nguyen-tuong_model_2009}. As regards their theoretical analysis, the most thorough investigation, framed in the general distributed non-parametric regression, is provided in~\cite{szabo_asymptotic_2019}. We also mention that MoE and PoE have been recently combined and analysed in~\cite{trapp_deep_2020}.

\subsection{Methods based on inducing variables}\label{sec:methodsinducing}
The rationale underlying this class of approaches is to augment the Gaussian model presented in~\eqref{eq:bayesprior} with additional function evaluations on a set $\indupo$ of inducing inputs (also known as pseudo-inputs, or active/support set) of cardinality $M \ll N$. The resulting vector of function evaluations, denoted according to the previous convention as $\induvar$, follows the same distribution of the latent function $g(\cdot)$, and therefore has the prior
\begin{equation}\label{eq:priorinduvar}
    \induvar \sim \mathcal{N}(0, \Ker_{\indupo, \indupo}).
\end{equation}
The vector $\induvar$ is then to be treated as a sufficient statistic: that is, $g_Z$ and $g_{Z^*}$ become conditionally independent given $\induvar$; in other words, $g_{Z}$ and $g_{Z^*}$ can communicate only through the inducing variables $\induvar$.

The approaches belonging to this framework can be divided into two main groups: in the first, the joint prior $p(g_Z,g_{Z^*})$ is approximated and exact inference is performed; in the second, one starts from the original prior and seeks an approximation for the posterior $p(g_{Z^*}|Y)$. These two classes of methods have been presented in unified frameworks in~\cite{quinonero-candela_unifying_2005} and~\cite{bui_unifying_2017}, respectively; moreover, they have been compared in~\cite{bauer_understanding_2016}.\\ 

We also mention a generalized approach, called inter-domain Gaussian processes~\cite{lazaro-gredilla_inter-domain_2009}, where pseudo-inputs can belong to a different space with respect to the input domain~$\mathcal{Z}$. Such a generalization has been applied to convolutional Gaussian processes~\cite{van_der_wilk_convolutional_2017} and can in principle allow for a more compact inducing variables representation. 
However, the specification of such an alternative domain might not be trivial; an exception using spherical harmonic features can be found in~\cite{dutordoir_sparse_2020}. In the method therein proposed, data are projected on a unit hyper-sphere, and Gaussian process regression is performed on this manifold, where kernels with known eigendecomposition are defined. In general, within the class of inter-domain Gaussian processes, one can also include the methods on finite-dimensional approximations of the kernel operator reviewed later in Section \ref{sec:finitedim}. 
 
\subsubsection{Approximating the prior}\label{sec:induprior} 
Referring to the expression for inference in~\eqref{eq:bayespost}, the computational bottleneck is found in the conditioning operation involving the joint prior $p(g_Z,g_{Z^*})$ detailed in~\eqref{eq:bayesprior}. To address this issue, one can consider the following approximation: 
    \begin{align}
        p(g_Z,g_{Z^*}) &= \int p(g_Z,g_{Z^*}|\induvar)p(\induvar)d\induvar\\
        &\approx \int q(g_Z|\induvar)q(g_{Z^*}|\induvar)p(\induvar)d\induvar,
    \end{align}
where $q(\cdot)$ denotes an approximated density. Such a decomposition is made following the rationale that $\induvar$ plays the role of sufficient statistic for the model.

The different approaches in this class of methods differ by the choices of the covariance matrices in $q(g_Z|\induvar) = \mathcal{N}(\Ker_{Z,\indupo}\Ker_{\indupo,\indupo}^{-1}\induvar, \tilde{Q}_{Z,Z})$ and $q(g_{Z^*}|\induvar) = \mathcal{N}(\Ker_{Z^*,\indupo}\Ker_{\indupo,\indupo}^{-1}\induvar, \tilde{Q}_{Z^*,Z^*})$. In particular, $\tilde{Q}$ will be a low-rank matrix, leading to a computational complexity of $\mathcal{O}(NM^2)$. This class of approaches is surveyed in~\cite{quinonero-candela_unifying_2005}, where also the terminology by which they have been known in the literature is introduced. Presented in chronological order, as well as with increasing design complexity, these are:
\begin{itemize}
    \item Subset of Regressors (SoR)~\cite{silverman_aspects_1985, wahba_bias-variance_1998, smola_sparse_2000}. The low-rank approximation of the kernel matrix, $\Ker_{Z,Z}\approx \Ker_{Z,\indupo}\Ker_{\indupo,\indupo}^{-1}\Ker_{\indupo,Z} = \tilde{Q}_{Z,Z}$, is obtained by postulating a deterministic relation between $g_Z$ and $\induvar$. This implies that the conditionals $q(g_Z|\induvar)$ and $q(g_{Z^*}|\induvar)$ become deterministic, possibly leading to overconfident predictions. 
    \item Deterministic Training Conditional (DTC)~\cite{csato_sparse_2002,seeger_fast_2003}. This approach leads to the same predictive mean of SoR, but comes with a more sensible predictive covariance because $q(g_{Z^*}|\induvar)$ is taken as the exact conditional; however, this results in an inconsistent Gaussian process (in the sense of Kolmogorov's existence Theorem~\cite{billingsley_probability_2012}), as the priors differ between training and testing points.  
    \item Fully Independent Conditional (FIC)~\cite{snelson_local_2007} presents a refined covariance, assuming that all the components of both $g_Z$ and $g_{Z^*}$ are independent given $\induvar$. This might result in a crude approximation; alternatively, the Fully Independent Training Conditional (FITC)~\cite{snelson_sparse_2005} admits the factorization on the training conditional only, at the price of having again an inconsistent Gaussian process model. FIC and FITC coincide if prediction is to be performed on a single point.
    \item Partially Independent (Training) Conditional~(PI(T)C)~\cite{quinonero-candela_unifying_2005} generalizes FI(T)C by introducing a block structure in the covariance. Even if the approximation is more accurate, it is claimed in~\cite{snelson_local_2007} that there is no significant impact on the performance with respect to FI(T)C.
\end{itemize}

These methods lead to an approximated marginal likelihood
\begin{equation}\label{eq:marg_lik_ind_pri_a}
q(Y) = \mathcal{N}(0, \tilde{Q}_{Z,Z} + \Ker_{Z,\indupo}\Ker_{\indupo,\indupo}^{-1}\Ker_{\indupo,Z} + \sigma_w^2 I_N)
\end{equation}
to be maximized to compute the hyper-parameters. As far as pseudo-inputs selection is concerned, originally in SoR and DTC they are chosen from the training data-set using similar rationales as in the subset of data method: that is, using, e.g., maximum information gain~\cite{seeger_fast_2003}, online learning~\cite{csato_sparse_2002}, and greedy posterior maximization~\cite{smola_sparse_2000}. An alternative approach consists in placing inducing points in the testing data-set $Z^*$, leveraging the transductive learning paradigm~\cite{tresp_bayesian_2000, schwaighofer_transductive_2002,deisenroth_distributed_2015}. In the FITC approach introduced in~\cite{snelson_sparse_2005}, pseudo-inputs are treated as additional hyper-parameters and are estimated maximizing~\eqref{eq:marg_lik_ind_pri_a}: this allows them to be arbitrarily placed in the domain $\mathcal{Z}$, and thus enables more flexibility~\cite{rossi_sparse_2021}. The advantages of this aspect have been highlighted in~\cite{snelson_sparse_2005}; still, the resulting optimization problem is hard to solve, and deterministic routines might lead to local optima and over-fitting. This problem can be overcome using Markov Chain Monte Carlo schemes, at the cost of longer training times~\cite{rossi_sparse_2021,scampicchio_markov_2023}. Similarly to SoR and DTC, one can select pseudo-inputs from the training data-set, as proposed in~\cite{cao_efficient_2015}.\\

Overall, the approaches outlined above display models that trade off simplicity with performance quality. The most popular approach in practice has been the FITC, which was extended to non-Gaussian likelihoods in~\cite{yuan_sparse-posterior_2012}, and has found successful applications in, e.g., human motion generation~\cite{wei_physically_2011}, power forecast~\cite{ahmad_methodological_2021}, battery state of charge~\cite{fan_state--charge_2022}, and ship maneuvering~\cite{xue_online_2022}; however, rigorously assessing its performance is a difficult task. One of the aspects highlighting such a difficulty is the fact that the full prior obtained by setting $\indupo = Z$ is not the global optimizer of the marginal likelihood~\eqref{eq:marg_lik_ind_pri_a}~\cite{bui_unifying_2017}. On the other hand, the SoR and DTC methods have not been applied much in practice, as the involved approximations negatively impact the predictions~\cite{hoang_unifying_2015}; yet, there could be some potential in providing bounds on their performance by drawing inspiration from results available for the related Nystr\"om method, which we discuss in detail in Section~\ref{sec:nystrom}.

\subsubsection{Approximating the posterior}\label{sub:indupost}
An alternative route to approximating the joint distribution $p(g_Z,g_{Z^*})$, which was proposed in~\cite{titsias_variational_2008, titsias_variational_2009}, is known as Variational Free Energy (VFE). It consists in directly tackling the predictive distribution 
\begin{align}
    &p(g_{Z^*}|Y) = \\ & = \iint p(g_{Z^*}|g_{Z},\induvar)p(g_Z|\induvar, Y)p(\induvar | Y)dg_Zd\induvar, \notag
\end{align}
where $p(g_{Z^*}|g_Z,\induvar,Y) = p(g_{Z^*}|g_Z,\induvar)$ because $Y$ is just a noisy version of $g_Z$. Again, leveraging the assumption that $\induvar$ is a sufficient statistic, we can approximate $p(g_{Z^*}|Y)$ with a distribution $q(g_{Z^*})$ such that
\begin{align}
    q(g_{Z^*}) &= \iint p(g_{Z^*}|\induvar)p(g_Z|\induvar)p(\induvar|Y)dg_Z d\induvar \notag \\ &= \int p(g_{Z^*}|\induvar)p(\induvar|Y)d\induvar \notag \\
    &\approx \int p(g_{Z^*}|\induvar)q(\induvar)d\induvar.
\end{align}
The key idea is then to choose the approximate distribution $q(\induvar)$ and the pseudo-input locations $\indupo$ via \textit{variational inference}~\cite{blei_variational_2017,zhang_advances_2019} as those minimizing the Kullback-Leibler~(KL) divergence
\begin{align}\label{eq:elbo}
&\infdiv[\bigg]{q(g_{Z^*},g_Z)}{p(g_{Z^*},g_Z|Y)} \\
& = \log p(Y) - \mathbb{E}_{q(\induvar,g_Z)}\Bigg[\frac{p(Y,g_Z,\induvar)}{q(\induvar,g_Z)}\Bigg],\notag
\end{align}
where the second term is called \textit{evidence lower bound}. 
Calculations reported in~\cite{titsias_variational_2008} show that
the optimizing distribution on $\induvar$ is such that $q(\induvar) = \mathcal{N}(\mu_q, \Sigma_q)$, with 
${\mu_q = \sigma_w^{-2}\Ker_{\indupo,\indupo}(\Ker_{\indupo,\indupo}+ \sigma_w^{-2}\Ker_{\indupo,Z}\Ker_{Z,\indupo})^{-1}\Ker_{\indupo,Z}Y}$ 
and 
${\Sigma_q = \Ker_{\indupo,\indupo} (\Ker_{\indupo,\indupo}+ \sigma_w^{-2}\Ker_{\indupo,Z}\Ker_{Z,\indupo})^{-1}\Ker_{\indupo,\indupo}}$.
Thus, hyper-parameters are found by optimizing 
\begin{align}\label{eq:marglikelbo}
    &\log p(Y) - \log[\mathcal{N}(0, \sigma_w^2 + \Ker_{Z,\indupo}\Ker_{\indupo,\indupo}^{-1}\Ker_{\indupo,Z})] + \notag \\ &+ \frac{1}{\sigma_w^2}\Tr(\Ker_{Z,Z} - \Ker_{Z,\indupo}\Ker_{\indupo,\indupo}^{-1}\Ker_{\indupo,Z}).
\end{align}
From these calculations, it results that the predictive distribution is the same as the one obtained in the DTC approach outlined in Section \ref{sec:induprior}, but the optimization problem yielding hyper-parameters and pseudo-inputs differs by the last addendum, which plays the role of a regularizer and acts against over-fitting. Pseudo-inputs can be either arbitrarily set in the domain $\mathcal{Z}$; searched over a grid in the so-called structured kernel interpolation (SKI) methods~\cite{wilson_kernel_2015,wilson_thoughts_2015,gardner_product_2018, evans_scalable_2018,izmailov_scalable_2018,wesel_tensor-based_2023} (which lead to $\mathcal{O}(N + M\log M)$, but suffer from the curse of dimensionality because they scale exponentially in $n_z$); or selected among the training inputs: see~\cite{tran_sparse_2021} for a Bayesian treatment of the problem, where sparsity in the pseudo-inputs is enforced. We also mention~\cite{lalchand_sparse_2022}, where a suitable choice of the variational distribution allows to overcome the sampling of the pseudo-targets, and hyper-parameters are trained with a Markov Chain Monte Carlo scheme. \\

The approach outlined above 
has found numerous extensions. For instance, it has been generalized to non-Gaussian likelihoods leveraging a Markov Chain Monte Carlo scheme~\cite{hensman_mcmc_2015}, and to non-stationary processes~\cite{paun_stochastic_2023}. Then, it was analyzed in the general framework of KL divergence between stochastic processes~\cite{matthews_sparse_2016,cheng_incremental_2016}, and has been included in the more general framework of Power Expectation Propagation, admitting also a unified representation with the FITC approach~\cite{bui_unifying_2017}: for an additional connection of variational inference and FITC, see~\cite{jankowiak_parametric_2020}. Moreover, providing a relaxed bound for~\eqref{eq:elbo},~\cite{hensman_gaussian_2013} propose to optimize $\mu_q$ and $\Sigma_q$ using stochastic gradient methods, thus further improving scalability and allowing for mini-batch training~\cite{hoffman_stochastic_2013}, attaining a computational complexity of $\mathcal{O}(M^3)$; see also~\cite{cheng_variational_2017}, where the parametrizations of $\mu_q$ and $\Sigma_q$ are decoupled to increase the pseudo-inputs capacity, and~\cite{salimbeni_orthogonally_2018,shi_sparse_2020}, which leverage an orthogonal decomposition of the original Gaussian process. \\ 

Overall, sparse Gaussian processes using variational inference have been successfully applied to, e.g., state-space models~\cite{frigola_variational_2014}, wind turbine power curves (thus, accounting for heteroscedastic noises)~\cite{rogers_probabilistic_2020}, Bayesian optimization~\cite{mcintire_sparse_2016}, material discovery~\cite{ament_autonomous_2021}, and deep learning~\cite{damianou_deep_2013,salimbeni_doubly_2017}. However, the only theoretical results quantifying the error introduced by these approximations are presented in~\cite{burt_rates_2019,burt_convergence_2020}, where a-priori bounds on the KL divergence are computed in view of an asymptotic analysis in terms of the number of inducing inputs $M$, and in~\cite{nieman_contraction_2022}, which provides minimax contraction rates. Finally, we mention the work~\cite{huggins_scalable_2019}, where a different metric is used: rather than the KL divergence, the authors discuss the Wasserstein 2-distance and the Fisher distance, and derive bounds on finite-data accuracy of posterior mean and covariance function estimates.

\subsection{Finite-dimensional representations of the kernel operator}\label{sec:finitedim}
While the approaches mentioned in Section \ref{sec:methodsinducing} revolve around the concept of eigen-decomposition of Gram matrices, we now consider the eigen-decomposition of the kernel operator $\Ker \colon \mathcal{Z}\times \mathcal{Z} \to \R$.

\subsubsection{Sparse Spectrum Gaussian processes (SSGP)}\label{sec:ssgp} The approach reviewed in this subsection was introduced in~\cite{lazaro-gredilla_sparse_2010} and builds upon the theory of Random Fourier features introduced in~\cite{rahimi_random_2007}, which in turn is recast in the research thread on fast kernel expansions for support vector machines~\cite{rahimi_uniform_2008,le_fastfood_2013, yang_carte_2015}.  

The proposed method holds for kernels that are stationary (also known as shift-invariant) -- i.e., that are a function of the distance between inputs, rather than their single values. This property is satisfied by the class of radial basis kernels, and specifically by the Gaussian one reported in~\eqref{eq:gaussker}.
The idea consists in studying the kernel in the frequency domain
and in considering the power spectral density $S(s)$ such that,  
for any arbitrary $z_a$, $z_b$ in $\mathcal{Z}$,
\begin{equation}\label{eq:wienerkhintch}
    \Ker(z_a,z_b) = \int \exp\{2\pi i s^{\top}(z_a - z_b)\}S(s)ds.
\end{equation}
By Bochner's Theorem, $S(s)$ is proportional to a probability measure $p_S(s)$, which for the Gaussian kernel~\eqref{eq:gaussker} can be computed as
\begin{align}\label{eq:bochnerpr}
    & p_S(s) = \frac{1}{\lambda}\int \exp\{-2\pi i s^{\top}(z_a - z_b)\}\Ker(z_a,z_b)dz \notag \\ &= \sqrt{2\pi \eta^{n_z}}\exp\{-2\pi^2\eta\|s\|^2 \},
\end{align}
and is a multivariate Gaussian. The approach proposed in~\cite{rahimi_random_2007} consists in interpreting~\eqref{eq:wienerkhintch} as an expected value with respect to the density $p_S(s)$, which is to be approximated by a Monte Carlo sum. Drawing samples $\{s_r, -s_r\}_{r=1}^{M/2}$ to respect symmetry and exploiting trigonometric identities, one obtains
\begin{align}
    \Ker(z_a,z_b) \approx 
    \frac{\lambda}{M}\sum_{r=1}^{M/2} & \Big(\cos(2\pi s_r^{\top}z_a)\cos(2\pi s_r^{\top}z_b) + \notag \\ &+ \sin(2\pi s_r^{\top}z_a)\sin(2\pi s_r^{\top}z_b)\Big).\notag
\end{align}
By inspection, it results that the approximated kernel can be written as $\Ker(z_a,z_b) = \langle \phi(z_a),\phi(z_b)\rangle$, denoting by $\langle \cdot, \cdot \rangle$ the inner product in $\R^{M}$, and by $\phi(z) = [\cos(2\pi s_1^{\top} z ), \cdots,\, \cos(2\pi s_{M/2}^{\top} z), \cdots,\, \sin(2\pi s_{M/2}^{\top} z)]^{\top}$ the feature vector inducing the finite-dimensional model $g(z) \approx \phi(z)^{\top}\alpha$. This reduces the original non-parametric estimation problem to trigonometric linear regression, which can be solved in $\mathcal{O}(N M^2)$.

While~\cite{rahimi_random_2007} treats the resulting approximation problem with deterministic kernel-based methods,~\cite{lazaro-gredilla_sparse_2010} recasts it in a Bayesian framework, endowing the vector $\alpha$ with a Gaussian prior $\alpha \sim \mathcal{N}(0, \lambda/M I_M)$. Hyper-parameters, together with the frequencies $\{s_r\}_{r=1}^{M/2}$, are then estimated via marginal likelihood optimization. Motivated by the fact that such an approach might be prone to over-fitting, further works leverage variational inference to retrieve the frequencies from data: see, e.g.,~\cite{gal_improving_2015,tan_variational_2016} and~\cite{hensman_variational_2018}, which recasts random Fourier features in the realm of inter-domain Gaussian processes~\cite{lazaro-gredilla_inter-domain_2009}.\\

Available bounds on the performance of these methods have been derived for random Fourier features in~\cite{sriperumbudur_optimal_2015,rudi_generalization_2017}; see also~\cite{scampicchio_error_2023} for an analysis of the resulting regularized trigonometric regression problem. The method has been applied, e.g., in robotics with online model updates ~\cite{gijsberts_real-time_2013,arcari_bayesian_2023}.

\subsubsection{Covariance series expansion}\label{sub:covserexp}
Instead of seeking a sparse-spectrum representation, this class of methods aims for a series expansion of the type
\begin{equation}\label{eq:genK-L}
    \Ker(z_a,z_b) = \sum_{k=1}^{+\infty} \gamma_k \varphi_k(z_a)\varphi_k(z_b),
\end{equation}
such that $\int \Ker(z_a,z_b)\varphi_{k}(z_a)p_z(z_a)dz_a = \gamma_k \varphi_k(z_b)$, with $p_z(\cdot)$ the input distribution; furthermore, $\{\varphi_k\}_{k=1}^{+\infty}$ is a family of orthonormal functions with respect to the measure induced by $p_z(\cdot)$, and $\{\gamma_k\}_{k=1}^{+\infty}$ is a set of decreasing, non-negative values. When $p_z(\cdot)$ is the uniform distribution and $\mathcal{Z}$ is compact, the representation in~\eqref{eq:genK-L} is known to lead to the Karhunen-Loève expansion~\cite{levy_karhunen_2008,picci_optimality_2006}, and exists if $\Ker^2(z_a,z_b)$ is integrable over $\mathcal{Z} \times \mathcal{Z}$~\cite{mercer_xvi_1997}. Once an expression as~\eqref{eq:genK-L} is obtained, then the series can be truncated, and a finite-dimensional model can be retrieved along the lines of the approach presented in Section~\ref{sec:ssgp}: see, e.g.,~\cite{zhu_gaussian_1997}, and also~\cite{ferrari-trecate_finite-dimensional_1998} for a theoretical analysis. There are few cases in which the eigen-decomposition in~\eqref{eq:genK-L} has a closed-form expression: notable examples are the Gaussian kernel~\eqref{eq:gaussker} with Gaussian-distributed inputs~\cite{fasshauer_positive_2011}, and the first-order spline kernel on an interval endowed with the uniform measure~\cite{wong_stochastic_1985}. These are derived assuming that the inputs are scalar; the multi-dimensional case is obtained by tensor products. 

When the eigen-decomposition is not known, available approximation techniques tend to be computationally heavy and scale unfavorably for highly-dimensional inputs; these typically involve Monte Carlo integration and a Nystr\"om approximation of the kernel -- see, e.g.,~\cite{marzouk_dimensionality_2009} and~\cite{peng_eigengp_2015,evans_scalable_2018}. This fact has limited the applicability of such an approach in real experiments.

An alternative solution was proposed in~\cite{solin_hilbert_2020} for compact domains and shift-invariant kernels, whose power spectral density (see Section~\ref{sec:ssgp}) admits a power series expansion $S(\|s\|^2) = \sum_{k=1}^{+\infty}a_k (\|s\|^2)^k$, where $\{a_k \}_{k=1}^{+\infty}$. The idea consists in approximating~\eqref{eq:genK-L} by means of an eigen-function expansion of the Laplace operator, which can be computed without using the expression of the kernel, thus speeding up hyper-parameter tuning. 
Notably, such an approach is complemented by a theoretical error analysis, and has been further studied in~\cite{riutort-mayol_practical_2022}. An application of such an approximation method can be found in~\cite{solin_modeling_2018}, where the task of magnetic-field estimation is considered.

\begin{table*}[h]
\begin{center}
    \resizebox{1.0\textwidth}{!}{
\begin{tabular}{|p{3.6em}|p{3em}|p{6em}|p{6em}|p{7em}|p{4em}|p{7.5em}|p{3em}|}
\hline
 & Full GP & Subset of Data (Sec.~\ref{subsec:subsetdatasingle}) & Expert-based methods (Sec.~\ref{sec:multiple_models}) & FITC \cite{snelson_sparse_2005}/VFE \cite{titsias_variational_2009} (Sec.~\ref{sec:methodsinducing})  & SSGP (Sec.~\ref{sec:finitedim}) & SKI \cite{wilson_kernel_2015} & SVGP \cite{hensman_gaussian_2013}\\
 \hline
 Training &$\mathcal{O}(N^3)$ &$\mathcal{O}(M^3)$ &$\mathcal{O}(NM_e^2)$ &$\mathcal{O}(NM^2)$ &$\mathcal{O}(Np^2)$ &$\mathcal{O}(N+ M\log M)$ &$\mathcal{O}(M^3)$ \\
 \hline
 Inference &$\mathcal{O}(N^2)$ &$\mathcal{O}(M^2)$ &$\mathcal{O}(M_e^2)$ &$\mathcal{O}(M^2)$ &$\mathcal{O}(p^2)$ &$\mathcal{O}(M\log M)$ &$\mathcal{O}(M^2)$\\
 \hline
\end{tabular}}
\vspace{1em}
\caption{Summary of computational complexities of scalable Gaussian process regression methods reviewed in Section \ref{sec:scalableGPs}.  \textbf{Acronyms:} FITC=Fully Independent Training Conditional; VFE=Variational Free Energy; SSGP=Sparse Spectrum Gaussian Processes; SKI=Structured Kernel Interpolation; SVGP=stochastic variational Gaussian Processes.
\textbf{Notation:} $N$ is the number of training data points; $M$ denotes the subset of data or inducing points; $p$ is the number of features in the finite-dimensional representations of the kernel operator -- we present it here for sparse-spectrum Gaussian processes, but it holds also for the covariance series expansion; $M_e$ is the number of data points allocated to each expert.  As regards the latter method, we report only the most favorable computational complexity.}
    \label{tab:scalableGPs}
\end{center}
\end{table*}

\subsection{Computational techniques}\label{sec:gpcomptricks}


The line of work presented in the following considers numerical techniques to efficiently reduce the computational cost incurred by the linear-system solve involving the Gram matrix for predictions~\eqref{eq:postparams}, as well as evaluation of the log-determinant and its gradient for hyper-parameter optimization~\eqref{eq:negloglik}.
Most of the presented works are complementary to the scalable Gaussian process approximation techniques mentioned in the preceding sections, as they may be similarly applied to the reduced-size linear-system solves required by, e.g., inducing-point methods.

\subsubsection{Nyström approximation}
\label{sec:nystrom}
Approximating the
$N\times N$ Gram matrix $\Ker_{Z,Z}$ by means of a
low-rank factorization based on $M$ of its columns $\Ker_{Z,\indupo} \in \mathbb{R}^{N \times M}$,
the standard Nyström method,
introduced in~\cite{williams_using_2000} and reviewed in~\cite{sun_review_2015},
yields the kernel matrix approximation $\Ker_{Z,Z} \approx \Ker_{Z,\indupo}\Ker_{\indupo,\indupo}^{-1}\Ker_{\indupo,Z}$.
Relying on the rapidly decaying eigenspectrum of kernel matrices~\cite{williams_using_2000,smola_sparse_2000},
a numerically accurate low-rank approximation is commonly achieved for a subset of $M \ll N$ data points,
leading to favorable computation (training and inference) and storage complexities of $\mathcal{O}(N M^2)$ and $\mathcal{O}(NM)$, respectively.
Note that
the Nyström approximation
can be efficiently obtained from an
incomplete (block) column-based Cholesky decomposition when the same (block) columns are selected\footnote{The connection between incomplete (row-based) Cholesky decomposition and Gaussian process regression with a subset of data points has also been explored by~\cite{bartels_kernel-matrix_2023,bartels_adaptive_2023} to accelerate hyper-parameter optimization using probabilistic early-stopping criteria for the approximate factorization.}%
,
which has been
pioneered
by~\cite{fine_efficient_2001, bach_kernel_2002} for approximate kernel matrix inversion.
It is furthermore strongly related to the subset-of-regressors (SoR) approximation proposed by~\cite{smola_sparse_2000} (see Section~\ref{sec:induprior}), which
leads to
the same kernel matrix approximation 
when the degenerate prior, i.e., low-rank kernel matrix approximation,
is obtained
from
the same deterministic relation
imposed by a linear combination of columns.
Still, in contrast to SoR, the Nyström method cannot be framed directly in terms of Gaussian process regression, because the
sole approximation of the kernel \emph{matrix}
-- in contrast to the kernel \emph{function} ---
does not guarantee 
positive semi-definiteness of
the
posterior covariance~\cite{williams_observations_2002,quinonero-candela_unifying_2005}.\\

The approximation quality of the Nyström 
method hinges on the selection of columns defining the low-rank approximation.
Beyond uniform sampling~\cite{williams_using_2000},
various column selection strategies have been proposed,
including
greedy selection strategies~\cite{smola_sparse_2000},
pivoting strategies used for the incomplete Cholesky factorization~\cite{fine_2001_efficient,bach_kernel_2002,bach_predictive_2005,harbrecht_low-rank_2012,schafer_compression_2021},
random sampling~\cite{drineas_nystrom_2005,kumar_sampling_2012},
k-means clustering~\cite{zhang_improved_2008},
and
sampling strategies based on
statistical leverage scores~\cite{drineas_fast_2012,alaoui_fast_2015,musco_recursive_2017}.
Besides column selection strategies, random projections have gained popularity for determining Nyström approximations with statistical guarantees, see~\cite{halko_finding_2011,martinsson_randomized_2020} for an overview, and have been employed for efficient Gaussian process regression by~\cite{banerjee_efficient_2013}.
The resulting approximation error
is quantified by a number of results, e.g.~\cite{%
    drineas_nystrom_2005,%
    zhang_improved_2008,%
    kumar_sampling_2012,%
    gittens_revisiting_2013,%
    jin_improved_2013,%
    alaoui_fast_2015,%
    rudi_less_2015,%
    musco_recursive_2017%
}; see also~\cite{cortes_impact_2010} and~\cite{bach_sharp_2013} for error bounds in prediction.
For a thorough discussion on the Nyström method we refer the interested reader to~\cite{sun_review_2015}.\\

Besides the direct application of 
the Nyström method
to construct an easy-to-invert low-rank approximation of the kernel matrix, 
it has
also 
successfully 
been  
employed as a preconditioner for other iterative linear-system solvers~\cite{%
    cutajar_preconditioning_2016,%
    rudi_falkon_2017,%
    gardner_gpytorch_2018,%
    wenger_preconditioning_2022%
}, 
such as the conjugate gradient method, which we review in the following.

\subsubsection{Conjugate gradient method}\label{sub:conjugategradient}

Arguably, the most prominent iterative linear-system solver for positive definite matrices is the conjugate gradient~(CG) method~\cite{hestenes_methods_1952}. By interpreting the solution of the linear system $A x = b$, with A being positive definite, as a necessary and sufficient optimality condition for the optimization problem
$\min_x \frac{1}{2} x^\top A x - x^\top b$,
CG iteratively refines its solution estimate by performing the minimization successively in a growing subspace of orthogonal search directions%
\footnote{Based on this optimization-based reformulation, also 
the quasi-Newton BFGS method
has
been applied for efficient Gaussian-process inference~\cite{leithead_on_2007}.%
}%
. As the algorithm's
computational complexity is dominated by
 matrix-vector multiplications,
it
scales with $\mathcal{O}(P N^2)$, where $P$ is the number of CG iterations. While the algorithm, in exact arithmetic, is guaranteed to recover the exact solution after $P = N$ iterations, which would still lead to a cubical complexity for the linear-system solve, it commonly converges to numerically accurate solutions after $P \ll N$ iterations, leading to a quadratic computational complexity in the data-set dimension.\\

For its use in Bayesian numerical analysis, CG has been suggested by~\cite{skilling_bayesian_1993},
based on which it has been successfully applied for efficient predictions and computation of the log-marginal likelihood (gradient) for hyper-parameter estimation~\cite{gibbs_efficient_1997,gibbs_bayesian_1997}.
As CG's key computational savings stem from the fact that its most expensive computations are mere matrix-vector multiplications~(MVMs),
therein,
the required computations for training are similarly expressed using MVMs.
In particular,
\mbox{$\frac{\partial}{\partial \xi} \log \det \left( \Ker_{Z,Z} \right) = \Tr \left( \Ker_{Z,Z}^{-1} \frac{\partial \Ker_{Z,Z}}{\partial \xi} \right)$}
can be approximated using Hutchinson's stochastic trace estimator~\cite{girard_algorithme_1987,hutchinson_stochastic_1989}, which finds an unbiased estimate for
\mbox{$\Tr\left( A \right) = \mathbb{E}_{W_i} \left[ W_i A W_i \right] \approx
        t^{-1}
        \sum_{i=1}^t W_i^\top A W_i$}
using $t$ additional CG solves with \mbox{(sub-)Gaussian} i.i.d.~probe vectors $W_i$ of unit variance.
The value for \mbox{$\log \det \left( \Ker_{Z,Z} \right)$} can be efficiently estimated from the partial Lanczos tri-diagonalization of the kernel matrix~\cite{ubaru_fast_2017,dong_scalable_2017}, which is constructed implicitly during the CG iterations~\cite[Chapter 11.3]{golub_matrix_2013};
or, alternatively,
using a power-series approximation of the log-determinant in conjunction with 
Hutchinson's trace estimator~\cite{zhang_approximate_2007}.

To save computation time, the successive refinement of the CG solution estimate lends itself to early-stopping approaches, particularly when only rough estimates of the solution are required, e.g., during hyper-parameter optimization.
While the approximation error can be efficiently tracked using upper and lower bounds on the quadratic objective~\cite{skilling_bayesian_1993,davies_effective_2015,artemev_tighter_2021}, terminating the iteration prematurely generally leads to biased estimates of the solution as well as the log-marginal likelihood (gradient);
as a remedy, unbiased estimators based on randomized truncations have been proposed by~\cite{filippone_enabling_2015,potapczynski_bias-free_2021}.\\

In recent years, the CG method has been successfully coupled and investigated together with other Gaussian process approximation methods, notably,
by exploiting the particularly efficient MVMs of structured kernel interpolation (SKI)~\cite{gardner_product_2018,pleiss_constant-time_2018,yadav_faster_2021} (see Section~\ref{sub:indupost}),
or scaling up the Nyström approximation by using CG to solve the lower-dimensional linear systems~\cite{meanti_kernel_2020}.
Another approach has been taken by~\cite{bartels_conjugate_2020} and~\cite{wenger_posterior_2022}, who use the search space generated by CG to construct a low-rank kernel approximation and estimate the computational uncertainty inherited by the iterative solver, respectively.

\subsubsection{Parallelization of computations}\label{sub:parallel}

In addition to the improved scaling properties of Gaussian process approximation methods in terms of computation and storage, distribution of the computational workload and memory allocation provides another layer of scalability for regression on increasingly large data sets,
particularly motivated by the rapidly increasing capability of modern graphical processing units (GPUs).
For most scalable methods mentioned in this section, parallelized implementations have been proposed:
\begin{itemize}
    
    \item \emph{Mixture-of-experts models:}
          Due to their distributed nature, mixture-of-experts models reviewed in Section~\ref{sec:multiple_models} allow for straightforward parallelization: as such, importance-sampled MoE models have been employed in~\cite{zhang_embarrassingly_2019}. 
    \item \emph{Inducing-point methods:}
          A parallel implementation that exploits the block-wise conditional independence assumption of the PI(T)C approximation mentioned in Section~\ref{sec:induprior} has been put forward by~\cite{chen_parallel_2013}.
          As a generalization of the PIC method and its parallelized implementation, the family of low-rank-cum-Markov approximations (LMAs)~\cite{low_parallel_2015} provides a spectrum of approximations between full Gaussian process inference and the PI(T)C approximation by relaxing the (block-wise) conditional independence assumption into a reduced-order Markov property on the residual process.
          The conditional independence assumption has also been
          utilized
          for distributed
          hyper-parameter optimization in the variational inference framework~\cite{hensman_gaussian_2013}:
          by noticing that the variational lower bound is decomposed into $N$ independent addends over each latent function value,
          the gradient computations can be efficiently parallelized~\cite{gal_distributed_2014}.
          Unifying the remaining inducing point methods of~\cite{quinonero-candela_unifying_2005} within the LMA framework~\cite{hoang_unifying_2015},
          they are also readily parallelized~\cite{hoang_distributed_2016}.
          Note that the unified variational inference procedure~\cite{hoang_unifying_2015} also allows for \emph{anytime} hyper-parameter optimization, i.e., using stochastic gradient descent with an unbiased natural gradient estimate based on fixed-size mini-batches of training data.
        
    \item \emph{Nyström approximation:} For low-rank kernel matrix approximations, parallelization has been proposed based on the (incomplete) \emph{row-based} Cholesky decomposition~\cite{george_parallel_1986,zhu_parallelizing_2007,chen_parallel_2013}, see also~\cite{bientinesi_families_2008} for an overview of (distributed) factorization algorithms for positive definite matrices.
    \item \emph{Conjugate Gradients:} As a matrix-vector-multiplication~(MVM) based solver, the computational workload of the most expensive
          CG operations
          can be straightforwardly parallelized.
          This idea has been taken further by~\cite{gardner_gpytorch_2018},
          who developed a batched version of CG to solve the linear systems for the Gaussian process prediction~\eqref{eq:bayespost} and stochastic trace estimate simultaneously. 
          Memory requirements of MVM-based methods can be addressed by working sequentially with input partitions~\cite{wang_exact_2019}, or by lazily evaluated reduction operations~\cite{charlier_kernel_2021}, which only query the entries of the kernel matrix during computation of the matrix-vector product; see also~\cite{srinivasan2009scaling,srinivasan2010gpuml,musizza2010accelerated} for early applications of these ideas.
\end{itemize}

\begin{addendumbox*}[ht!]
\begin{mdframed}[backgroundcolor=gray!10]
\subsection*{ADDENDUM: Scalable Gaussian process regression in the geostatistics literature}\label{sec:geostats}%
    As mentioned in Section~\ref{subsub:frameworkGP}, Gaussian process regression belongs to the non-parametric kernel-based methods that have been extensively studied in the geostatistics community (originally focusing on kriging~\cite{matheron_principles_1963,journel_mining_1978,cressie_statistics_1993}). In such a context, it is of key importance to balance the accuracy in estimation with scalability, as the problems under investigation typically encompass very large data-sets. Even though the results available in the geostatistics literature do not dramatically differ from the ones that 
    we thoroughly review in this paper, in this textbox we provide an overview of the available sources to highlight the  research in that community and point out some approaches that have been overlooked in machine learning and control. For a more thorough presentation of the available results, we refer the interested reader to~\cite{sun_geostatistics_2012} and~\cite{camps-valls_survey_2016}.\\

    The main approaches for scalable Gaussian process regression, as outlined in~\cite{katzfuss_general_2021}, are the ones leading to (i) sparse covariance matrices; (ii) sparse inverse covariance matrices; (iii) low-rank matrices. 
    
    As regards scenario (i), the main approach is the so-called tapering, which consists in thresholding the entries of the covariance matrix~\cite{furrer_covariance_2006,kaufman_covariance_2008,furrer_asymptotic_2016}.
    The idea is thus to limit the correlation to a local neighborhood, and this can be exploited to speed up computation using tools from sparse linear algebra and parallel computing (see Section \ref{sec:gpcomptricks}). It has not been much used in the machine learning or engineering communities, except for the works~\cite{vanhatalo_modelling_2008} (combined in the FIC reviewed in Section \ref{sec:induprior}), and~\cite{ranganathan_online_2011}, where it has been deployed for online learning. Following a similar tapering rationale, also close to the expert-based methods reviewed in \ref{sec:multiple_models}, nearest-neighbor Gaussian processes have been studied in~\cite{datta_hierarchical_2016,datta_nearest-neighbor_2016,datta_nonseparable_2016}.
    
    The best-known method of the class (ii) is the Vecchia approximation~\cite{vecchia_estimation_1988}, which is one of the first methods available to scale up kernel-based methods and has been recently generalized in~\cite{katzfuss_general_2021} for latent variable models. Sparsifying the precision matrix is not very common in the statistics and machine learning literature and it might deserve further investigations, possibly combined with the numerical methods reviewed in Section~\ref{sec:gpcomptricks}. An exception is~\cite{wu_variational_2022}, which combines sparse approximation of the precision matrix with stochastic variational inference~\cite{hensman_gaussian_2013} mentioned in Section~\ref{sub:indupost}. 
    
    As regards low-rank matrices (iii), an approach related to the subset of regressors/Nystr\"om approximation reviewed in Sections \ref{sec:induprior} and \ref{sec:nystrom}, respectively, is presented in~\cite{banerjee_gaussian_2008,cressie_fixed_2008,banerjee_efficient_2013}.\\
    
    All of the methods mentioned above have been taken into consideration in comparative studies in~\cite{heaton_case_2019}, and their computational aspects are investigated in~\cite{gramacy_local_2015,guhaniyogi_meta-kriging_2018}. 
    We point out that the references presented so far focus on spatial processes -- however, a very large body of literature concerns spatio-temporal ones, which we review in Section \ref{sub:spatiotemporal}. The first work combining kriging and Kalman filter is~\cite{mardia_kriged_1998}, followed by, e.g.,~\cite{wikle_dimension-reduced_1999}; further developments can be found in the book~\cite{cressie_space-time_2014}. We bring to attention~\cite{banerjee_high-dimensional_2017}, where approaches from (i) -- specifically, the nearest-neighbour Gaussian processes -- and (iii) for spatio-temporal processes are compared.
\end{mdframed}
\end{addendumbox*}

\subsection{Online learning: dealing with streaming data}\label{sec:online}
    
The approaches for scalable Gaussian process regression reviewed in the previous subsections deal with the batch setting, i.e., they operate on the whole data-set to obtain a set of representative quantities in training (active set of data/pseudo-points/features), which then allow to speed up performance in prediction. Here we instead consider the streaming data scenario, i.e., the case in which the data-set grows at each time-step. We structure the discussion according to the methods surveyed for the batch case in Sections~\ref{sec:subsetdata}, \ref{sec:methodsinducing}, \ref{sec:finitedim} and~\ref{sec:gpcomptricks}. 

\paragraph*{Active-data selection} Most of the contributions available regard methods that select an active set of representative data points — and this aligns with the rationales of the subset of data, subset of regressors, and DTC for the batch setting. The complete pipeline is summarised, e.g., in~\cite{petelin_control_2011}, and it works as follows: the new datum is observed; if the number of data points exceeds the budget, then an acceptance/rejection mechanism is run to keep the most informative ones; finally, hyper-parameters are updated. Many approaches have been proposed to deal with the acceptance/rejection of data points, and perhaps the most impactful work in this direction is given by~\cite{csato_sparse_2002}, where the RKHS inner product is used to define a projection-based rule; note, however, that the approach therein presented, despite not accounting for hyper-parameter updates, is still computationally heavy~\cite{le_gogp_2017}. The work~\cite{csato_sparse_2002} has been used for inverse robot dynamics learning in~\cite{de_la_cruz_online_2012}, which also performs hyper-parameter updates, and in~\cite{li_learning_2022}, where the data selection rule also entails a temporal forgetting factor. Similarly, using sample windowing,~\cite{ranganathan_online_2011} combines~\cite{csato_sparse_2002} with a tapering principle~\cite{furrer_covariance_2006} (see the textbox \textit{Scalable Gaussian process regression in the geostatistics literature}) to sparsify the covariance matrix and further enable a computational speed-up. Temporal correlation is taken into account in~\cite{soh_incrementally_2014}, where~\cite{csato_sparse_2002} is applied with kernels encompassing a recurrent term to deal with tactile robot tasks. Other applications can be found in the so-called warped Gaussian process regression~\cite{snelson_warped_2003} for wind forecasting~\cite{kou_sparse_2013} and in model reference adaptive control~\cite{chowdhary_bayesian_2015}.  Finally, we also mention the method proposed in~\cite{nguyen-tuong_incremental_2011}, which updates a fixed-budget dictionary with an insertion/removal strategy using an independence measure based on the RKHS metric, and that extends the method proposed in~\cite{engel_kernel_2004}. 

Other projection-based approaches to select the active set of data can be found in~\cite{schreiter_fast_2015,schreiter_efficient_2016}, building upon the method based on KL divergence of~\cite{smola_sparse_2000} for fast insertion/deletion of data points. Another method using an information loss between a new candidate input and the set of features centered at the active data-set is given in~\cite{le_gogp_2017}, which is also complemented with a rigorous theoretical analysis. Finally, we mention~\cite{koppel_consistent_2021}, where the projection rule is based on the Hellinger metric, preserving the distance between the current empirical distribution and its population counterpart following the rationale of matching pursuit~\cite{keerthi_matching_2005}, and allowing for a dynamic size of the active data-set. 

\paragraph*{Expert-based methods} The rationale of expert-based methods finds application, for the streaming setting, in the use of localized models, inspired by~\cite{snelson_local_2007, vijayakumar_incremental_2005} and investigated in~\cite{nguyen-tuong_learning_2008,nguyen-tuong_local_2008,nguyen-tuong_model_2009}. The proposed scheme consists in associating each new datum to a cluster of points, with distance defined thanks to the Gaussian kernel, and then updating the corresponding local model. Prediction on query points is computed using a weighted average of all the predictions given by the local models. Insertion/deletion can be dealt with windowing or using the information gain criterion. The direct combination of localised models and the approach in~\cite{csato_sparse_2002} is proposed in~\cite{park_learning-based_2013}. 

Another usage of expert-based models is found in~\cite{urtasun_sparse_2008}, where each local model captures a mode of the robot pose to be reconstructed, and inference is performed by selecting the training points that are closer to the testing datum. Finally, we mention that a survey on methods that incrementally split the input space into regions and perform regression on local models associated with these regions, with a specific focus on robotics applications, is given by~\cite{sigaud_-line_2011}.

\paragraph*{Inducing-point methods} There has been a big interest in extending the methods based on inducing points from the batch to the streaming data scenario. The FITC approach studied in~\cite{snelson_sparse_2005} has been adapted to online learning in~\cite{bijl_online_2015}, updating pseudo-inputs at each iteration. Such an approach has been combined with the NIGP method of~\cite{mchutchon_gaussian_2011} (see Section~\ref{sub:alternativegaussmod}) and applied to system identification~\cite{bijl_system_2017} and ship maneuvering in~\cite{xue_online_2022}. Another approach, which treats Gaussian process regression as a filtering problem and provides a good trade-off between accuracy and runtime, is presented in~\cite{huber_recursive_2013,huber_recursive_2014}. There, hyper-parameters are updated online thanks to a strategy deploying sigma-points; pseudo-inputs are kept fixed, and new data points are used to update the pseudo-targets. The rationale therein presented has also been deployed, combined with the Bayesian committee machine and particle filtering, for target tracking in a distributed architecture in~\cite{yin_distributed_2017}.

A large body of work has also been devoted to extending the VFE method to data arriving online. A first approach is presented in~\cite{meier_drifting_2015,meier_drifting_2016} for robotics applications. There, the sparsified model is re-trained at each time-step using data windows of fixed size, using previous estimates to initialize the new ones, and multiple models with growing window sizes are trained in parallel and then combined. A work that strongly resonated in the community is~\cite{bui_streaming_2017}, which subsumes~\cite{csato_sparse_2002,opper_bayesian_1999} as well as the streaming variational Bayes framework for parameter learning of~\cite{broderick_streaming_2013}, which in turn builds upon~\cite{honkela_-line_2003,sato_online_2001}. It is more suitable for dealing with the streaming setting compared to~\cite{cheng_incremental_2016}, and similarly to the latter, it performs online hyper-parameter tuning, thus outperforming~\cite{hensman_gaussian_2013,hoang_unifying_2015}. It has also been extended in~\cite{maddox_conditioning_2021}, proposing an alternative to the resampling heuristic used in~\cite{bui_streaming_2017}. We also refer to~\cite{nguyen_online_2017}, where a similar approach is presented and theoretically analysed. While~\cite{bui_unifying_2017} updates hyper-parameters by considering each data point once, the approach in~\cite{schurch_recursive_2020} proposes to use mini-batches several times to improve accuracy and ensures computational feasibility by a novel Kalman-filter formulation of the problem. The approach of~\cite{bui_streaming_2017} was also combined with local models to further speed up computation time and possibly allow for parallelization: works in this direction are, e.g., the product-of-experts approach of~\cite{wilcox_solar-gp_2020}, which however does not learn the expert weights jointly with the local model parameters;~\cite{kepler_wasserstein-splitting_2021}, proposing a data splitting strategy ruled by Wasserstein distance allowing for online hyper-parameters update; and~\cite{watson_learning_2022}, which presents generalized product-of-experts, leveraging variational inference. The latter approach allows for simultaneous online optimization of local model parameters, a clear and interpretable data partitioning strategy, and the principled fusion of multiple model predictions without the need for joint training.
Finally, we mention that, for direct application to state-space models, VFE has been studied in~\cite{park_online_2022}, where stochastic variational inference is deployed, and in~\cite{liu_gaussian_2021}, in which the case of time-varying systems (and hyper-parameters) is considered.

\paragraph*{Finite-dimensional approximations of kernel operator} As regards methods that focus on sparsifying the kernel operator, results available outside the batch setting are not numerous. We mention~\cite{gijsberts_real-time_2013} for online updates of sparse-spectrum Gaussian processes~\cite{lazaro-gredilla_sparse_2010}, and the works~\cite{berntorp_recursive_2019,berntorp_online_2020,berntorp_online_2021}, where the kernel approximation given by the Laplace operator is sequentially updated deploying a particle filter. 

\paragraph*{Computational methods} Among the computational methods for scalable and online Gaussian process regression, we would like to point out that a very promising approach to deal with the online setting, leveraging the structured kernel interpolation (SKI) of~\cite{wilson_kernel_2015} (see also Section~\ref{sub:indupost}), is given by~\cite{stanton_kernel_2021}. Moreover,~\cite{bopardikar_sequential_2016} employs a sequential, randomized low-rank matrix factorization approach of~\cite{halko_finding_2011} to compute the matrix inverse entering the posterior distribution in $\mathcal{O}(pN^2 + p^2N)$, where $p$ is the number of eigenfunctions considered in the truncation. 
\begin{remark}[On nomenclature] 
Consistently with the majority of the literature, we use the word ``online", which can be regarded as the most general term denoting approaches that deal with data-sets of increasing size, with or without hyper-parameter update. However, ``online" is mostly used for learning in the regret minimization framework~\cite{koppel_projected_2019,gijsberts_real-time_2013}, and this is the kind of analysis that is carried out in~\cite{le_gogp_2017}. The pipeline of data acceptance/rejection and hyper-parameter update was named ``evolving Gaussian processes" in~\cite{petelin_control_2011,kocijan_modelling_2016} and was also used in~\cite{maiworm_online_2021}. In the robotics literature the term that is often encountered is ``incremental" (see, e.g.,~\cite{gijsberts_real-time_2013,nguyen-tuong_incremental_2011}), mostly encompassing methods that do perform hyper-parameters selection online. Another term, usually encountered in (but not limited to) the computational algebra literature, is ``sequential"~\cite{bopardikar_sequential_2016}. Finally, methods deploying update steps that mimic the ones of the Kalman filter typically use the adjective ``recursive", as done in~\cite{huber_recursive_2014,schurch_recursive_2020}. 
\end{remark}

\section{UNCERTAINTY PROPAGATION}\label{sec:uncertaintyprop}
\begin{figure*}[h!]
\begin{mdframed}[backgroundcolor=gray!5, hidealllines=true]
    \centering
\begin{tikzpicture}[mindmap, concept color=blue!20]
\begin{scope}[scale=0.7,transform shape]
 \node [concept] {\ref{sec:uncertaintyprop} Uncertainty propagation}
    child[concept color=red!10,grow=180] {node[concept] {\ref{sec:gaussapproxup}  Independent one-step ahead predictions}
    child[grow=35] {node[concept] {\ref{subsubsec:linearization} Linearization}}
    child[grow=75] {node[concept] {\ref{sub:exactmomentmatch} Moment matching}}
    child[grow=120] {node[concept] {\ref{sub:sigmapoint} Sigma-point propagation}}
    child[grow=166] {node[concept] {\ref{sub:numapproxuncprop} Numerical approximations}} 
    child[grow=210] {node[concept] {\ref{sub:alternativegaussmod} Alternative Gaussian models}}      
    }
    child[concept color=red!10,grow=25]  {node[concept] {\ref{subsec:no_indep_ass} Overcoming the independence assumption}}
    child[concept color=red!10,grow=-10]  {node[concept] {\ref{subsec:robust_prop} Robust uncertainty propagation}};

    \end{scope}
\end{tikzpicture}
\caption{Graphical overview of Section~\ref{sec:uncertaintyprop}.}
\label{fig:mindmap_sec5}
\end{mdframed}
\end{figure*}
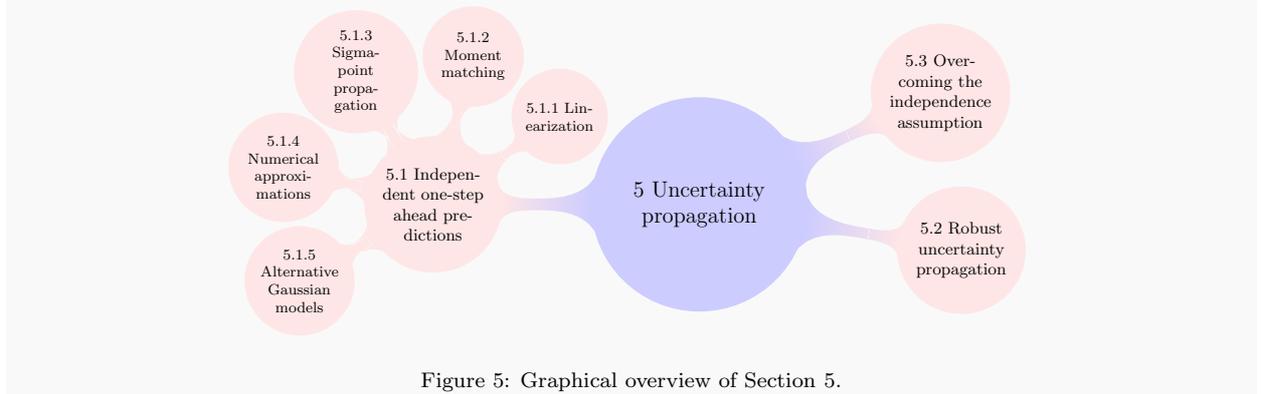

This section addresses the question of performing multi-step time-series predictions, as required in problem~\eqref{eq:optimization_problem}, deploying the Gaussian process estimate entering the term $g(\cdot)$ in the dynamics model. This is typically tackled in two ways. 
The first is by learning a dedicated model for each multi-step prediction  
(\textit{direct method}); the second is by covering the horizon length of interest via repeated one-step-ahead predictions (\textit{indirect method})~\cite{farmer_exploiting_1989}.

Employing the direct method in an MPC framework 
involves learning a model for each $i$-steps-ahead prediction to be made, where $i = 1,\dots,T$ and $T$ is the length of the MPC prediction horizon (see~\eqref{eq:optimization_problem}). As $T$ increases, so does the complexity of the nonlinear mapping, 
as well as the computational cost of the underlying predictive model since it depends on longer input sequences. However, the main advantage is that, at each time step in the horizon, direct models evaluate an exact Gaussian distribution~\cite{hachino_time_2007,hachino_multiple_2011, pfefferkorn_exact_2022}.

In contrast, for the indirect method, a single one-step-ahead model is sufficient for any horizon length due to its recursive nature. This also
makes such a model easier to train and more data-efficient, not having to deal with high-dimensional inputs as in multi-step predictors for large time-horizons. 
These factors have established the indirect method as the main choice for Gaussian process-based MPC approaches, which we discuss further in the remainder of this section, and for which an extensive overview is provided in~\cite{landgraf_probabilistic_2023}. However, despite their computational appeal, recursive evaluations of the Gaussian process model result in the propagation of stochastic uncertainties over the prediction horizon, i.e., a random output predicted at the current time step becomes a random input of the function $g$ at the successive time step. In general, exact expressions of the obtained distributions do not exist, as these are typically non-Gaussian and analytically intractable. An additional disadvantage of breaking down a $T$-steps-ahead prediction into an iteration of one-step-ahead predictions is the implicit assumption that correlations between non-consecutive states in the overall predicted trajectory can be neglected. Depending on the nature of the uncertainty affecting the system dynamics, this assumption can severely deteriorate the quality of the uncertainty estimated along the trajectory (see examples in~\cite{hewing_simulation_2020}). \\

This section, whose structure is summarized in Figure~\ref{fig:mindmap_sec5}, is structured as follows. We first focus on methods that have been developed under this \textit{independence} assumption in Section~\ref{sec:gaussapproxup}. 
Within this discussion, in Section~\ref{sub:alternativegaussmod} we further review the issue of evaluating Gaussian process predictions at random inputs, distinguishing between methods that either deal with input uncertainty only at test time or acknowledge its presence already at training time. In Section~\ref{subsec:robust_prop} we mention works that use ideas from established robust control methods and leverage a quantification of the propagated uncertainty by means of confidence regions. Finally, we present alternative methods that overcome the independence assumption in Section~\ref{subsec:no_indep_ass}.

Note that in the following subsections we assume that the nominal dynamics $g_{\text{nom}}$ is equal to 0 and set the matrix $B_d$ as the identity. The general case requires careful consideration of both propagating a random input through a nonlinear map (that does not preserve Gaussianity) and computing correlations -- we defer the investigation of such a set-up to, e.g.,~\cite{hewing_cautious_2020}.

\subsection{Independent one-step-ahead predictions}\label{sec:gaussapproxup}

\begin{figure*}[t]
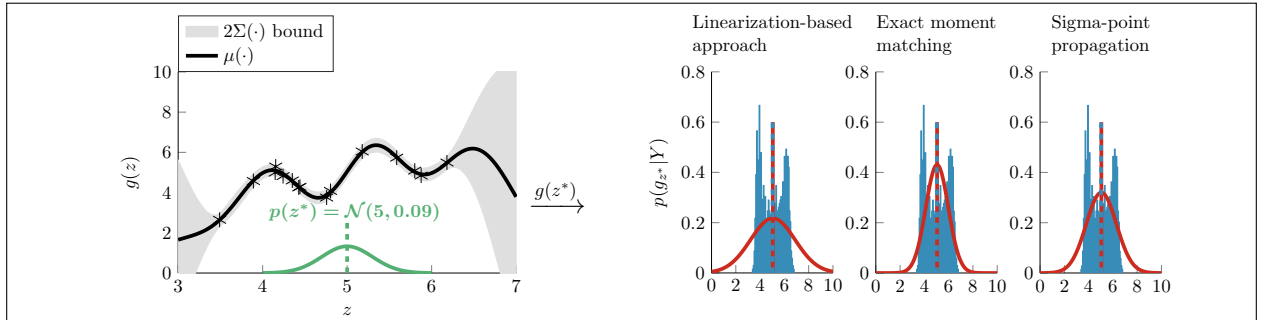

\begin{mdframed}
    \hspace{0.5in}
    \begin{minipage}[t]{0.5\textwidth}
    \includegraphics[scale=0.7]{figure1.tex}
    \end{minipage}%
    \hspace{-0.5in}
    \begin{minipage}[t]{0.5\textwidth}
    \includegraphics[scale=0.7]{figure2.tex}
    \end{minipage}    
    
\caption{Independent one-step-ahead predictions: (Left panel) Gaussian process trained on 15 data points (black stars) and evaluated on a uniform grid in the interval $z\in[3,7]$. The plots show the predictive mean in black and the bounds obtained by the predictive variance in gray. The random input $z^* \sim \mathcal{N}(5,0.09)$ on which the Gaussian process is evaluated is shown in green. (Right panel) Blue histograms depict the numerical approximation of $p(g_{z^*}|Y)$ (see Section~\ref{sub:numapproxuncprop}) that is compared against its Gaussian approximations obtained via (from left to right):  linearization-based approach (Section~\ref{subsubsec:linearization}), exact moment matching (Section~\ref{sub:exactmomentmatch}), and sigma-point propagation (Section~\ref{sub:sigmapoint}), all shown in red; dotted lines represent the means. \\ Note that all the Gaussian approximations correctly identify the first moment of $p(g_{z^*}|Y)$, but cannot model bimodality. The linearization-based approach, in this case, tends to particularly overestimate uncertainty -- however, depending on the sign of the correction terms, it may also lead to underestimation. Exact moment matching, by definition~\eqref{eq:var_emm}, computes the exact second moment and sigma-point propagation shows a similar behavior.}
    \label{fig:one-step-prop}
    \end{mdframed}
\end{figure*}

In Section \ref{sec:background}, single predictions were computed with~\eqref{eq:allpost} and by assuming that the input locations $Z^*$ are deterministic. However, when performing a series of prediction steps, in which the successively evaluated input corresponds to the currently predicted output, then $Z^*$ is random. In this case, assuming $Z^* \sim p(Z^*)$, the predictive distribution is obtained as
\begin{equation}
    p(g_{Z^*} | Y) = \int p(g_{Z^*} | Z^*, Y) p(Z^*)dZ^*, 
    \label{eq:random_input}
\end{equation}
where $p(g_{Z^*} | Z^*, Y) = \mathcal{N}(\mu(Z^*), \Sigma(Z^*))$ is the Gaussian posterior distribution specified in~\eqref{eq:allpost}.
Note that, even if the current predicted output is Gaussian distributed, i.e.,  when the distribution is evaluated at a deterministic input location, the next prediction will generally lose Gaussianity. In fact, in this case, $p(Z^*) = \mathcal{N}(\mu^*, \Sigma^*)$: therefore, the computation of the predictive distribution in equation~\eqref{eq:random_input} requires integrating the product of two Gaussian distributions, for which in general no closed-form solution exists. 
One option to overcome this intractability issue is to apply successive Gaussian approximations of $p(g_{Z^*} | Y)$, thereby requiring only the derivation of the first and second moments. Using the law of iterated expectations and conditional variance, we obtain the following~\cite{girard_gaussian_2002-1}: 
\begin{subequations}
    \begin{align}
    \mathbb{E}_{g_{Z^*}}[g_{Z^*} | Y] & = \mathbb{E}_{Z^*}[\mathbb{E}_{g_{Z^*}}[g_{Z^*} | Z^*, Y]] \notag \\ & \overbrace{=}^{\eqref{eq:allpost}} \mathbb{E}_{Z^*}[\mu(Z^*)], \label{eq:mean_approx_gp} \\       \text{Var}_{g_{Z^*}}[g_{Z^*} | Y] & = \mathbb{E}_{Z^*}[\text{Var}_{g_{Z^*}}[g_{Z^*} | Z^*, Y]] \notag \\ &\quad + \text{Var}_{Z^*}[\mathbb{E}_{g_{Z^*}}[g_{Z^*} | Z^*, Y]], \nonumber \\
         \overbrace{=}^{\eqref{eq:allpost}} \mathbb{E}_{Z^*}&[\Sigma(Z^*)] + \text{Var}_{Z^*}[\mu(Z^*)]. \label{eq:var_approx_gp}
    \end{align}\label{eq:gaussian_approximation}%
\end{subequations}%
These expressions can be approximated using techniques discussed in the following subsections and depicted in Figure~\ref{fig:one-step-prop}. Finally, note that in Sections~\ref{subsubsec:linearization},~\ref{sub:exactmomentmatch}, and~\ref{sub:sigmapoint}, we consider a single test point $z^*$, since the case of a collection of test points $Z^*$ is notationally more involved.  

\subsubsection{Linearization-based approach}
\label{subsubsec:linearization}
Assume that the query input $z^*$ is a Gaussian random variable $z^* \sim \mathcal{N}(\mu^*, \Sigma^*)$. 
The linearization-based approach proposed in~\cite{girard_gaussian_2002-1,girard_gaussian_2002} considers the first two moments of the predictive distribution $p(g_{z^*} | z^*, Y) = \mathcal{N}(\mu(z^*), \Sigma(z^*))$ entering~\eqref{eq:gaussian_approximation} and computes a first-order Taylor expansion of $\mu(z^*)$ and a second-order Taylor expansion of $\Sigma(z^*)$ around $\mu^*$, i.e.,
\begin{subequations}
\begin{align}
       & \mu(z^*) \approx \mu(\mu^*) + \nabla_z\mu(z)|_{z=\mu^*}^\top(z^* - \mu^*),  \\
       & \Sigma(z^*) \approx \Sigma(\mu^*) + \nabla_z\Sigma(z)|_{z=\mu^*}^\top(z^* - \mu^*) + \notag \\& \qquad + \frac{1}{2}(z^* - \mu^*)^\top\nabla_z^2\Sigma(z)|_{z=\mu^*}(z^* - \mu^*),\label{eq:covariance_expansion}
\end{align}%
\label{eq:linpropagationgen}%
\end{subequations}%
\begin{figure*}[h!]
\hrule
\begin{subequations}
    \begin{align}
        \mathbb{E}_{z^*}[\mu(z^*)] & \approx \mathbb{E}_{z^*}[\mu(\mu^*) + \nabla_z\mu(z)|_{z=\mu^*}^\top(z^* - \mu^*)] = \mu(\mu^*), \\
        \mathbb{E}_{z^*}[\Sigma(z^*)]  &\approx \mathbb{E}_{z^*}[\Sigma(\mu^*) + \nabla_z\Sigma(z)|_{z=\mu^*}^\top(z^* - \mu^*) \notag + \frac{1}{2}(z^* - \mu^*)^\top\nabla_z^2\Sigma(z)|_{z=\mu^*}(z^* - \mu^*)] \nonumber \\
        & = \Sigma(\mu^*) + \frac{1}{2}\text{Tr}\{ \nabla_z^2\Sigma(z)|_{z=\mu^*} \Sigma^*\}, \\
        \text{Var}_{z^*}[\mu(z^*)] & \approx \text{Var}_{z^*}[ \mu(\mu^*) + \nabla_z\mu(z)|_{z=\mu^*}^\top(z^* - \mu^*)] = \nabla_z\mu(z)|_{z=\mu^*}^\top \Sigma^* \nabla_z\mu(z)|_{z=\mu^*}.
    \end{align}%
\label{eq:linpropspec}%
\end{subequations}%
\hrule
\end{figure*}%
where $\nabla_z\mu(z)|_{z=\mu^*}$ and  $\nabla_z\Sigma(z)|_{z=\mu^*}$ are the gradients of the predictive mean and variance evaluated at $\mu^*$ respectively, and $\nabla_z^2\Sigma(z)|_{z=\mu^*}$ is the Hessian of the predictive variance evaluated at $\mu^*$. 
This approximation is then used to obtain expressions of the expectations and variances needed in~\eqref{eq:gaussian_approximation}: the expressions are given in~\eqref{eq:linpropspec}.
Note that in the context of MPC, often only a first-order Taylor expansion of the predictive variance is considered in order to simplify the expressions and speed up the computations~\cite{hewing_cautious_2020,hewing_cautious_2018}. This simplified linearization approach is also considered when propagating the uncertainty for sparse-spectrum Gaussian processes~\cite{pan_prediction_2017}.

\subsubsection{Exact moment matching}\label{sub:exactmomentmatch}
The idea of moment matching was originally studied in~\cite{opper_bayesian_1999}, where it is pointed out that for distributions belonging to the exponential family, equating the first two statistical moments (i.e., projecting the original distribution on the space of Gaussian ones) is the result of minimizing the Kullback-Leibler divergence between the two probability measures involved. \\

When the kernel function is the squared-exponential one given in~\eqref{eq:gaussker} and $z^* \sim \mathcal{N}(\mu^*, \Sigma^*)$, the expressions in~\eqref{eq:gaussian_approximation} admit a closed form, i.e., the associated integrals can be computed exactly~\cite{quinonero-candela_propagation_2003, quinonero-candela_prediction_2003, deisenroth_analytic_2009}. 

The predictive mean results in the following:
\begin{equation}
    \mathbb{E}_{z^*}[\mu(z^*)] = \int \mu(z^*)p(z^*)dz^*,
\end{equation}
which is the integral in $z^*$ of the product of Gaussian-shaped functions~\cite{quinonero-candela_propagation_2003}. This allows for computing  $\int \mu(z^*)p(z^*)dz^* = \beta^\top \mathbf{l}$, where $\beta = (\Ker_{Z,Z} + \sigma_w^2I_N)^{-1}Y$, and $\mathbf{l} = [l_1, \dots, l_N]^\top$, whose $i$-th component is a function of $z_i,\mu^*,\Sigma^*$ and of the kernel hyper-parameters~\cite[Equation (33)]{quinonero-candela_prediction_2003} for each~$i=1,\dots, N$. 

The predictive variance can be computed as
\begin{align}
    &\mathbb{E}_{z^*}[\Sigma(z^*)] + \text{Var}_{z^*}[\mu(z^*)]  \label{eq:var_emm}\\ = & \mathbb{E}_{z^*}[\Sigma(z^*)] + \mathbb{E}_{z^*}[\mu^2(z^*)] - \mathbb{E}_{z^*}[\mu(z^*)]^2 \nonumber \\
     = & \beta^\top L \beta - \text{Tr}\{ (\Ker_{Z,Z} + \sigma_w^2I_N)^{-1} L \} - (\beta^\top \mathbf{l})^2,\notag
\end{align}
where each $i,j$-th entry, $i,j = 1, \dots, N$, of the matrix $L$ is a function of $[\Ker_{\mu^*, Z}]_{i,j}, \: \mu^*, \: \Sigma^*$ and of the kernel hyper-parameters. \\

Closed-form expressions of the terms in~\eqref{eq:gaussian_approximation} are available also for multivariate outputs~\cite{kus_gaussian_2006}, as well as for polynomial kernels~\cite{deisenroth_robust_2012}, periodic kernels~\cite{hajighassemi_analytic_2014}, finite-dimensional kernel representations using Gaussian basis functions~\cite{quinonero-candela_prediction_2003}, and for sparse-spectrum Gaussian processes~\cite{pan_prediction_2017}.

\subsubsection{Sigma-point propagation}\label{sub:sigmapoint}
Another approach to simulate the state trajectory via iterative one-step-ahead predictions is presented in~\cite{hardy_multiple-step_2014, ostafew_robust_2016, ko_gp-bayesfilters_2008}. It builds upon the sigma-point transformation, which is typically deployed in the unscented Kalman filter~\cite{julier_unscented_2004} (see also Section~\ref{subsubsec:alt_funct_learning_state_inf}). 
Its key idea is that of approximating a continuous probability distribution by means of a finite set of (sigma-)points, whose empirical mean and covariance mimic their continuous counterparts: in other words, moment matching is performed between the original distribution and a set of suitably chosen, weighted points. In our context, the procedure is the following: first, we compute the sigma-points for $z^*\sim(\mu^*,\Sigma^*)$, which we denote by $\{\bar z_j^*\}_{j=0}^{2n_z}$, where $n_z$ is the dimensionality of $z$. These are typically set as:
\begin{subequations}
\begin{align}
& \bar z^*_0 = \mu^* \\
& \bar z^*_j = \mu^* + \sqrt{n_z + \lambda_{mm}}[\text{chol}(\Sigma^*)]_j \\& \qquad \text{for } j=1,\dots,n_z, \notag \\
& \bar z^*_j = \mu^* - \sqrt{n_z + \lambda_{mm}}[\text{chol}(\Sigma^*)]_{j-n_z}, \\ & \qquad \text{for } j=n_z+1,\dots,2n_z, \notag
\end{align}
\end{subequations}
where $[\text{chol}(\Sigma^*)]_j$ is the $j$-th column of the Cholesky factorization of matrix $\Sigma^*$, and $\lambda_{mm}$ is a user-defined parameter representing how far the sigma-points are spread from the mean. At this point, the sigma-points must be propagated through the Gaussian distribution $p(g(\cdot)|Y)$; this is possible by viewing the Gaussian process $g(\cdot)$ as the sum of two components $g(\cdot) = \mu(\cdot) + \tilde w(\cdot)$ with $\tilde w(\cdot) \sim \mathcal{N}(0,\Sigma(\cdot))$ -- i.e., treating $\tilde w(\cdot)$ as ``process noise". Therefore, we can approximate $\mathbb{E}_{g_{z^*}}[g_{z^*} | Y]$ by evaluating the predictive mean $\mu(\cdot)$ on the sigma-points and then computing the propagated sigma-point weighted sample mean:
\begin{equation}
    \mathbb{E}_{g_{z^*}}[g_{z^*} | Y] \approx \mu_{\text{sp}} = \sum_{j=0}^{2 n_z} W^m_j \mu(\bar z^*_j).
\end{equation}
Regarding $\text{Var}_{g_{z^*}}[g_{z^*} | Y]$, we approximate it as the sum of the propagated sigma-point weighted sample variance and $\Sigma(\cdot)$ evaluated at $\mu^*$, i.e.,
\begin{align}
    &\text{Var}_{g_{z^*}}[g_{z^*} | Y] \\ &\approx \sum_{j=0}^{2 n_z} W^v_j (\mu(\bar z^*_j)- \mu_{\text{sp}})(\mu(\bar z^*_j) - \mu_{\text{sp}})^T + \Sigma(\mu^*).\notag
\end{align}
The weights $W^m_j, W^v_j,$ for $j=0,\dots,2n_z$, are typically chosen such that they sum to 1. Some examples on how to tune them can be found in~\cite{julier_unscented_2004}.

\subsubsection{Numerical Approximation}\label{sub:numapproxuncprop}
The integral in~\eqref{eq:random_input} can also be computed via a sampling-based approximation, e.g., via Monte Carlo integration. This allows for obtaining a more accurate representation of the predictive distribution, at the price of an increased computational overhead.
An approach that effectively tackles this issue, based on~\cite{haylock_inference_1996}, was developed in~\cite{oakley_bayesian_2002}, where the sampling process is guided by a measure of correlation to the previous evaluations. More recently, the work in~\cite{vinogradska_stability_2016} proposes a numerical quadrature to approximate the predictive distribution, which allows for computing an additional success probability of the predicted outputs to belong to a target set. This can be useful for constructing reliable confidence regions, which are a crucial ingredient for incorporating Gaussian processes in MPC. 

\subsubsection{Alternative Gaussian Models}\label{sub:alternativegaussmod}
The methods previously described do not exploit the underlying distribution of the input data in a principled manner, i.e., the uncertainty is only handled during inference at test time via practical approximations (analytical or sampling-based). An alternative consists of directly modeling the input uncertainty in the Gaussian process at training time~\cite{mchutchon_gaussian_2011}. The idea is to exploit an input latent variable space and a variational approach for approximately propagating the densities, which connects with the approaches outlined in Section~\ref{subsubsec:latent_state_opt}. Within such a class of methods,~\cite{damianou_semi-described_2015}  proposes a semi-described learning procedure (i.e., taking into account noise in the input locations) that builds on the variational approach presented in~\cite{titsias_bayesian_2010}, and shows that the proposed latent variable Gaussian process models can be used to obtain predictions in an iterative manner. 

Another approach based on~\cite{quinonero-candela_prediction_2003} proposes to use Gaussian mixture models to refine the approximation mainly when a simple unimodal (Gaussian) distribution is too far from the actual predictive distribution. As a measure for detecting non-Gaussianity, the authors in~\cite{hardy_multi-step_2015} use a sampling-based evaluation of the kurtosis.

\begin{SCfigure*}
    \centering 
    \includegraphics[scale=0.55]{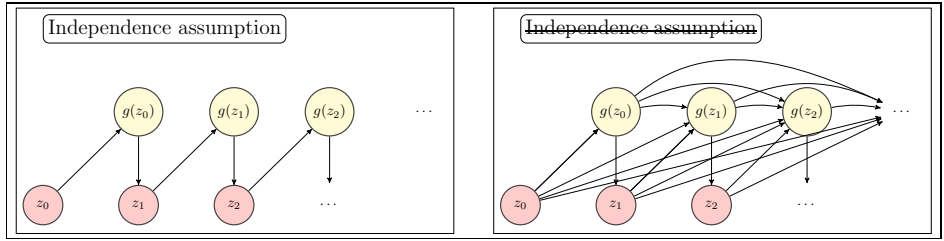}
\caption{Visualization of the dependencies among states and Gaussian process model in the case in which there is Markovianity (left panel) and in the one where all states enter the evaluation of the Gaussian process (right panel).}
\label{fig:dag_independence}
\end{SCfigure*}

\subsection{Robust uncertainty propagation}
\label{subsec:robust_prop}
Most of the literature covered in the previous subsections proposes approaches for performing multi-step-ahead time-series predictions with Gaussian processes, with the aim of exploiting the intrinsic uncertainty quantification directly available from the underlying distribution associated with $g$. This idea of maintaining a strict relation to the probabilistic nature of Gaussian process models poses a challenging scenario for safety-critical systems, for which a careful quantification of the uncertainty is crucial to determine control actions that fulfill constraints at all times. For this reason, an alternative approach that does not exploit the \textit{stochastic} nature of Gaussian processes for propagating the associated uncertainty was proposed: its rationale consists in leveraging an assumed deterministic structure on the function to be estimated, thus enabling a \textit{robust} uncertainty propagation by means of confidence bounds.\\ 

The paper that follows such a rationale is~\cite{koller_learning-based_2018}, where the idea is to assume that the unknown function $g(\cdot)$ has bounded norm in the RKHS $\mathcal{H}$ (see the textbox titled \textit{Kernel-based methods: the view from RKHSs} on page~\pageref{sec:boxkernel}), induced by the continuously differentiable kernel $\Ker(\cdot,\cdot)$ (see Section~\ref{subsec:overview}). Under these assumptions, the authors devise a strategy for generating robust multi-step predictions by first characterizing one-step-ahead predictions that exploit such function properties. Defining the dynamics of one state dimension as $g_{\text{tot}}(z) = \fnom(z) + g(z)$ (note that the constants that characterize the bound in~\eqref{eq:robust_bound} are different for each state dimension), the first step consists in linearizing the nominal model $\fnom$ around $z = \bar z$ as $\fnom(z) \approx \fnom(\bar z) + \nabla_z \fnom(z)|_{z=\bar{z}}^\top (z - \bar z)$; the second step involves a locally constant approximation of the Gaussian process residual model, i.e., $g(z) \approx g(\bar{z})$. 
Then, in combination with the Gaussian process confidence interval $|g(z) - \mu(z)| \leq \beta_{\mathcal{H}} \Sigma(z)$ (where $\beta_\mathcal{H}$ is chosen according to the results in~\cite{srinivas_information-theoretic_2012,chowdhury_kernelized_2017,fiedler_practical_2021} -- see also~\eqref{eq:boundKrause} in Remark~\ref{rmk:uncertainty}), the authors provide an approximation error bound that depends on the scaling $\beta_\mathcal{H}$, $\Sigma(\cdot)$, the Lipschitz constant $L_{\nabla \fnom}$ of $\nabla_z \fnom$, and the Lipschitz constant $L_{g}$ of $g$. This reads as
\begin{align}
    &|g_{\text{tot}}(z) - \widetilde{g}_{\text{tot}}(z)| \label{eq:robust_bound}\\ & \leq \beta_\mathcal{H} \Sigma(\bar z) + \frac{L_{\nabla_z \fnom}}{2}|| z - \bar z||^2_2 + L_{g}|| z - \bar z||_2, \notag
\end{align}
where $\widetilde{g}_{\text{tot}}(z) = \fnom(\bar z) + \nabla_z \fnom(z)|_{z=\bar{z}}^\top (z - \bar z) + \mu(\bar z)$. This bound implicitly characterizes a confidence hyper-box whose dimension depends on the right-hand side of the bound in~\eqref{eq:robust_bound}. 

Assume now that $z \in \mathcal{E}(z_p, Q_p)$, i.e., $z$ belongs to an ellipsoidal confidence region centered at $z_p$ and whose shape depends on the matrix $Q_p$. Then, propagating this region through the mapping $\widetilde{g}_{\text{tot}}(\cdot)$ preserves its ellipsoidal shape, 
being $\widetilde{g}_{\text{tot}}(\cdot)$ an affine transformation. 
Additionally, the residual uncertainty characterized by the hyper-box induced by~\eqref{eq:robust_bound} can be over-bounded by an ellipsoid as well. The sum of these two ellipsoidal regions ultimately determines the final approximate ellipsoid containing all possible one-step-ahead predictions starting from~$\mathcal{E}(z_p, Q_p)$.

Overall, the approach presented in~\cite{koller_learning-based_2018} is rigorous and computationally practical, as its propagated uncertainty quantification relies on the calculation of ellipsoids. Nevertheless, it suffers from the weaknesses of the bounds~\eqref{eq:boundKrause}, relying on constants that are difficult to evaluate, and tends to be overly conservative  due to the subsequent ellipsoidal over-approximations based on global Lipschitz bounds of the unknown function. 

\subsection{Overcoming the independence assumption}
\label{subsec:no_indep_ass}
Recursive one-step-ahead predictions implicitly assume that correlations between non-consecutive states in the overall predicted trajectory can be neglected, i.e., encompass an underlying notion of Markovianity (see Figure~\ref{fig:dag_independence}). 
In the last years, several approaches have addressed this issue~\cite{lambert_learning_2021}. For instance,~\cite{conti_gaussian_2009, umlauft_scenario-based_2018} present a way for forward-sampling the whole trajectory 
by iteratively reconditioning the Gaussian-process model,
enabling a scenario-based~\cite{calafiore_scenario_2006} control design. 
An efficient implementation of the forward-sampling strategy has been proposed by~\cite{prajapat_towards_2024}, 
which enables parallel-in-time sampling using a simultaneous implementation of the optimal control problem within a sequential quadratic programming framework. The proposed strategy recovers the true trajectory distribution in the asymptotic regime for all candidate functions in the RKHS: thus, it guarantees robust-in-probability safety 
(see Section~\ref{subsec:RIP})
as the number of samples approaches infinity.
The work in~\cite{bradford_nonlinear_2019, bradford_stochastic_2020} develops a Gaussian processes-based MPC approach where explicit back-offs (i.e., constraint tightenings) are designed along the lines of~\cite{paulson_nonlinear_2018} by using 
scenario-based uncertainty quantification. 

While sampling from the trajectory distribution overcomes the independence assumption, it is affected by the well-known 
``curse of dimensionality", particularly for large time horizons. Alternative approximate (but computationally cheaper) approaches were proposed in~\cite{bradford_efficient_2018,hewing_simulation_2020}, where trajectories are sampled from a finite-dimensional approximation of the Gaussian process, and corrective correlation terms are introduced following the linearization-based approach discussed in Section~\ref{subsubsec:linearization}, respectively. 
An additional analytical approach that explicitly models the correlations in the state trajectories can be found in~\cite{ialongo_overcoming_2019}, which makes use of the Gaussian process state-space model framework derived in~\cite{frigola_variational_2014} (for additional references, see Section~\ref{subsubsec:alt_funct_learning_state_inf}). A similar framework is employed in~\cite{beckers_prediction_2022} and addresses the computational aspect by proposing tractable memory management.

\section{CLOSED-LOOP SAFETY GUARANTEES}\label{sec:safetyguar}
\begin{figure*}
\begin{mdframed}[backgroundcolor=gray!5, hidealllines=true]
    \centering
\begin{tikzpicture}[mindmap, concept color=blue!20]
\begin{scope}[scale=0.7,transform shape]
\node [concept] {\ref{sec:safetyguar} Closed-loop safety guarantees}
    child[concept color=red!10,grow=190] {node[concept] {\ref{sub:boundedsupport}  Bounded support assumption}}
     child[concept color=red!10,grow=160] {node[concept] {\ref{subsec:RIP} Robust-in-probability. }}
    child[concept color=red!10,grow=20] {node[concept] {\ref{sub:samplingclosed} Sampling-based approaches}}
    child[concept color=red!10,grow=-10] {node[concept] {\ref{sub:additionalclosed} Additional results}};

\end{scope}
\end{tikzpicture}
\caption{Graphical overview of Section~\ref{sec:safetyguar}.}
\label{fig:mindmap_sec6}
\end{mdframed}
\end{figure*}
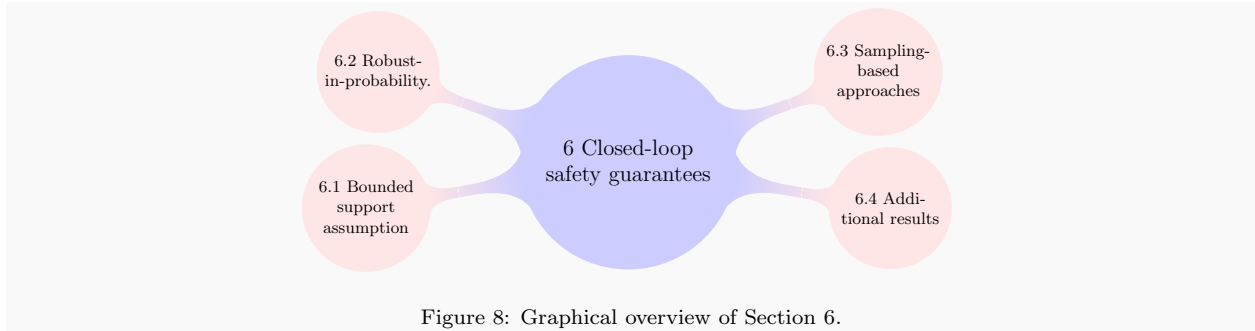

As specified in problem~$(\mathcal{P})$ in Section \ref{sec:problem}, the goal is to design a control law ensuring that the closed-loop states and inputs satisfy chance-constraints 
within the specified probability levels. The sequence of policies $\pi_i$ specified in the optimal control problem~\eqref{eq:SOCP} can be obtained via the approximate solution~\eqref{eq:optimization_problem}, based on a receding-horizon strategy. In this section, we provide an overview of the proposed tools to build the constraint sets in~\eqref{eq:constr}. 
Note that, in general, one aims at obtaining chance constraints that hold jointly in time -- however, computing the corresponding constraint sets directly can be difficult. Thus, some of the reviewed approaches rely on taking the union of constraint sets that hold point-wise in time, trading off computational tractability with increased conservatism. \\

The remainder of this section, whose structure is pictorially summarized in Figure~\ref{fig:mindmap_sec6}, unfolds as follows. We first discuss approaches that 
are oblivious to
unlikely model error realizations 
of
the underlying Gaussian process distribution (Section~\ref{sub:boundedsupport}). 
In this way, one can work with compact confidence sets, thereby allowing the use of established \textit{robust} MPC tools. While the bounded-support assumption is accurate for one-step-ahead predictions, its validity is lost as soon as one considers multi-step predictions: therefore,
chance-constraint satisfaction at all times with a pre-specified probability, i.e., as defined in~\eqref{eq:OCconstr},  
cannot be provided. 
To guarantee this definition of safety, one possibility is to first construct an appropriate high-probability region for the model uncertainty that explicitly takes into account the probability of Gaussian process outliers and then to analyze the control framework for all function realizations inside that region: the resulting \emph{robust-in-probability} method is reviewed in Section~\ref{subsec:RIP}. We additionally mention in Section~\ref{sub:samplingclosed} sampling-based approaches that are generally useful to reduce the conservativeness of such a chance-constraint reformulation. Finally, in Section~\ref{sub:additionalclosed} we discuss alternative approaches that neither exploit sampling-based nor robust tools for analyzing Gaussian process-based MPC schemes.

\subsection{Bounded support assumption}\label{sub:boundedsupport}
An established approach in the learning-based control literature consists in handling performance and safety separately~\cite{aswani_provably_2013}. The learned model is used for computing the cost function, while a fixed-uncertainty model based on either parametric or non-parametric estimation is assumed to be known before operation and is used to perform robust control design. In view of this, it is then key to assume that the model error $g(\cdot)$ lies in a compact set -- a heuristic based on the practical assumption that unbounded disturbances (such as the ones modeled by the Gaussian variables in $g(\cdot)$) usually do not occur. Such a compact set is typically computed from the Gaussian distribution employing the quantile function of the $\chi^2$-distribution, defining a sufficiently inclusive range for the overall model uncertainty.

Based on this framework,~\cite{bonzanini_learning-based_2022} proposes to compute state-dependent back-off parameters by using the empirical cumulative distribution function, similarly to~\cite{paulson_nonlinear_2018}, to reformulate individual (i.e., point-wise in time) chance constraints. Under the assumption that the model errors (i.e., Gaussian process realizations) lie within a state-dependent compact set
$\mathcal{G}(x,u)$, 
the authors apply offline a non-conservative and time-varying tightening of the reformulated chance-constraints based on data, as in~\cite{soloperto_learning-based_2018}. Ultimately, they establish recursive feasibility, closed-loop stability, and robust constraint satisfaction of the nominal system. A similar analysis is carried out in~\cite{soloperto_guaranteed_2023}; however, in that work the learning aspect is treated by modifying the MPC problem. Specifically, the proposed approach includes a novel constraint and introduces a suitable discount factor in the cost guaranteeing uncertainty reduction while learning in closed-loop (which is generally not obvious in the Bayesian framework). 
Note that, in all these works, the nominal system is typically considered without loss of generality to be linear, since the linearization error can be easily included as an additive term to the model error $g(\cdot)$. In this linear setup, the typical approach in a robust MPC framework is to parametrize the control law as $\pi_{i|k}(x) = K x + c_{i|k}$, where $K$ is a stabilizing feedback gain for the linearized system, while the MPC problem optimizes over the sequence $\{ c_{i|k} \}_{i=0}^{T-1}$.\\

Within the bounded-support setting, it is also worth mentioning the line of work exploiting robust kernel-based bounds available under the assumption that the unknown function $g(\cdot)$ belongs to an RKHS and that the noise $w$ affecting the function is bounded~\cite{maddalena_deterministic_2021,wakulicz_informative_2022,scharnhorst_robust_2023}. Despite assuming structural function properties that are related to Gaussian processes, in these works $g(\cdot)$ is not a stochastic process, but a \textit{deterministic} member of the RKHS. The kernel-based bounds are used in an MPC framework exploiting multi-step predictors~\cite{maddalena_kpc_2021} while computing safe open-loop trajectories. 

\subsection{Robust-in-probability}
\label{subsec:RIP}
The bounded-support heuristic enables state-of-the-art robust analysis, but is reliable only for one-step-ahead predictions: because it lacks a-priori quantification of the confidence level with which the unknown function is actually included in the bound, such a scheme does not guarantee the notion of closed-loop safety in equation~\eqref{eq:OCconstr}. 
In this subsection, we focus on methods based on constructing a reliable confidence region for the model error $g(\cdot)$ to be satisfied with required probability $p$ of constraint satisfaction. Specifically, defining $\triangle$ as the event 
\begin{align}
    \triangle = 
    \Big\lbrace & g(x,u) \in \mathcal{G}(x,u) \;  \text{for all} \; (x,u) \in \mathcal{Z} \Big\rbrace, 
    \label{eq:triangleDef}
\end{align}
we consider methods which, leveraging concentration inequalities, guarantee that
\begin{equation}
    \mathbb{P}(\triangle) \geq p.
\label{eq:confidence_region} 
\end{equation}
The construction of a confidence region guaranteeing~\eqref{eq:confidence_region} allows for isolating the main stochastic uncertainty source, namely the model error described by a Gaussian process. At this point, a robust controller valid for all uncertainty realizations inside the confidence region can be designed, ensuring that
\begin{equation}
    \mathbb{P}( h_j(x_i,u_i) \leq 0, \; \forall i \geq 0, \; \forall \{j\}_{j=1}^{n_h}\; | \; \triangle ) = 1. \notag
\end{equation} 
By~\eqref{eq:confidence_region}, this implies
\begin{align}
&\mathbb{P}\Big(h_j(x_i,u_i) \leq 0, \; \forall i \geq 0, \; \forall \{j\}_{j=1}^{n_h}\Big) \label{eq:robust_in_prob} \\ &\geq \mathbb{P} \Big( h_j(x_i,u_i) \leq 0, \; \forall i \geq 0, \; \forall \{j\}_{j=1}^{n_h}\; | \triangle \Big) P(\triangle) \notag \\ &\geq p: \notag 
\end{align} 
that is, joint-in-time robust constraint satisfaction is ensured for all functions belonging to the set $\mathcal{G}(x,u)$, which by~\eqref{eq:confidence_region} spans a fraction $p$ of the whole hypothesis space. In other words, \textit{robustness-in-probability} is ensured.  
Note that $p$ can be chosen to satisfy the individual probabilities $p_j$ in problem~\eqref{eq:SOCP} by using Boole's inequality.\\

One possibility for constructing such confidence regions in~\eqref{eq:triangleDef} is provided in~\cite{srinivas_information-theoretic_2012} (see also~\eqref{eq:boundKrause} in Remark~\ref{rmk:uncertainty_paradigms}), where concentration bounds are derived based on the assumption that the function sample belongs to an RKHS. Most importantly, it requires mild assumptions on the function itself and it is independent of the Gaussian process training points, i.e., the bound still holds no matter if data are added or removed (as long as the number of data points remains constant). This result was improved in~\cite{chowdhury_kernelized_2017} and used for instance in model-based reinforcement learning~\cite{berkenkamp_safe_2017}. Assuming that the model errors have bounded norm in an RKHS, and therefore are Lipschitz-continuous, one can determine a policy that stabilizes a system while maintaining it safe at all times with pre-specified probability. This is possible by improving the knowledge of the system via safe model learning, which in turn continuously improves the policy and enlarges the estimated region of attraction~\cite{berkenkamp_safe_2017}. In~\cite{koller_learning-based_2018} the tools are similar, but the authors additionally introduce an uncertainty propagation scheme based on ellipsoids (see Section~\ref{subsec:robust_prop}); recursive feasibility is guaranteed thanks to the existence of a safe backup controller. Related work conducted in~\cite{wabersich_nonlinear_2021} is developed for nonlinear systems affected by nonlinear parametric uncertainty that can be modeled as a Gaussian process. The idea is to quantify the model uncertainty via a set-valued confidence map while conducting the analysis leveraging tube-based MPC concepts.

An alternative to these works is presented in~\cite{maiworm_stability_2018}, where the authors propose an output feedback nonlinear model predictive controller using the Gaussian-process mean as a prediction model for a NARX system. In this case, input-to-state robust stability is proven to hold in probability for the prediction error between the Gaussian process and the true dynamics, provided that the modeling error is bounded in probability as in~\eqref{eq:confidence_region}. However, in~\cite{maiworm_stability_2018} the existence of such a bound is only postulated to exist in theory, but no tool is provided for its computation. 
In~\cite{maiworm_online_2021} the authors extend the previous framework to account for data updates only if the updated prediction error is smaller than a certain threshold in order to preserve guarantees. \\

A complete overview of available uncertainty bounds for Gaussian processes is provided in~\cite{fiedler_practical_2021}, where the authors additionally propose a rigorous and practical bound that can be computed based on reasonable and established assumptions, and such that the obtained confidence regions are robust against model mis-specifications (at least to some extent). These results are for instance used in~\cite{nguyen_high-probability_2022}, where a Gaussian process-based MPC policy is computed for systems in which the nonlinearity depends only on the state and not on the input. The employed high-probability uncertainty bound holding for all times provides rigorous robust closed-loop guarantees in terms of safety and stability as formulated in~\eqref{eq:robust_in_prob}.

\subsection{Sampling-based approaches}\label{sub:samplingclosed}

The combined use of uncertainty bounds for Gaussian processes and robust tools enables on the one hand a systematic framework for discarding unlikely scenarios, and on the other the possibility of picking from a very rich literature of robust MPC approaches. The drawback of this strategy is mainly determined by the confidence region constructed for the model error, which is often based on Gaussian process uncertainty bounds that are either hard to obtain or too large for practical use. To overcome such an issue, sampling-based approaches, which we now review, have been proposed.

In~\cite{bradford_nonlinear_2019,bradford_stochastic_2020,ma_model-and_2023} the authors propose a constraint tightening method based on Gaussian-process samples obtained using the exact sampling method given in~\cite{umlauft_scenario-based_2018} (see also Section~\ref{subsec:no_indep_ass}). Chance-constraints are reformulated using the empirical cumulative distribution function as in~\cite{paulson_nonlinear_2018}, devising both an offline and an online strategy. The latter requires more computational capacity since it involves online learning. Additionally, 
joint-in-time 
chance-constraint satisfaction with pre-specified probability can be proven, provided a shrinking-horizon formulation of the Gaussian process-based MPC problem, where the unknown model plant follows exactly the identified Gaussian process distribution; this result is generally referred to as \textit{feasibility probability}. 

Note that, while these designs typically enable a non-conservative constraint tightening that 
guarantees closed-loop chance constraint satisfaction according to~\eqref{eq:OCconstr}, 
they 
require an expensive offline design phase for each initial condition
and
are restricted to shrinking-horizon implementations and feasibility-probability guarantees. 

\subsection{Additional results}\label{sub:additionalclosed}

In the previous sections, we have discussed methods that can 
be used to satisfy
joint-in-time 
chance-constraints, 
which is typically possible either by constructing appropriate confidence regions or by approximating them via samples. However, there exists a rich literature in both linear~\cite{farina_stochastic_2016} and nonlinear stochastic MPC~\cite{mcallister_nonlinear_2023} providing analytical tools that are not based on either approach, but which generally consider point-wise-in-time chance constraints. For instance, the work in~\cite{gruner_recursively_2022} considers a linear model with additive time-dependent disturbance modeled as a Gaussian process. The latter is formulated via a temporal state-space representation both to avoid the computational burden and to leverage the concept of indirect feedback in~\cite{hewing_recursively_2020} for providing closed-loop guarantees. In~\cite{matschek_safe_2023}, the authors use Gaussian processes to learn a static mapping from states to forces and moments, therefore avoiding the issue of uncertainty propagation. This setup simplifies the analysis for determining guarantees since the learning component does not interfere either with the states or with the dynamics. The constraints are tightened based on posterior bounds on the estimated forces and moments, and the analysis is carried out by leveraging the path-following framework given in~\cite{faulwasser_nonlinear_2016}.\\

Recent advances have been made also in the analysis of Gaussian process state-space models discussed in Section~\ref{subsubsec:latent_state_opt}. For instance, in~\cite{beckers_equilibrium_2016} the concept of equilibrium for a stochastic system description is defined, and its stability properties are analyzed. In the case of a squared-exponential kernel, mean-square boundedness of the model is established. In~\cite{umlauft_learning_2017}, stochastic stability is directly enforced in the learning phase of the model by using sum-of-squares as a control Lyapunov function. The question of how to exploit stochastic notions for direct analysis of Gaussian process-based MPC frameworks is still open for future endeavors.

\section{DISCUSSION}\label{sec:discussion}
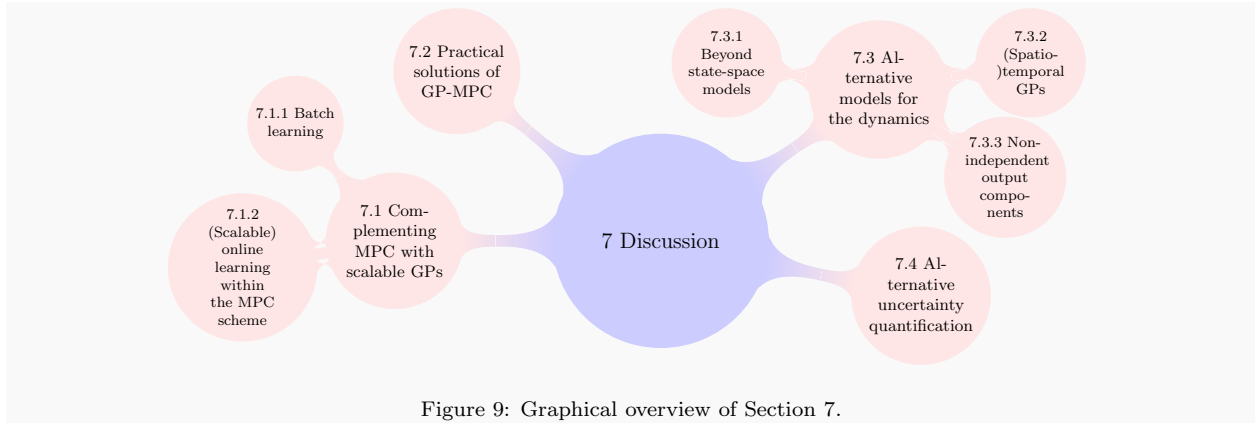
\begin{figure*}[h!]
\begin{mdframed}[backgroundcolor=gray!5, hidealllines=true]
    \centering
\begin{tikzpicture}[mindmap, concept color=blue!20]
\begin{scope}[scale=0.7,transform shape]
 \node [concept] {\ref{sec:discussion} Discussion}
    child[concept color=red!10,grow=180] {node[concept] {\ref{sub:pipeline} Complementing MPC with scalable GPs}
    child[grow=130] {node[concept] {\ref{sub:batchpipeline} Batch learning}}
    child[grow=190] {node[concept] {\ref{subsec:scalableonlinempc} (Scalable) online learning within the MPC scheme}}      
    }
     child[concept color=red!10,grow=140]  {node[concept] {\ref{sub:solvingMPC} Practical solutions of GP-MPC}}
    child[concept color=red!10,grow=35]  {node[concept] {\ref{sec:alternative_models} Alternative models for the dynamics}
     child[grow=170] {node[concept] {\ref{sub:beyondssm} Beyond state-space models}}
      child[grow=10] {node[concept] {\ref{sub:spatiotemporal} (Spatio-)temporal GPs}}
      child[grow=-35] {node[concept] {\ref{sub:multioutputGP} Non-independent output components}}
    }
    child[concept color=red!10,grow=-12]  {node[concept] {\ref{sub:alternativeuncertaintyprop} Alternative uncertainty quantification}};
\end{scope}
\end{tikzpicture}
\caption{Graphical overview of Section~\ref{sec:discussion}.}
\label{fig:mindmap_sec7}
\end{mdframed}
\end{figure*}

Sections~\ref{sec:scalableGPs}-\ref{sec:safetyguar} have provided an extensive overview of the main choices that can be made regarding the reformulation of problem~\eqref{eq:optimization_problem}, in particular concerning computational aspects (Section~\ref{sec:scalableGPs}), model propagation along the prediction horizon dealing with the presence of uncertainty (Section~\ref{sec:uncertaintyprop}), and constraint handling and available safety guarantees (Section~\ref{sec:safetyguar}).
This section, whose structure is summarized by the mindmap of Figure~\ref{fig:mindmap_sec7}, aims to provide a discussion of both state-of-the-art and possible novel Gaussian process-based MPC approaches combining all of these aspects. We first concentrate in Section~\ref{sub:pipeline} on the scalability-aware methods reviewed in Section~\ref{sec:scalableGPs}, surveying existing learning-based MPC methods that use them, and pointing out some opportunities. Next, Section~\ref{sub:solvingMPC} focuses on practical considerations for actually implementing the Gaussian process-based MPC problem. Then, in Section~\ref{sec:alternative_models} we discuss alternative uses of Gaussian processes to model the dynamics, focusing in particular on latent variable models and spatio-temporal processes. Finally, we lay out possible alternative uncertainty bounds that could be included in the stochastic MPC pipeline in Section~\ref{sub:alternativeuncertaintyprop}.

\subsection{Complementing MPC with scalable Gaussian processes}\label{sub:pipeline}

We now discuss how the ``blocks" presented in Sections~\ref{sec:scalableGPs},~\ref{sec:uncertaintyprop} and ~\ref{sec:safetyguar} can be combined together to formulate the MPC problem of interest. We first focus on Gaussian process regression using batch data, providing a comparison of the main techniques for scalable Gaussian process regression and a review of their combination with uncertainty propagation schemes. Next, we focus on online model updates, highlighting their practical effectiveness against the theoretical challenge they introduce.
We provide a summary of the available papers on applications of Gaussian process-based MPC in Table~\ref{tab:GPMPC_Applications} on page~\pageref{tab:GPMPC_Applications}, where we highlight which kind of Gaussian process model is used, the uncertainty propagation scheme, the availability of closed-loop guarantees, and the kind of application considered. 

\subsubsection{General considerations for batch learning}\label{sub:batchpipeline}
We first discuss the methods reviewed in Section~\ref{sec:scalableGPs} in terms of computational complexity, effectiveness, and theoretical properties.

Overall, the most widely adopted approximation methods in the literature have been (in chronological order): the subset of data (Section~\ref{subsec:subsetdatasingle}), the FITC proposed in~\cite{snelson_sparse_2005} (Section~\ref{sec:induprior}), the variational method (VFE) proposed in~\cite{titsias_variational_2009} (Section~\ref{sub:indupost}), and the sparse-spectrum Gaussian processes~\cite{lazaro-gredilla_sparse_2010} (Section~\ref{sec:ssgp}). A thorough comparison of the two methods based on pseudo-inputs, namely the FITC and VFE, has been carried out in~\cite{bauer_understanding_2016}. The results show that FITC tends to underestimate the noise variance, while VFE overestimates it; when adding pseudo-inputs, the performance analysis shows that FITC might not benefit from it -- and this is also reflected in the large data scenario, in which the FITC, differently from the VFE, is not able to recover the full Gaussian process. 
Overall, as claimed in~\cite{huggins_scalable_2019}, the VFE method is the one that returns the best empirical performance. Its advantage consists in effectively facing over-fitting, as (inducing) parameters are chosen already considering the performance in prediction. However, it is worth noting that the optimization problem needed to retrieve all the parameters tends to be intricate to solve, and practically FITC performance, even if difficult to certify theoretically, turns out to be more effective in practice. Furthermore, as pointed out in~\cite{rossi_sparse_2021}, in principle the predictive distribution of FITC is expected to perform better because the one in VFE is the same as DTC (which is weaker due to the strong structural assumption discussed, e.g., in~\cite{snelson_local_2007}) -- so more reliable hyper-parameter estimation schemes, for instance relying on a full Bayesian treatment as done in~\cite{rossi_sparse_2021,scampicchio_markov_2023}, may be key in improving FITC performance. Another work that frames FITC in the context of Expectation Propagation and aims at contrasting overfitting is~\cite{qi_sparse-posterior_2010}.

FITC has been also compared with subset of data, mixture-of-experts and conjugate-gradient speedups in~\cite{chalupka_framework_2013}, where one of the results is that FITC tends to achieve better prediction accuracy when the number of pseudo-inputs is equal to the cardinality of the data subset. The reported analysis shows that the subset of data is still a valuable option when hyper-parameter tuning time is a priority. The same holds also for the ``hybrid" method that combines subset of data and FITC.\\

Regarding works in which scalable methods are combined with the iterated one-step-ahead uncertainty propagation methods reviewed in Section \ref{sec:uncertaintyprop}, the first results were given in~\cite{groot_multiple-step_2011}, where the computations for moment matching are reported for the FITC approximation method. The analysis was then extended, encompassing also the linearization-based approach, in~\cite{gutjahr_sparse_2012}, where SoR, DTC, and sparse spectrum Gaussian processes were additionally considered. As regards the latter, since the kernel is not the square-exponential, moment matching was not analyzed. Such a gap was then filled in~\cite{pan_prediction_2017}, where also a promising application to MPC is provided, but constraints are not considered. \\

All these considerations hold for models that have been trained offline with batch data. In this scenario, when deploying Gaussian process regression in an MPC scheme, closed-loop guarantees are easier to be obtained, because the model is not updated during operation. However, this set-up does not exploit the full potential of the learning-based approach. In the following, we discuss approaches dealing with streaming data -- consequently, with models that are updated during operation as new observations are gathered.

\subsubsection{(Scalable) online learning within the MPC scheme}\label{subsec:scalableonlinempc}

While online learning improves the accuracy of the model 
as new data are gathered, their change can undermine both control-theoretic, closed-loop guarantees and recursive feasibility of the MPC scheme (see, e.g.,~\cite{bonzanini_learning-based_2022}). As regards the first, in principle, uncertainty bounds based on RKHSs of the type of~\cite{srinivas_information-theoretic_2012} can be still deployed in the streaming setting, given that they depend on the number of data points and not on their value: this was pointed out in~\cite{koller_learning-based_2018} and used with updating full Gaussian processes, but the same argument holds if working with scalable methods using a fixed budget of data. The second is a delicate issue that has to be typically addressed with additional assumptions (e.g., the need for a safe backup strategy in~\cite{koller_learning-based_2018}). Some results for online learning in the context of safe tracking control deploying feedback linearization can be found in~\cite{umlauft_feedback_2020,umlauft_smart_2020}.\\

We now review the MPC approaches in which new trajectory observations are used to update the system description.
The deployment of the full Gaussian process update has been used for systems with slow dynamics, as the computational complexity is a limiting factor in this case. The works~\cite{bradford_nonlinear_2019,bradford_stochastic_2020}, for instance, deal with semi-batch reactors for photosynthetic microorganisms. For the proposed approach, trajectories are sampled  and, at each time-step, the posterior of the Gaussian process is updated using all of the previously observed data. As the method is sampling-based, guarantees are derived for the feasibility probability (see Section~\ref{sub:samplingclosed}), which in turn characterizes the reliability of the resulting back-offs. The only approach using online Gaussian process regression and deriving theoretical guarantees is~\cite{maiworm_online_2021}, where input-to-state stability is proved, and online learning is performed using the evolving Gaussian processes paradigm:  specifically, data inclusion is ruled by a threshold on the prediction error, removal is performed on a temporal basis, and the recursive update of the Cholesky factor is performed according to the method reviewed in~\cite[Appendix B]{osborne_bayesian_2010} (see also~\cite[Chapter 6]{golub_matrix_2013}). All other available MPC approaches entailing \textit{scalable} online Gaussian process regression have been successfully applied, both in simulation and in real-world experiments, but are not complemented with a rigorous theoretical analysis. Furthermore, many of them do not perform uncertainty propagation and consider just the nominal Gaussian process prediction: this is the case of~\cite{bergmann_nonlinear_2022,boedecker_approximate_2014,gandhi_model_2017} for simulation of systems such as heavy-duty Diesel engines and inverted pendula, while~\cite{kim_path_2017} deals with a real skid-steer vehicle (with no constraints).
Overall, the main scalable online methods that have been deployed for MPC (and reviewed in Section \ref{sec:online}) are the evolving Gaussian processes with different data inclusion/removal strategies~\cite{kocijan_modelling_2016}; online FITC~\cite{bijl_online_2015} or its version where pseudo-points are transductively selected~\cite{hewing_cautious_2020}; and the online sparse-spectrum Gaussian processes of~\cite{gijsberts_real-time_2013} -- less used is the variational approach of~\cite{titsias_variational_2009}. We defer a full overview of the available applications combining (online) Gaussian processes with MPC to Table~\ref{tab:GPMPC_Applications} on page \pageref{tab:GPMPC_Applications}.

\subsection{Towards real-time solutions of Gaussian process-based MPC}\label{sub:solvingMPC}

Solving the Gaussian process-based MPC problem~\eqref{eq:optimization_problem} incurs additional computational challenges that complicate its real-time implementation. In this subsection we will discuss the available approaches that address this problem.\\

We start by 
presenting in~\eqref{eq:optimization_problem_solvable} the MPC formulation that is most common among the applications presented in Table~\ref{tab:GPMPC_Applications}. It enables computational tractability and is encapsulated in the following optimal-control problem structure to be solved in a receding-horizon fashion.
\begin{figure*}[h!]
\hrule 
\begin{mini!}|l|[3]{\substack{\{u_{i|k}, \mu^x_{i|k}, \Sigma^x_{i|k} \}}}{\ell_T(\mu^x_{T|k}, \Sigma^x_{T|k}) + \sum_{i=0}^{T-1} \ell_i(\mu^x_{i|k},u_{i|k},\Sigma^x_{i|k}) \label{eq:solvable_cost}}{\label{eq:optimization_problem_solvable}}{}
\addConstraint{\mu^x_{i+1|k} = \rho(u_{i|k}, \mu^x_{i|k}, \Sigma^x_{i|k}), \> i=0,\ldots,T-1 \label{eq:solvable_dynamics_mean}}{}{}
\addConstraint{ \Sigma^x_{i+1|k} = \Phi(u_{i|k}, \mu^x_{i|k}, \Sigma^x_{i|k}), \> i=0,\ldots,T-1 \label{eq:solvable_dynamics_covar}}{}{}
\addConstraint{  h_j( \mu^x_{i|k},u_{i|k}) + \nu_j(u_{i|k}, \mu^x_{i|k}, \Sigma^x_{i|k}) \leq 0, \> i=0,\ldots,T-1, \, j=1,...,n_h \label{eq:solvable_constraints}}{}{}
\addConstraint{\mu^x_{0|k} = x_k \label{eq:solvable_initialcond_mean}}{}{}
\addConstraint{\Sigma^x_{0|k} = 0_{n_x,n_x}
.\label{eq:solvable_initialcond_covar} }{}{}
\end{mini!}
\hrule
\end{figure*}
In such a formulation, we customarily choose to optimize over a control sequence $ u_{0|k},\dots,u_{N-1|k} $ rather than policies. 
Equations~\eqref{eq:solvable_dynamics_mean} and~\eqref{eq:solvable_dynamics_covar} represent the evolution of the mean and covariance of the state obtained from Gaussian process regression along the prediction horizon, given the initial conditions~\eqref{eq:solvable_initialcond_mean} and~\eqref{eq:solvable_initialcond_covar}. 
Such deterministic dynamics are ruled by the functions $\rho \colon \mathbb{R}^{n_u} \times \mathbb{R}^{n_x} \times \mathbb{R}^{n_x \times n_x} \to \mathbb{R}^{n_x}$ and $\Phi \colon \mathbb{R}^{n_u} \times \mathbb{R}^{n_x} \times \mathbb{R}^{n_x \times n_x} \to \mathbb{R}^{n_x \times n_x}$, which
encapsulate the user's choice about what scalable Gaussian process regression technique (see Section~\ref{sec:scalableGPs}) and what uncertainty propagation paradigm (see Section~\ref{sec:uncertaintyprop}) to adopt. Note that the presented formulation relies on the independence assumption investigated in Section~\ref{sec:uncertaintyprop}, and that the general case of non-Markovian dependence (see Section~\ref{subsec:no_indep_ass} and Figure~\ref{fig:dag_independence}) has not yet been applied to receding-horizon MPC problems so far (except for shrinking horizon formulations, see Section~\ref{sub:samplingclosed}, and in particular~\cite{umlauft_scenario-based_2018,bradford_nonlinear_2019}). Finally, the constraint formulation~\eqref{eq:solvable_constraints} is given by a deterministic constraint~${h \colon \mathbb{R}^{n_x}\times \mathbb{R}^{n_u} \to \mathbb{R}^{n_h}}$ (see Section~\ref{sec:problem}, in particular~\eqref{eq:OCconstr}) based on the state mean $\mu^x_{i|k}$ and input $u_{i|k}$, as well as a tightening~$\nu \colon \mathbb{R}^{n_u}\times \mathbb{R}^{n_x}\times \mathbb{R}^{n_x\times n_x} \to \mathbb{R}^{n_h}$, which additionally depends on the predicted state uncertainty $\Sigma^x_{i|k}$. Typically,~$\nu(\cdot,\cdot,\cdot)$ is computed according to arguments based on the inverse cumulative Gaussian distribution function (which is the
typical choice when working with Gaussian variables),  or Chebychev's inequality (which tends to be very conservative, but can be used for any distribution). 
Note also that the constraint is expressed component-wise to allow the specification of different probability levels. \\

\begin{addendumbox*}
\begin{mdframed}[backgroundcolor=gray!10]
    \paragraph*{Example inspired by~\cite{kabzan_learning-based_2019}}
    We now provide the exact expressions involved in the MPC problem~\eqref{eq:optimization_problem_solvable} in the case in which  FITC~\cite{snelson_sparse_2005} is adopted as an approximate Gaussian process, uncertainty propagation is performed using linearization~\cite{girard_gaussian_2002}, and the inverse cumulative Gaussian distribution is used to tighten the constraints.\\

    We make the following assumptions. First, for ease of notation, we consider the case in which $n_x = 1$: for multi-dimensional Gaussian processes with independent components, one-step-ahead means and covariances can be computed  by considering the expressions derived in the remainder of this textbox for each state component. The more general case of correlated components discussed in Section~\ref{sub:multioutputGP} requires a more involved analysis.
    
    Furthermore, we assume that in the dynamics model~\eqref{eq:model} the nominal term $\fnom$ is set to zero and $B_d$ is the identity matrix. We consider this choice for ease of explanation, and refer the general (and notationally more involved) case to~\cite{hewing_cautious_2020}.\\ 
    
    Following Section \ref{sec:induprior}, denote with $\bar{Z}$ the set of (fixed) $M$ inducing points and with $Z$ the training data-set of cardinality $N$. The expressions for the predictive mean and covariance in FITC, evaluated at an arbitrary state-input pair $(x^*,u^*)=z^*$ and using the notation of Section~\ref{sec:background}, read as follows:
    \begin{align*}
&\mu^{{\scriptscriptstyle FITC}}(z^*)=\Ker_{z^*,\bar{Z}}Q^{-1}\Ker_{\bar{Z},Z}(\Lambda + \sigma_w^2I_N)^{-1}Y \notag\\
        &\Sigma^{{\scriptscriptstyle FITC}}(z^*)=\Ker_{z^*,z^*} - \Ker_{z^*,\bar{Z}}(\Ker_{\bar{Z},\bar{Z}}^{-1} - Q^{-1})\Ker_{\bar{Z},z^*} \notag + \sigma_w^2, 
    \end{align*}%
    where $Q = \Ker_{\bar{Z},\bar{Z}} + \Ker_{\bar{Z},Z}(\Lambda + \sigma_w^2 I_N)^{-1}\Ker_{Z,\bar{Z}}$ and $\Lambda$ is a diagonal matrix such that $[\Lambda]_{a,a} = \Ker_{z_a,z_a} - \Ker_{z_a,\bar{Z}}\Ker_{\bar{Z},\bar{Z}}^{-1}\Ker_{\bar{Z},z_a}$ for each $a=1,...,N$ ranging the training points in $Z$.\\

    We now derive the expressions for  $\mu_{i+1|k}^x$ and $\Sigma_{i+1|k}^x$ using the linearization-based approach of~\cite{girard_gaussian_2002} and reviewed in Section~\ref{subsubsec:linearization}. In particular, using Equations~\eqref{eq:linpropagationgen} and~\eqref{eq:linpropspec}, the predicted mean and covariance of the state are computed as follows:
    \begin{subequations}
        \begin{align*}
            &\mu_{i+1|k}^x = \mu^{{\scriptscriptstyle FITC}}(\mu_{i|k}^x,u_i) \\
             &\Sigma^x_{i+1|k} = \Sigma^{{\scriptscriptstyle FITC}}(\mu_{i|k}^x,u_i) 
             + \frac{1}{2}\Tr\Big(\nabla_x^2\Sigma^{{\scriptscriptstyle FITC}}(x,u_i)|_{x=\mu^x_{i|k}}\Sigma^x_{i|k}\Big) 
             + \nabla_x \mu^{{\scriptscriptstyle FITC}}(x,u_i)|_{x=\mu^x_{i|k}} ^{\top}\Sigma^x_{i|k}\nabla_x \mu^{{\scriptscriptstyle FITC}}(x,u_i)|_{x=\mu^x_{i|k}}.
        \end{align*}
    \end{subequations}
Therefore, the equations above yield the exact expressions for the functions $\rho$ and $\Phi$ entering~\eqref{eq:solvable_dynamics_mean} and~\eqref{eq:solvable_dynamics_covar}.\\

Finally, we consider the state and input chance-constraints. We start by linearizing the $j$-th component around the current values $(\mu_{i|k}^x,u_i)$ as 
\begin{align*}
    h_j(x,u_i) &\approx h_j(x,u_i)|_{x=\mu^x_{i|k}} + \nabla_x h_j(x,u_i)|_{x=\mu^x_{i|k}}^{\top}(x-\mu_{i|k}^x).
\end{align*}
    Its mean and covariance, taken with respect to $x$, read as $h_j(\mu_{i|k}^{x},u_i)$ and $\nabla_x h_j(x,u_i)|_{x=\mu^x_{i|k}}^{\top}\Sigma^x_{i|k}\nabla_x h_j(x,u)|_{x=\mu^x_{i|k}}$, respectively. Thus, one can enforce a chance-constraint with given probability level $p_j$ as 
\begin{align*}   
    &h_j(\mu_{i|k}^{x},u_i) + \bar{\alpha} \sqrt{\nabla_x h_j(x,u_i)|_{x=\mu^x_{i|k}}^{\top}\Sigma^x_{i|k}\nabla_x h_j(x,u)|_{x=\mu^x_{i|k}}} \leq 0
\end{align*}
where $\bar{\alpha}$ is chosen as $\bar{\alpha} = \mathbf{\Phi}^{-1}(p_j)$, with $\mathbf{\Phi}$ being the inverse of the cumulative distribution function of a standard (zero-mean and unit-variance) Gaussian random variable. 
By inspection of the inequality written above, we can retrieve the closed-form expression for $\nu_j$ in~\eqref{eq:solvable_constraints}.
\end{mdframed}
\vspace{1em}
\end{addendumbox*}
 
Finally, note that the formulation guaranteeing robustness-in-probability discussed in~\ref{subsec:RIP}, even if it requires set-based computations, can be captured by a problem structure that is similar to the one presented in~\eqref{eq:optimization_problem_solvable}. 
On the other hand,  since~\eqref{eq:optimization_problem_solvable} is a moment-based reformulation of problem~\eqref{eq:optimization_problem}, it does not capture the sampling-based approaches mentioned in  Section~\ref{sub:samplingclosed}. \\

Even if~\eqref{eq:optimization_problem_solvable} gives an implementable problem, it is still computationally intense, mainly due to (i)~the high number of optimization variables and to (ii)~the computation of gradients for the mean and the covariance of the Gaussian process. As regards issue (i), an alternative to the large and sparse problem consists in expressing the variables related to the state as a function of the inputs, resulting in a small and dense problem (scaling as $O(T^3 n_u^3)$ instead of $O(T(n_x+ n_x^2 + n_u)^3)$) -- see, e.g.,~\cite{cao_gaussian_2017-1} for the condensed formulation, and~\cite{cao_gaussian_2017,nghiem_data-driven_2017} for the non-condensed, simultaneous implementation of~\eqref{eq:optimization_problem_solvable}. However, a na\"ive Gaussian process-based MPC program remains computationally prohibitive for fast-sampled applications with longer prediction horizons in both cases, the reason being the high degree of nonlinearity of the condensed formulation, and the large number of optimization variables in the sparse one. 
As regards issue (ii), the evaluation of the first- and second-order derivatives of the Gaussian process posterior mean and covariance is typically very expensive and can constitute a major bottleneck in terms of computation time for Newton-type optimization methods. This hinders for instance approaches based on (probabilistic) differential dynamic programming~\cite{pan_probabilistic_2014,pan_prediction_2017}.

A first way to address these issues consists in simplifying the problem and neglecting the covariance propagation term. In this view, 
a large number of applications 
directly 
enforce
state constraints
on the approximate mean dynamics,
while neglecting the 
state covariance associated with
the Gaussian process model
~\cite{ostafew_learning-based_2014,gandhi_model_2017,torrente_data-driven_2021,picotti_lbmatmpc_2022}.
Improving upon the uncertainty-agnostic constraint formulation,~\cite{jain_learning_2018,maddalena_experimental_2022} employ a \emph{zero-variance method} in terms of a NARX dynamics model formulation, predicting the next state and associated covariance, 
assuming that the past states used for prediction are noise-free.
Another method that avoids propagating the uncertainty associated with state predictions is to consider a Gaussian process model for uncertain output prediction based on deterministic state dynamics~\cite{bergmann_nonlinear_2022,matschek_safe_2023}.

A computationally efficient way to incorporate approximate uncertainty propagation into the optimization problem is to model the state covariance as a constant parameter,
whose value is determined based on the optimal inputs at the last sampling time of the model predictive controller~\cite{hewing_cautious_2018,carron_data-driven_2019,kabzan_learning-based_2019,hewing_cautious_2020}.
In this way, uncertainty-aware constraints can be implemented while avoiding expensive derivative computations 
and retaining the same number of optimization variables as a nominal MPC method, allowing for larger prediction horizons within a simultaneous optimal control problem implementation. 

In a similar fashion, the covariances can also be updated in between solver iterations.
In~\cite{polcz_efficient_2023}, this idea is applied to a linear-parameter-varying reformulation of problem~\eqref{eq:optimization_problem_solvable};
using a linear-parameter-varying embedding of the mean dynamics,
the covariances are thereby treated as scheduling variables and updated between solver iterations. 
In~\cite{lahr_zero-order_2023}, 
updating the covariances between Newton-type solver iterations is formalized
as a tailored Jacobian approximation of the original problem~\eqref{eq:optimization_problem_solvable},
leading to feasible and up-to-date covariance estimates at convergence; 
variants of the resulting
\emph{zero-order optimization}
method 
have been  
implemented in~\cite{ostafew_robust_2016,vaskov_friction-adaptive_2024} for a robotic and autonomous driving application, respectively.

\begin{table*}
\label{tab:GPMPC_Applications}
\begin{center}
    \resizebox{1\textwidth}{!}{ \colorbox{blue!10}{
        \begin{tabular}{|p{2.2em}|p{9em}|p{6em}|p{8em}|p{14em}|p{4.5em}|p{10em}|}
            \hline\hline
            Work & GP model & Uncertainty propagation & Analysis of closed-loop guarantees & Applications & Simulation only & Online\\
            \hline 
            \cite{kocijan_gaussian_2004} & full GP & \cite{girard_gaussian_2002} & - & pH neutralization process & yes & no\\
            \hline
            \cite{likar_predictive_2007} & full GP & \cite{girard_gaussian_2002} & - & gas-liquid separation plant & yes & no \\
            \hline
            \cite{grancharova_explicit_2007} & full GP & \cite{girard_gaussian_2002} & (no constraints) & analytic function & yes & no\\
            \hline
            \cite{maciejowski_fault_2013} & \cite{snelson_sparse_2005} and \cite{hall_modelling_2012} & \cite{quinonero-candela_propagation_2003} & - & fault detection on aircraft benchmark data-sets & yes & no\\
            \hline
            \cite{boedecker_approximate_2014} & SoR \cite{smola_sparse_2000} & (none) & (no constraints) & inverted pendulum & yes & yes (posterior + hyper-parameter updates)\\
            \hline
            \cite{ostafew_learning-based_2014} &  full GP & (none) & (no constraints) & outdoor robot & no & no \\
            \hline
            \cite{ostafew_robust_2016}  & full GP & \cite{julier_unscented_2004} & - & outdoor robot & no & no \\
            \hline
            \cite{gandhi_model_2017} & sparse-spectrum GPs \cite{gijsberts_real-time_2013} & (none) & - & simulation & yes & yes (posterior update)\\
            \hline
            \cite{kim_path_2017} & \cite{bijl_online_2015} & (none) & (no constraints) & skid-steer vehicle & no & yes (posterior updates)\\
            \hline       
            \cite{nghiem_data-driven_2017} & full GP & \cite{girard_gaussian_2002} & - & building energy management & no & no\\
            \hline
            \cite{pan_prediction_2017} & sparse-spectrum GPs \cite{gijsberts_real-time_2013} & \cite{quinonero-candela_propagation_2003} & (no constraints) & Puma-560 robotic arm, quadrotor, autonomous car & no & yes (posterior updates)\\ 
            \hline
            \cite{jain_learning_2018} & subset of data & (none) & - & building energy management & yes & yes (every two weeks)\\
            \hline
            \cite{bradford_nonlinear_2019, bradford_stochastic_2020} & full GP  & trajectory sampling & $\checkmark$ (feasibility probability) & semi-batch reactor & yes & yes (posterior updates) \\
            \hline       
            \cite{carron_data-driven_2019} & \cite{titsias_variational_2009} & \cite{girard_gaussian_2002} & - & robotic arm & no & yes (kalman filter)\\
            \hline
            \cite{kabzan_learning-based_2019} & FITC \cite{snelson_sparse_2005}, with dictionary selected as in \cite{csato_sparse_2002} & \cite{girard_gaussian_2002} & - & autonomous racing & no & yes (posterior + hyper-parameters updates)\\
            \hline      
            \cite{hewing_cautious_2020} & transductive FITC \cite{snelson_sparse_2005} & \cite{girard_gaussian_2002} & - & autonomous racing & no & no\\
            \hline 
            \cite{bradford_combining_2021} & Polynomial Chaos Expansion & \cite{quinonero-candela_propagation_2003} & - & semi-batch reactor & yes & no \\
            \hline
            \cite{maiworm_online_2021} & evolving GPs \cite{kocijan_modelling_2016} & - & $\checkmark$ (ISS, not on constraint satisfaction) & simulation (stirred-tank reactor) & yes & yes (rule-based - decreasing Lyapunov function)\\
            \hline       
            \cite{mannel_iterative_2021} & full GP | \cite{sarkka_gaussian_2019} & \cite{girard_gaussian_2002}, Kalman filter & - & (simulation) pressure control in mechanical ventilator & yes & yes (posterior updates)\\
            \hline
            \cite{ota_data-efficient_2021} & full GP & (none) & - & circular maze system & no & no\\
            \hline 
            \cite{torrente_data-driven_2021} & FITC \cite{snelson_sparse_2005} & \cite{girard_gaussian_2002} & (no constraints) & quadrotor & no & no \\ 
            \hline
            \cite{wabersich_nonlinear_2021} & full GP & (none) & \checkmark (robust in probability) & approximately linear 10-dimensional quadrotor & yes & no \\
            \hline
            \cite{bergmann_nonlinear_2022} & evolving GPs \cite{kocijan_modelling_2016} & (none) & - & heavy-duty Diesel engine & yes & yes (tailored)\\
            \hline
            \cite{bonzanini_learning-based_2022} & full GP & (none) & \checkmark (bounded support) & atmospheric pressure plasma jet & yes & no \\
            \hline
            \cite{gruner_recursively_2022} &  \cite{hartikainen_kalman_2010} & Kalman filter & $\checkmark$ & quadrotor & yes & yes (posterior updates)\\
            \hline
            \cite{maddalena_experimental_2022,di_natale_lessons_2022} & full GP & (none) & - & building energy management & no & no\\  
            \hline
            \cite{picotti_lbmatmpc_2022} & subset of data, \cite{gardner_gpytorch_2018} & (none) & - & Furuta pendulum & no & no \\ 
            \hline
            \cite{schmid_real-time_2022} & \cite{hartikainen_kalman_2010} & Kalman filter & $\checkmark$ & UAVs & yes & yes (posterior + hyperparameter updates)\\
            \hline
            \cite{vaskov_friction-adaptive_2022,vaskov_friction-adaptive_2024} & \cite{solin_hilbert_2020} & \cite{girard_gaussian_2002} & - & CarSim, trajectory tracking with unknown tire-friction function & yes & yes (posterior update - particle filter)\\
            \hline
            \cite{chen_gaussian-process-based_2023} & full GP & \cite{girard_gaussian_2002} & (no constraints, stability analysis) & Furuta pendulum, bikebot & no & no \\ 
            \hline
            \cite{ma_model-and_2023} & full GP & trajectory sampling & $\checkmark$ (feasibility probability) & simulation (tank system) & yes & no\\ 
            \hline
            \cite{panetsos_gp-based_2024} & SoR~\cite{smola_sparse_2000} with time-varying weights & (none) & - & UAVs with suspended loads & no & yes (posterior updates) \\
            \hline       
        \end{tabular}
    }}
\end{center}
\caption{Applications of Gaussian process-based MPC}
\end{table*}

\subsection{Alternative models for the dynamics}
\label{sec:alternative_models}
Up to this point, the dynamics model that has been considered for application in MPC is the one presented in~\eqref{eq:model}, treating all of the output components of the map $g$ independently (i.e., each of them as a separate Gaussian process). We now discuss some alternatives that could be considered for MPC formulations. Specifically, we discuss the use of Nonlinear Output Error (NOE) structures~\cite{kocijan_modelling_2016} and latent variable models mentioned in Section~\ref{subsubsec:latent_state_opt}; the case in which the map $g$ is time-dependent, giving thus an overview of the so-called \textit{spatio-temporal} Gaussian processes; and ways for overcoming the independence assumption in the output components of $g$.

\subsubsection{Beyond state-space models}\label{sub:beyondssm}
Section~\ref{sec:gp4dyn} provided an extensive overview regarding the possibilities for representing nonlinear system dynamics with Gaussian processes. However, the setup considered in this paper -- and in most of the presented works -- is the one described in Remark~\ref{rmk:gp_digression}, i.e.,  assuming that the states can be perfectly observed, the dynamics can be formulated as a state-space model~\eqref{eq:statespacegen} with emission map equal to the identity and without measurement noise, or as a NARX model~\eqref{eq:narx} with lags $\tau_u = \tau_y = 1$. This choice is in line with the typical use of a model within an MPC framework, which serves solely for open-loop predictions, thus remaining unaffected by any potential measurement noise present in the closed-loop system. There are, however, two issues with this choice: first, despite using Gaussian processes for estimating open-loop predictive models, these are generally trained on closed-loop data, and in practice, it is difficult to find applications where state measurements are not affected by noise at all. Second, even in the case of perfect state measurements, propagating the uncertainty over the prediction horizon may still require evaluating Gaussian processes at uncertain inputs, which, as observed in Section~\ref{sec:uncertaintyprop}, is typically considered only in the evaluation phase and not during training. Both these issues could be solved by considering NOE models~\cite[Chapter 2.3.1]{kocijan_modelling_2016} with no process noise but only measurement noise, and training the Gaussian process on the \textit{simulation} error rather than the \textit{prediction} error~\cite{kocijan_output-error_2011}. In contrast to NARX models, which focus on one-step-ahead prediction, NOE models perform multi-step predictions, 
hence falling in the category of \textit{direct methods} (see Section~\ref{sec:uncertaintyprop}). While this avoids the typical error accumulation of NARX models due to iterative one-step-ahead predictions~\cite{kocijan_output-error_2011}, the drawback is a more complex training procedure~\cite{beckers_prediction_2022} and poor scalability to longer prediction horizons.\\

A computationally cheaper, alternative choice that can deal with the issues mentioned above is to consider latent, i.e. unobserved, variables already in the modeling phase, which is compatible with both auto-regressive~\cite{souza_learning_2021} or state-space~\cite{ialongo_overcoming_2019} representations. These models are inherently designed to deal with uncertainty in the input locations by employing approximate training and inference techniques. It would be interesting to deepen the combination of Gaussian process state-space models such as~\cite{curi_structured_2020, fan_free-form_2023, buisson-fenet_joint_2021} that exploit variational inference with an MPC framework, which, to the best of our knowledge, until now has been only studied in the context of meta-learning for MPC~\cite{saemundsson_meta_2018}.

\subsubsection{(Spatio-)temporal Gaussian processes}\label{sub:spatiotemporal}
This paper has focused on static maps — that is, the function $g(x_i, u_i)$ in the dynamics equation \eqref{eq:model} (compactly denoted as $g(z_i)$ as in Section \ref{sec:background}) is assumed not to depend on time but only on “spatial” information. This is not sufficient in contexts where the environment is subject to change, as for example in traffic or pollution data. 
The theory of spatio-temporal processes~\cite{cressie_statistics_2011} allows for including time dependency, i.e., by considering $g(z_i;t_k)$.  
A na\"ive application of Gaussian process regression in this context would impractically scale cubically in the product of the number of both temporal and spatial data-points. 
In the following, we review the approaches that have been proposed to address this issue.\\

The first attempts have been mainly studied within the geostatistics community and rely on particular model structures. For example, in~\cite{mardia_kriged_1998} and~\cite{cressie_space-time_2014} the time-varying nature of the spatial field is conveyed by expressing it as a linear combination of common fields, and the coefficients follow Markovian temporal dynamics. An alternative formulation is presented in~\cite{reece_introduction_2010}, where the hyper-parameters entering the spatial kernel are time-varying, following again Markovian dynamics. Spatio-temporal models are treated with factor analysis in~\cite{luttinen_variational_2009}, where variational inference is deployed to address the intractability of the posterior. This set-up has also been considered for traffic-data imputation and kriging in~\cite{lei_bayesian_2022}, where an MCMC scheme is used for more reliable hyper-parameters learning.\\

An alternative, elegant approach consists in reformulating spatio-temporal Gaussian regression as a Kalman filtering/smoothing problem. This is done by building upon the temporal-only case in~\cite{hartikainen_kalman_2010}, where a representation of the Gaussian process by means of a stochastic differential equation is presented. Such a result is obtained by performing spectral factorisation of the power spectral density computed from the Gaussian process covariance. If the power spectral density is rational (and this is the case, e.g., for kernels of the Matérn class), the obtained state-space model is a stochastic differential equation of evolution type; otherwise, the model would be a stochastic pseudo-differential equation or a fractional stochastic equation~\cite{sarkka_infinite-dimensional_2012}. Thanks to such a reformulation, Gaussian process regression becomes equivalent to Kalman filtering/smoothing. Works that deploy such an approach for the spatio-temporal case are, e.g.,~\cite{sarkka_infinite-dimensional_2012,sarkka_spatiotemporal_2013}. The reformulation of Gaussian process regression as a stochastic differential equation can be already traced back to~\cite{leith_gaussian_2004} for power load forecasting, and is analysed in~\cite{lindgren_explicit_2011}; see also the recent review~\cite{lindgren_spde_2022} that contains an account of both theoretical and applicative advances in this topic.\\

The approaches mentioned above consider kernels that are non-separable in space and time. Since they are operating with infinite-dimensional Kalman filters and smoothers, these require approximations based, e.g., on Laplace eigenfunctions. On the other hand, as pointed out in~\cite{sarkka_spatiotemporal_2013}, if the kernel is separable, then more efficient schemes can be devised. A work presented in this direction is~\cite{todescato_efficient_2020}, where an exact and computationally feasible solution for the space-time Gaussian process regression is derived and analysed. The approach therein presented has also been extended in~\cite{zhang_efficient_2023}, where computational complexity is further improved by leveraging the Kronecker product structure of the kernel. Covariance factorization is also exploited in~\cite{solin_modeling_2018}, where spatio-temporal Gaussian processes are deployed to estimate time-varying magnetic fields, and computations for the spatial component are further sped up thanks to the Laplace operator kernel eigen-decomposition studied in~\cite{solin_hilbert_2020}. Other approaches to speed up computations applied to factorized covariance matrices are tapering~\cite{luttinen_efficient_2012} and subset of data~\cite{xu_mobile_2011}. Spatio-temporal Gaussian processes with kernel representation as stochastic differential equations have also been complemented with other scalable methods for the spatial component, also using expectation propagation~\cite{minka_family_2001} to deal with non-Gaussian likelihoods: for instance,~\cite{hartikainen_sparse_2011}, deploying the sparse prior approximation in~\cite{snelson_local_2007}, and~\cite{hamelijnck_spatio-temporal_2021}  (which builds upon~\cite{wilkinson_state_2020}), deals with the spatial component using variational inference and a further mean-field assumption of independence. Notably, this last work provides a further speed-up also in the temporal component thanks to natural gradients and parallel computing.\\

So far, there has been no direct application of spatio-temporal Gaussian process regression to (model predictive) control, and the derivation of formal guarantees appears to be a challenging open task. This is not the case when one is interested in \textit{temporal} processes only, where the dynamics take the form
\begin{align}
    x_{i+1} = \fnom(x_i,u_i) + g(t_i), \label{eq:temporalGPlfm}&\\   \text{with }\fnom(x_i,u_i) \text{ linear.}& \notag
\end{align}
This kind of model has been incorporated in stochastic MPC schemes in~\cite{klenske_gaussian_2016} to deal with periodic disturbances, and in~\cite{wang_stochastic_2016} for an application to drinking water networks. When model~\eqref{eq:temporalGPlfm} is combined with the Gaussian process representation of~\cite{hartikainen_kalman_2010}, one speaks of \textit{latent force models}, and the recursive feasibility of the resulting stochastic MPC scheme is studied in~\cite{gruner_recursively_2022} (see also Section~\ref{sub:additionalclosed}).

\subsubsection{Non-independent output components}\label{sub:multioutputGP}

As pointed out in Section~\ref{sec:problem}, the most common modeling choice for the function $g \colon \mathbb{R}^{n_x+n_u} \to \mathbb{R}^{n_d}$ in the dynamics~\eqref{eq:model} consists in treating each of the $n_d$ output dimensions independently, namely to consider $n_d$ separate Gaussian processes. In this subsection, we review some of the available alternatives that have been derived mostly in the context of geostatistics and multi-task learning~\cite{dowd_many_2024,bonilla_multi-task_2007}. Note that these enable the generalization of the dynamics model~\eqref{eq:model} to the case of noises presenting component-wise correlation.\\

To account for possible correlation among outputs, a first option consists in modelling each component as a linear combination of latent functions. This approach is known as ``linear coregionalization"~\cite{journel_mining_1978} in geostatistics, and has been further investigated in the machine learning community, e.g., in~\cite{teh_semiparametric_2005}. A further development consists in the so-called convolution processes, where the linear combination is substituted by a convolution with a smoothing kernel~\cite{higdon_process-convolution_1998,higdon_space_2002}. The advantage of convolution processes is that they allow the integration of prior information from physical models, such as ordinary differential equations, into the covariance function~\cite{alvarez_latent_2009}.

The methods mentioned above are parametric, meaning that correlation is given by a particular user-chosen, hard-coded model. An alternative would be to resort to the theory of vector-valued Gaussian processes, or multi-output RKHS~\cite[Chapter 4]{alvarez_kernels_2012}. These amount to encoding correlation by suitable choices of the kernels.\\

The methods mentioned above are typically encumbered by high computational complexity; furthermore, when correlation is encoded in a non-parametric way, it is typically ruled by a large number of hyper-parameters, whose tuning might become difficult to perform. These are the main factors that limit the application of multi-output Gaussian process regression to MPC problems. 
Promising works to scale up these models can be found in~\cite{alvarez_sparse_2008,alvarez_efficient_2010,alvarez_computationally_2011} for the parametric options; as for the non-parametric case, any (combination) of the methods reviewed in Section~\ref{sec:scalableGPs} could be applied.

\subsection{Alternative uncertainty quantification approaches}\label{sub:alternativeuncertaintyprop}
The overview in Section~\ref{sec:safetyguar} focused on the approaches proposed in the literature for handling (and ultimately satisfying) the chance constraints in problem ($\mathcal{P}$). Their main differences depend on the perspective adopted when quantifying the uncertainty associated with Gaussian processes (see Remark~\ref{rmk:uncertainty_paradigms}). Whether outliers are disregarded, or probabilities are factored in, is a decision that dictates the outcome of the closed-loop analysis. The goal of this section is to provide alternative directions for quantifying the uncertainty that were not explored yet in the context of MPC, but that are showing promise in other control frameworks making use of Gaussian process regression.\\

Many of the results discussed in Section~\ref{subsec:RIP} are based on the use of an RKHS uniform error bound. One key challenge is determining an upper bound for the RKHS norm of the unknown function $g$, which is an assumption that is hard to fulfill: while there have been efforts to address this issue, it remains largely unsolved, and verifying the RKHS norm bound assumption using data is unclear~\cite{lederer_gaussian_2023}. For this reason, this bound has been employed with heuristics, such as using a constant scaling factor for the Gaussian process standard deviation~\cite{berkenkamp_safe_2017}. However, this approach raises concerns about the lack of theoretical guarantees for safety-critical control applications. An alternative is proposed in~\cite{lederer_uniform_2019}, where a new Bayesian uniform error bound is computed by exploiting the probabilistic nature of Gaussian processes rather than functional properties, i.e., the unknown function is not deterministic and does not belong to an RKHS. 
The primary benefit of considering this viewpoint lies in the fact that the effectiveness of RKHS bounds diminishes as the kernel becomes smoother, since the RKHS associated with a covariance kernel is typically smaller compared to the support of the prior distribution of a Gaussian process~\cite{vaart_information_2011}. However, the assumption of knowing the prior distribution containing the unknown function to be learned can be challenging to satisfy. This can be addressed, for instance, thanks to the availability of calibration methods, e.g.~\cite{kuleshov_accurate_2018}. 

Several extensions to the work in~\cite{lederer_uniform_2019} were proposed. For instance,~\cite{lederer_gaussian_2021} addresses the question of scalability by using an aggregation of local Gaussian processes, which inherit the safety guarantees of full Gaussian processes, but can be processed more efficiently in model-based control frameworks. The results obtained in~\cite{capone_gaussian_2022} consider the case in which the 
model hyper-parameters are not known a-priori, while still being able to determine guarantees for backstepping control~\cite{capone_backstepping_2019}. A reformulation of the bound is proposed in~\cite{papadimitriou_control_2022} via a Gaussian process generalization based on Student-t process regression, which is looser than~\cite{lederer_uniform_2019} but is more robust to outliers. Additionally, the work in~\cite{lederer_gp3_2020} addresses the complexity of estimating Lipschitz constants for Gaussian process mean functions. \\

Both RKHS-based and Bayesian uniform error bounds have their limitations, and their suitability depends on application-specific requirements: for instance,  
when robust-(in-probability) safety guarantees are required, then non-Bayesian uncertainty bounds should be employed~\cite{fiedler_practical_2021}. We now review some promising results available in the literature for the noise-free, bounded noise, and additive unbounded noise cases.

 In the noise-free case, a promising kernel-based approach worth mentioning is~\cite{thorpe_data-driven_2022}, where the authors make use of distributional kernel embeddings~\cite{muandet_kernel_2017} for reformulating the joint chance constraint relative to the entire state trajectory as a linear operation in an RKHS. The challenge consists in finding closed-form expressions of such kernel embeddings, which are generally not available. For this reason, a sampling-based approximation is introduced to allow a practical implementation. Since this is the only approximation involved, it would be interesting to quantify the finite-sample error between the true and empirical embeddings. 
 
 When the system is affected by bounded noise, a promising approach in this context is the kernel-based MPC approach in~\cite{maddalena_kpc_2021}. To propagate uncertainty, the authors employ multi-step predictors, 
which can be considered as an example of a direct uncertainty propagation method (see Section~\ref{sec:uncertaintyprop}). The main benefit of this choice lies in reduced conservatism of the constructed bounds, however, it remains unclear how to provide guarantees in a receding-horizon implementation, in particular recursive feasibility, due to incompatible open-loop models, as well as addressing scalability for long prediction horizons. Furthermore, it is not obvious if reduced open-loop conservatism translates into better closed-loop performance (for a thorough analysis of multi-step predictors see~\cite{kohler_state_2022}). 

In the presence of additive unbounded noise, some results are available for linear systems, e.g., the work in~\cite{kohler_state_2022} provides a constraint tightening for stochastic predictive control affected by parametric uncertainty, while~\cite{fiedler_probabilistic_2023} proposes probabilistic multi-step predictors for a linear system with noisy state measurements and unknown process and measurement noise covariances. However, the case of general nonlinear models is quite challenging to address. One could resort to distribution-free results leveraging conformal prediction (see Remark~\ref{rmk:uncertainty_paradigms}). Works that deploy such a tool for MPC are~\cite{chen_reactive_2021,lindemann_safe_2023}. They address the problem of motion planning, with the goal of designing a control action for a system, whose dynamics are perfectly known, that is interacting with some external agents. In this context, the behavior of these agents is predicted with neural networks, and confidence bounds are computed with conformal prediction and included in collision avoidance constraints. Overall, the context is thus different from the stochastic MPC considered in this paper, but the tool is interesting and could have some potential. \\

In this section, we have mostly considered works dealing with quantifying the uncertainty associated with the learned model, with few exceptions also addressing potential model mismatch~\cite{fiedler_practical_2021, capone_gaussian_2022}. However, another error source derives from indirect uncertainty propagation methods leveraging one-step-ahead predictions. The authors in~\cite{polymenakos_safety_2020}  provide a formal bound considering any deterministic mapping employed for approximating one-step-ahead predictions such as, e.g., moment matching. In the numerical simulations, such bound is added to enhance Safe PILCO~\cite{polymenakos_safe_2019}, i.e., a safety framework already acting upon PILCO~\cite{deisenroth_pilco_2011}, showing that for systems where moment matching is severely underestimating the propagated uncertainty, this additional bound improves safety. It would be interesting to see how it can be exploited in Gaussian process-based MPC approaches. 

\vspace{-5ex}

\begin{strip}
\section*{CONCLUSIONS}


\begin{mdframed}[backgroundcolor=yellow!10,innerleftmargin=0.7em,innerrightmargin=0.7em]
    \vspace{-1.7ex}
   \subsection*{Summary}
    \begin{itemize}
    \item Synopsis on Gaussian process regression (static and dynamical systems);
    \item Problem formulation: stochastic optimal control and its approximation via MPC;
    \item Survey on methods to scale up Gaussian processes (batch- and streaming data);
    \item Review on uncertainty propagation paradigms;
    \item Presentation of set-ups addressing closed-loop safety guarantees;
    \item Focus on available approaches combining scalable Gaussian processes, uncertainty propagation, and closed-loop guarantees, and their implementation in a control pipeline;
    \item Outline of alternative choices for the dynamics model;
    \item Presentation of other promising uncertainty quantification paradigms.
\end{itemize}
\end{mdframed}



\begin{mdframed}[backgroundcolor=orange!10,innerleftmargin=0.7em, innerrightmargin=0.7em]
\vspace{-1.7ex}
\subsection*{Challenges}
\begin{itemize}
\item Complementing the empirical effectiveness of Gaussian process-based MPC with more theoretical results on closed-loop guarantees, possibly relying on yet unused and available alternative uncertainty bounds for Gaussian process estimates.

\item Rigorously addressing the issue of performing online learning within the control loop.


        \item Integrating uncertainty bounds
        derived from the approximations in scalable Gaussian process regression and uncertainty propagations, and deploying them
        for back-off computation.

        \item Providing new insights and results in stochastic MPC, potentially allowing for an easier analysis of the control scheme, especially in view of closed-loop guarantees. A possible route can be given by modelling the dynamics with latent variables models.
    \end{itemize}
\end{mdframed} 
\end{strip}


\bibliographystyle{elsarticle-num-names} 
\bibliography{full_references}

\onecolumn
\tableofcontents

\twocolumn





\end{document}